\documentclass[11pt]{amsart}
\usepackage[margin=1in]{geometry}
\usepackage{amssymb}
\usepackage{amsthm}
\usepackage{amsfonts}
\usepackage{float}
\usepackage{booktabs}
\usepackage{amsmath}
\usepackage{graphicx}
\usepackage{caption}
\usepackage[ruled,vlined]{algorithm2e}
\usepackage{tikz}
\usepackage{hyperref}
\hypersetup{
    colorlinks=true,       
    linkcolor=red,        
    citecolor=blue,       
}
\usepackage{placeins} 
\usepackage{subcaption}
\definecolor{nodeblue}{HTML}{83cbeb}
\definecolor{edgegreen}{HTML}{00b050}
\definecolor{nodeoutline}{HTML}{3c697e}

\newtheorem{theorem}{Theorem}
\newtheorem{corollary}{Corollary}
\newtheorem{lemma}{Lemma}
\newtheorem{definition}{Definition}
\newtheorem{assumption}[theorem]{Assumption} 

\newtheorem{proposition}{Proposition}

\newtheorem{example}{Example}
\newtheorem{remark}{Remark}

\newcommand{\skel}[1]{\operatorname{Skel}(#1)}

\newcommand{\G}{\mathcal{G}}
\newcommand{\Gc}{\mathfrak{G}}

\newcommand{\coverby}{\lessdot}
\newcommand{\cover}{\gtrdot}
\DeclareMathOperator*{\argmin}{arg\,min}
\DeclareMathOperator*{\argmax}{arg\,max}
\usepackage[comma, sort&compress]{natbib} 

\let\oldsection\section
\renewcommand{\section}{\vspace{0.2in}\oldsection}

\let\oldsubsection\subsection
\renewcommand{\subsection}{\vspace{0.05in}\oldsubsection}

\let\oldsubsubsection\subsubsection
\renewcommand{\subsubsection}{\vspace{0.05in}\oldsubsubsection}

 \usepackage[foot]{amsaddr}
\title[Confidence regions over Bayesian networks]{Interpretable confidence regions for causal discovery}

\date{}

\author[J. Kang]{Jian Kang $^{1,*}$}
\address{$^1$ Department of Statistics, University of Washington}
\email{jian004@uw.edu}

\author[A. Mazumder]{Arnav Mazumder $^{2,*}$}
\address{$^2$ Department of Computer Science, University of Washington }
\email{am2107@uw.edu}

\author[G. C. Lembo]{Giovanni Carter Lembo $^3$}
\address{$^3$ Department of Applied Mathematics, University of Washington }
\email{gclembo@uw.edu}

\author[A. Taeb]{Armeen Taeb $^{1}$}
\email{ataeb@uw.edu}

\address{$^*$ These authors contributed equally to this work.}
\begin{document}

\begin{abstract}
Most work in causal discovery focuses on point estimation of the underlying causal graph, while uncertainty in the learned structure has received substantially less attention. We develop confidence regions for the completed partially directed acyclic graph (CPDAG) identified from observational data. Our approach builds on perturbed PC \citep{chang2026post}, which generates plausible CPDAGs by perturbing the conditional independence tests underlying the widely used PC algorithm. The threshold used in these tests controls how perturbations translate into variation across the resulting CPDAGs. We propose a new thresholding strategy that suppresses spurious variation across runs while preserving asymptotic coverage, producing a substantially tighter collection of plausible CPDAGs. This collection is summarized through confidence balls and model confidence intervals constructed from the partially ordered set of CPDAGs under model inclusion. We also develop new theory and algorithms for the exact computation of these regions. Simulations show that our confidence regions attain good coverage while being substantially smaller and more interpretable than the original perturbed PC collections.

\end{abstract}
\maketitle
\vspace{-0.1in}

\section{Introduction}
A directed acyclic graph (DAG) provides an interpretable representation of causal relationships among a set of variables. Nodes represent variables, while directed edges represent direct causal relationships. The DAG also encodes conditional independence relationships implied by the causal structure \citep{spirtes2000causation,pearl2009causality,PetersJanzingSchoelkopf2017}. Causal DAGs have been used in many scientific applications, including genomics and molecular biology \citep{Friedman2000UsingBN,Sachs2005CausalPN}, neuroscience \citep{Smith2011NetworkMM}, epidemiology \citep{Greenland1999CausalDF,Hernan2020CausalInference}, and climate science \citep{Runge2019InferringCausation}.

In practice, the underlying DAG is typically unknown and must be learned from data, a task known as causal discovery. Observational data generally identify only a Markov equivalence class of DAGs rather than a unique DAG. A completed partially directed acyclic graph (CPDAG) represents this equivalence class, with directed edges for identifiable orientations and undirected edges for unresolved ones \citep{Verma1990EquivalenceAS}. Thus, causal discovery from observational data often targets the CPDAG.

A large body of work focuses on obtaining point estimates of the underlying CPDAG. Constraint-based methods, most prominently the PC algorithm and its variants, recover the graph through conditional independence testing  \citep{spirtes2000causation,kalisch2007estimating,Colombo2012OrderindependentCC,Harris2013PCAF,Ramsey2006Conservative}. Score-based methods instead optimize a score over DAGs or Markov equivalence classes, with prominent examples including greedy equivalence search and its variants \citep{chickering2002optimal,Chickering2002FindingOB,Huang2018Generalized}, as well as exact search methods based on dynamic programming and shortest-path search \citep{silander2012simple,Yuan2013ShortestPath}. Hybrid methods combine constraint-based screening with score-based search \citep{tsamardinos2006max,nandy2018high}. Other score-based approaches formulate this optimization problem explicitly using linear or integer programming \citep{Bartlett17discrete,jaakkola10a,cussens2017bayesian,cussens2017polyhedral,Manzour21,kucukyavuz2022consistent,xu2025integer,Zhang2025ConvexMP}, or continuous optimization \citep{zheng2018dags,yu2019dag}. See \citet{glymour2019review} for a broader review of causal discovery methods.

Despite this extensive literature on point estimation, quantifying uncertainty in the learned graph remains relatively understudied. Causal discovery procedures can be pointwise consistent but not uniformly consistent under the usual Markov and faithfulness assumptions \citep{robins2003uniform,zhang2003strong}. Uniform guarantees require stronger separation conditions, such as strong faithfulness, that rule out distributions arbitrarily close to unfaithfulness but can be highly restrictive in practice \citep{uhler2013geometry,sondhi2019reduced}. A single estimated graph may therefore mask substantial uncertainty about the underlying causal structure, motivating confidence regions with formal coverage guarantees.

Several approaches quantify uncertainty in graphical structure. Bayesian methods retain collections of high posterior graphical models \citep{friedman1999data,friedman2003being,kuipers2017partition,Niinimaki2016StructureDI}, but these summaries depend on the prior and model specification and do not provide frequentist coverage guarantees for the underlying graph. Frequentist approaches have largely focused on confidence regions for causal effects that account for uncertainty in the underlying causal structure \citep{strieder2021confidence,strieder2023confidence,strieder2024dual}. Relatedly, \citet{gradu2025valid} develop post-selection inference for a causal effect defined relative to the selected graph rather than the true graph. More directly related to structural uncertainty, \citet{wang2026confidence} construct confidence sets for causal orderings and summarize them through lower and upper bounds on ancestral relations. Their analysis assumes that the underlying DAG is identifiable. None of these works construct frequentist confidence regions with coverage guarantees for the true CPDAG.

The \emph{perturbed PC} procedure of \citet{chang2026post}, which we build on in this paper, generates a collection of plausible CPDAGs with asymptotic coverage guarantees. The procedure constructs this collection by repeatedly perturbing the conditional independence tests used by PC. Each perturbed test compares a noisy Fisher-transformed sample partial correlation with a threshold to decide whether to reject conditional independence. In the original method, the CPDAG collection serves primarily as an intermediate object for constructing confidence intervals for causal effects rather than as a direct summary of structural uncertainty. Our focus is different: we study the CPDAG collection itself and seek confidence regions that provide coverage for the true CPDAG while remaining compact and interpretable. This shift in focus requires controlling spurious variation across the entire collection while still retaining runs that detect weak signals.

\subsection{Our contributions}
We first provide a theoretical analysis of perturbed PC across different thresholds, perturbation scales, and numbers of resamples. The threshold of \cite{chang2026post} is of order $\log n /M$, where $n$ is the sample size and $M$ is the number of perturbed runs, so it decreases as $M$ grows. This decreasing threshold helps weak signals survive in at least one run, but it also allows increasingly extreme null statistics to pass the threshold. Our proposed threshold instead grows on the order of $\sqrt{\log M}$, matching the scale of the largest perturbations under the null. This larger threshold filters out most null statistics across runs while still allowing weak nonzero signals to pass in at least one run. This tradeoff preserves asymptotic coverage without requiring a signal strength condition such as strong faithfulness and produces substantially tighter CPDAG collections.

We next develop interpretable summaries of the perturbed PC collection. A large collection of plausible CPDAGs can be difficult to interpret directly, so we seek regions that capture their structural extent in a concise form. We exploit the partially ordered set (poset) of CPDAGs under model inclusion for this purpose. The poset induces a shortest path distance, which we use to construct confidence balls centered at the original PC estimate with radius chosen to contain the collection. Its order structure also defines confidence intervals through common lower and upper CPDAGs that bound every graph in the collection. These regions provide concise and interpretable summaries of structural uncertainty while retaining the coverage of the underlying collection.

Finally, we develop new computational theory for constructing both confidence balls and intervals. The confidence ball calculations rely on the shortest path distance, for which we derive stronger lower and upper bounds than those in \cite{taeb2025distance}. We also show that the distance decomposes exactly across disconnected components, reducing large shortest path problems to smaller ones. For polytrees, we obtain an exact characterization that substantially simplifies the shortest path search. These structural results also sharpen the bounds and search procedures for general pairs of CPDAGs, even when the graphs are neither disconnected nor polytrees. The confidence interval calculations require common lower and upper CPDAGs, for which we derive new bounds. Together with new restrictions on shortest paths, these results substantially reduce the search space and speed up exact computation. These tools are motivated by confidence region construction but apply more broadly to CPDAG comparison, including benchmarking causal discovery methods and sensitivity analysis.

\begin{figure}[h!]
\centering

\includegraphics[width=0.43\textwidth]{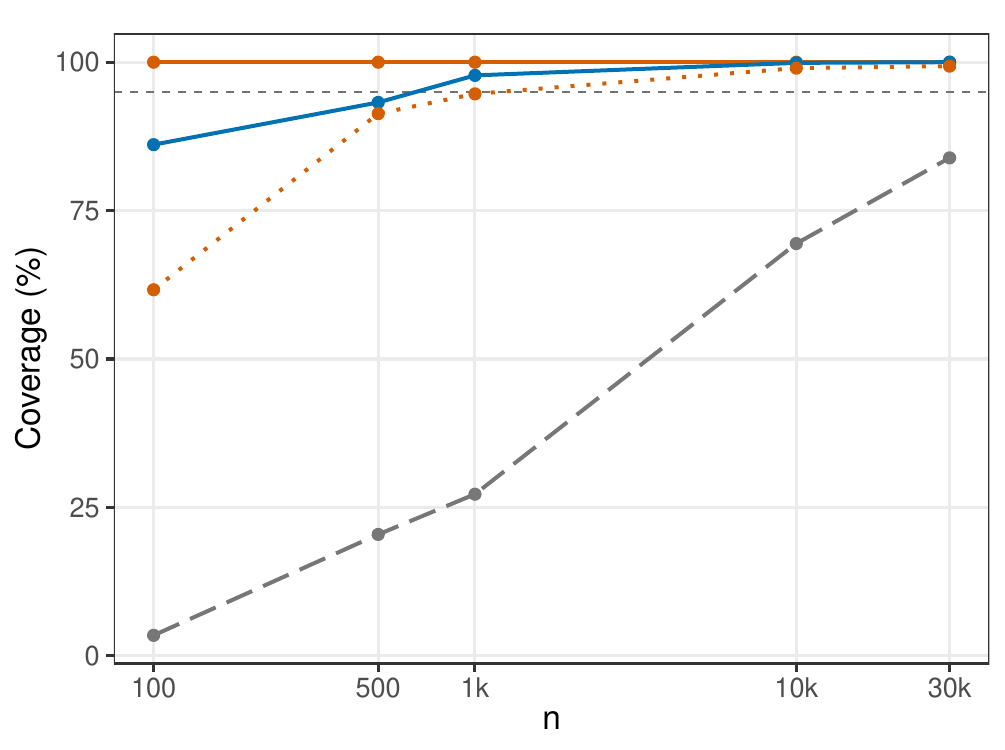}%
\includegraphics[width=0.42\textwidth]{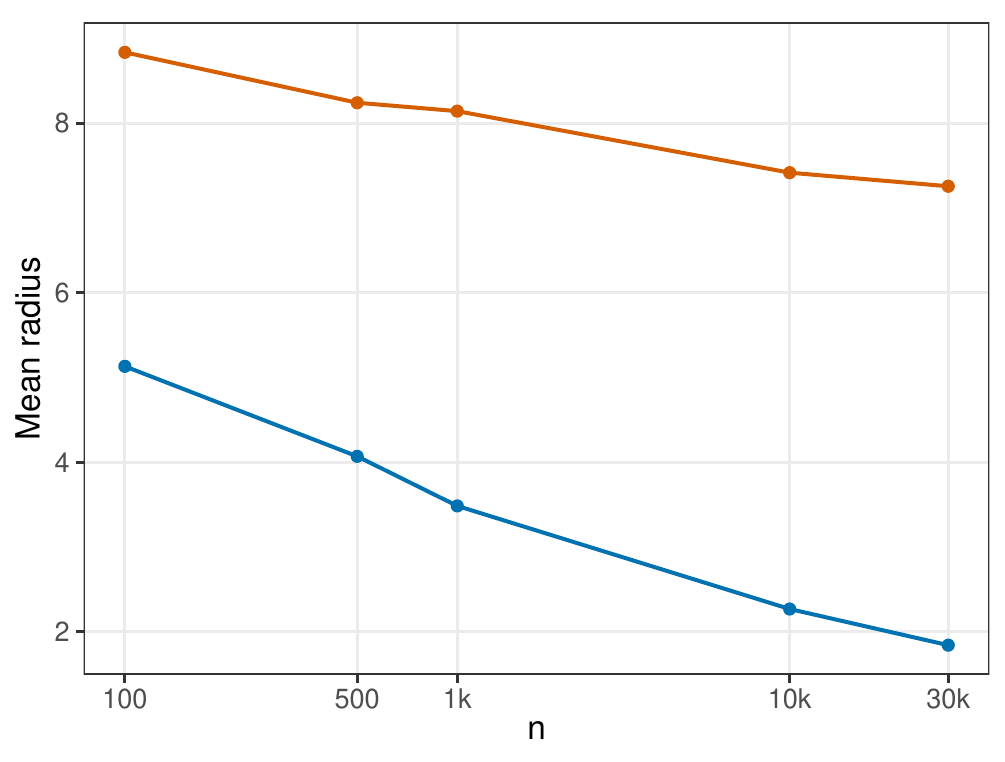}

\vspace{-0.15cm}
\includegraphics[width=0.95\textwidth]{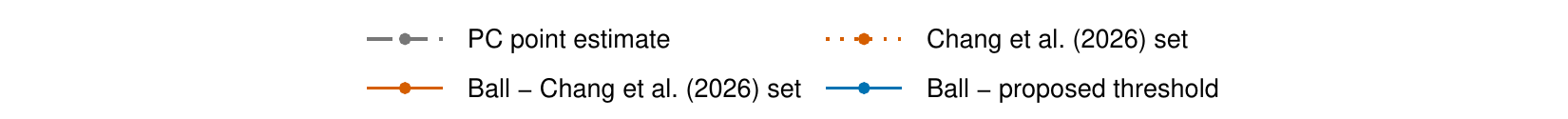}

\vspace{-0.2cm}
\caption{Empirical confidence ball coverage (left) and mean ball radius (right) for a synthetic setting with $p=10$. Results are obtained over $600$ instances. The data generating setup is described in Section~\ref{sec:confidence-experiments}.}
\label{fig:confidence-balls-sparse-p10-low-signal}
\end{figure}

We illustrate our approach in a synthetic experiment with $p=10$ variables. We generate 20 distinct linear Gaussian structural equation models (SEMs), each defined on an independently sampled random DAG with expected degree two, and simulate 30 independent datasets from each SEM. See Section~\ref{sec:confidence-experiments} for details on the setup. The resulting $600$ datasets are analyzed over a range of sample sizes $n$ using perturbed PC with $M = 1000$ runs. As a benchmark, we use the perturbed PC collection of \citet{chang2026post}. We construct confidence balls from both this collection and the collection obtained using our proposed threshold. We compare the coverage of these regions with that of the \citet{chang2026post} collection and the PC point estimate. We also report the ball radius. Figure~\ref{fig:confidence-balls-sparse-p10-low-signal} shows the results. 

The PC point estimate has poor coverage, as expected when structural uncertainty is ignored. The collection of \citet{chang2026post} improves coverage but undercovers at small sample sizes, while converting this collection into a confidence ball substantially improves coverage. The main advantage of our proposed threshold is a substantial reduction in radius while maintaining comparable coverage. The larger threshold suppresses more spurious null rejections across perturbed runs, reducing variation in the retained CPDAG collection and producing substantially smaller confidence balls. More broadly, confidence balls and model confidence intervals replace a large unstructured collection of plausible CPDAGs with a compact structural summary, given by a center and radius or by lower and upper graph bounds, and can further improve coverage in more difficult settings; see Section~\ref{sec:confidence-experiments}.

\vspace{-0.1in}
\captionsetup[figure]{font=footnotesize}

The rest of the paper is organized as follows. Section~\ref{sec:background} reviews CPDAGs and the CPDAG poset. Section~\ref{sec:confidence_regions} defines confidence balls and intervals. Section~\ref{sec:perturbed_pc} analyzes the generalized perturbed PC procedure and constructs the regions. Section~\ref{sec:distance-theory} develops bounds and exact results for the required computations, and Section~\ref{sec:computation} assembles them into search procedures. Sections \ref{sec:confidence-experiments} and \ref{sec:computational_performance} report simulations on the confidence regions and on computational performance.

\section{Background}
\label{sec:background}
\subsection{Graphical preliminaries}We consider graphs with nodes representing random variables and edges between them representing probabilistic dependencies between
these variables. Core graphical concepts are described below.

\subsubsection{Node, edge, subgraph, cycle} A graph $\mathcal{G} = (\mathrm{V}, \mathrm{E})$ consists of a finite set of vertices $\mathrm{V}$ and a set of edges $\mathrm{E}$. We consider two types of edges: directed ($u \to v$) and undirected ($u - v$). The \textit{skeleton} of $\mathcal{G}$, denoted $\operatorname{Skel}(\mathcal{G})$, is the undirected graph obtained by replacing all directed edges in $\mathcal{G}$ with undirected edges. A graph $\mathcal{G}' = (\mathrm{V}, \mathrm{E}')$ is a \textit{subgraph} of $\mathcal{G} = (\mathrm{V}, \mathrm{E})$ if $\mathrm{E}' \subseteq \mathrm{E}$. A \textit{cycle} is a sequence of distinct vertices $v_1, \dots, v_k$ ($k \ge 3$) such that there is an edge between every two adjacent vertices in the sequence and between $v_k$ and $v_1$. A \textit{directed cycle} is a cycle where all edges are directed $v_1 \to v_2 \to \dots \to v_k \to v_1$.

\subsubsection{Directed acyclic graphs (DAGs)} A \textit{directed acyclic graph (DAG)} is a graph containing only directed edges and no directed cycles. For a DAG $\mathcal{D}$, let $\mathrm{Pa}_{\mathcal{D}}(v)$ denote the set of parents of vertex $v$, i.e., $\{u \in \mathrm{V} \mid u \to v \in \mathrm{E}\}$. A triple of vertices $(u, v, w)$ forms a \textit{v-structure} (or unshielded collider) if $u \to v \leftarrow w$ exists in the graph and there is no edge between $u$ and $w$. DAGs can be interpreted as probabilistic models via a concept known as d-separation \cite{pearl1995causal,drton2017structure}, which provides a graphical criterion for determining conditional independence relationships implied by the graph.

\subsubsection{Completed partially directed acyclic graphs (CPDAGs)}
\begin{figure}[t]
    \centering
    \includegraphics[width=1\columnwidth]{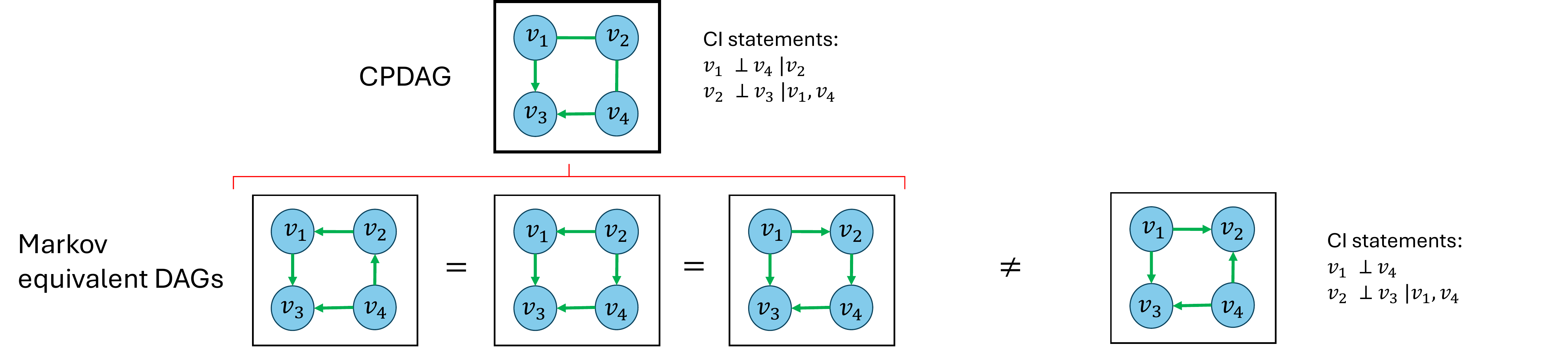}
    \caption{Example CPDAG representing DAGs in the same equivalence class. These DAGs have the same skeleton and same v-structures, and thus encode the same conditional independence (CI) statements. We also present a DAG in the right that does not have the same v-structure as the remaining and is hence not Markov equivalent with the others.}
    \label{fig:markov_equivalent}
\end{figure}

Different DAGs can encode the same conditional independencies; see Figure~\ref{fig:markov_equivalent}. As such, from observational data alone, a DAG is only identifiable up to a Markov equivalence class. Two DAGs are \textit{Markov equivalent} if and only if they possess the same skeleton and the same set of v-structures \cite{andersson1997characterization, Verma1990EquivalenceAS}. Each equivalence class can be uniquely represented by a \emph{completed partially directed acyclic graph} (CPDAG), which has directed edges where all DAGs in the class agree and undirected edges where the direction is not identifiable. The \textit{skeleton} of a CPDAG $\G$, denoted $\skel{\G}$, is the undirected graph obtained by replacing all directed edges with undirected edges.

\captionsetup[figure]{font=footnotesize}

\subsection{CPDAGs as both graphs and models}
A CPDAG represents a Markov equivalence class of DAGs and therefore encodes a set of conditional independence relationships. We associate each CPDAG with the statistical model consisting of distributions that satisfy these relationships.

Let $V$ be a finite vertex set indexing random variables \(X=(X_v)_{v\in V}\) taking values in a product space \(\mathcal X\). Let \(\mathcal M^\circ\) denote the collection of distributions over \(\mathcal X\) that are absolutely continuous with respect to a product measure.

Each DAG \(\mathcal D\) defines a statistical model consisting of distributions that factorize according to its structure:  $\mathcal{M}_\text{dag}(\mathcal{D}) := \left\{ P \in \mathcal{M}^{\circ} \mid p(x) = \prod_{v \in \mathrm{V}} p(x_v \mid x_{\mathrm{Pa}_{\mathcal{D}}(v)}) \right\}$, where $p$ denotes the density of $P$. A CPDAG $\G$ represents a Markov equivalence class and defines the statistical model: $\mathcal{M}(\mathcal{G}) = \bigcup_{\mathcal{D} \in [\mathcal{G}]} \mathcal{M}_\text{dag}(\mathcal{D})$, where $[\mathcal{G}]$ denotes all DAGs in the equivalence class represented by the CPDAG $\mathcal{G}$. 

Let $\mathfrak{G}$ denote the collection of all CPDAGs over $V$. The model map $\mathcal M$ satisfies the following properties.
\begin{proposition}
For any $\mathcal G_1,\mathcal G_2\in\mathfrak G$:
\begin{enumerate}
\item \textbf{Model containment:} \(\mathcal M(\mathcal G_1)\subseteq\mathcal M(\mathcal G_2)\) if and only if every conditional dependence encoded by \(\mathcal G_1\) is also encoded by \(\mathcal G_2\).
\item \textbf{Injectivity:} \(\mathcal G_1\neq\mathcal G_2\) implies \(\mathcal M(\mathcal G_1)\neq\mathcal M(\mathcal G_2)\).
\item \textbf{Extremal models:} For every \(\mathcal G\in\mathfrak G\), $\mathcal M(\mathcal G_{\mathrm{empty}})
\subseteq
\mathcal M(\mathcal G)
\subseteq
\mathcal M(\mathcal G_{\mathrm{full}})$, where \(\mathcal G_{\mathrm{empty}}\) and \(\mathcal G_{\mathrm{full}}\) are the empty and complete undirected CPDAGs.
\end{enumerate}
\label{prop:background}
\end{proposition}

\subsection{Organizing CPDAGs via posets}
\begin{definition}[Partially ordered set]
A partially ordered set (poset) is a tuple $(\mathcal{S}, \preceq)$ consisting of a set $\mathcal{S}$ and a relation $\preceq$ that is reflexive ($a \preceq a$), antisymmetric ($a \preceq b$ and $b \preceq a \implies a=b$), and transitive ($a \preceq b$ and $b \preceq c \implies a \preceq c$).
\end{definition}

Letting $\mathfrak{G}$ be the collection of all CPDAGs over a fixed vertex set $\mathrm{V}$, we organize this collection according to a poset, ordered based on model inclusion, following \cite{taeb2024model}. 

\begin{definition}[CPDAG poset]
The CPDAG poset is $\mathcal{L} = (\mathfrak{G}, \preceq)$, ordered according to model inclusion, i.e., the pair of CPDAGs $\mathcal{G}_1,\mathcal{G}_2\in\mathfrak{G}$ satisfy $\mathcal{G}_1 \preceq \mathcal{G}_2$ if and only if $\mathcal{M}(\mathcal{G}_1) \subseteq \mathcal{M}(\mathcal{G}_2)$.
\end{definition}
The second and third results of Proposition~\ref{prop:background} establish that \(\mathcal L\) is a poset with least and greatest elements, given by the fully disconnected CPDAG \(\mathcal G_{\mathrm{empty}}\) and fully connected CPDAG \(\mathcal G_{\mathrm{full}}\), respectively. By the first result, \(\mathcal G_1\preceq\mathcal G_2\) if and only if every conditional dependence encoded by \(\mathcal G_1\) is also encoded by \(\mathcal G_2\).

The CPDAG poset $\mathcal{L}$ has a special structure known as \emph{gradedness}, which we describe using the notion of neighboring graphs.
\begin{definition}[Cover and neighbor] For distinct CPDAGs $\G_1,\G_2 \in \mathfrak{G}$, we say that $\G_2$ covers $\G_1$, written  $\mathcal{G}_1 \coverby \mathcal{G}_2$, if $\mathcal{G}_1 \preceq \mathcal{G}_2$, and no $\mathcal G\in\mathfrak G\setminus\{\mathcal G_1,\mathcal G_2\}$ satisfies: $\mathcal{G}_1 \preceq \mathcal{G} \preceq \mathcal{G}_2$. Two CPDAGs are neighbors if one covers the other.
\label{defn:covering}
\end{definition}

\begin{definition}[Graded poset] A poset is graded if it admits a rank function $\mathrm{rank}: \mathfrak{G} \to \mathbb{Z}_+$ such that $\G_1 \coverby \G_2$ means $\mathrm{rank}(\G_2)=\mathrm{rank}(\G_1)+1$. 
\end{definition}
The CPDAG poset \(\mathcal L\) is graded with rank function $\mathrm{rank}(\G) = |\mathrm{Skel}(\G)|$, where $\mathrm{Skel}(\G)$ denotes the edge set of its skeleton. Thus, rank measures graphical complexity, and neighboring CPDAGs differ in rank by one.

\begin{figure}[h]
    \centering
    \includegraphics[width=.5\columnwidth]{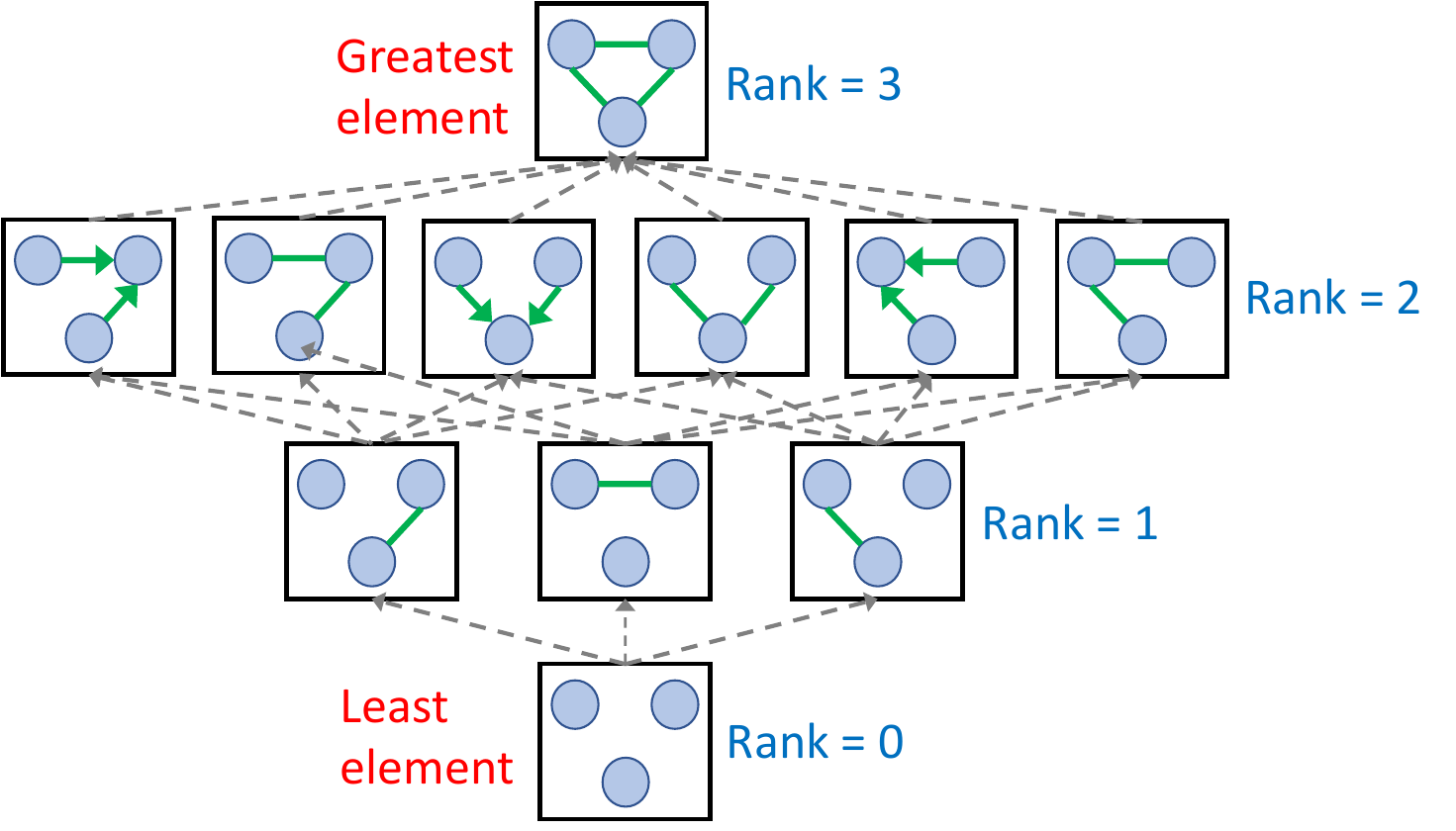}
    \caption{Hasse diagram of the CPDAG poset on 3 nodes. Each box represents a CPDAG, and each dashed arrow represents a covering relation, i.e., we draw $\G_1  \dashrightarrow  \G_2$ pointing upwards if $\G_2$ covers $\G_1$. We can determine whether $\G_1 \preceq \G_2$ by checking whether there is a directed path from $\G_1$ to $\G_2$. The poset has a least and a greatest element, the empty graph and the full graph, respectively. The rank, which is the number of edges in the graph, is indicated.}
    \label{fig:placeholder}
\end{figure}

The poset of CPDAGs can be visualized by a \emph{Hasse diagram} shown in Figure~\ref{fig:placeholder}. Each box is a CPDAG; if $\G_2$ covers $\G_1$, we then place $\G_2$ above $\G_1$ and draw an upward arrow $\G_1  \dashrightarrow \G_2$.
Partial order can be read off from the diagram: $\G_1 \preceq \G_2$ if there exists a directed path (can be of length zero) from $\G_1$ to $\G_2$.

\section{Confidence region over CPDAGs}
\label{sec:confidence_regions}
\subsection{Setup}
Let $X=(X_1,\ldots,X_p)$ denote a collection of $p$ jointly Gaussian random variables with joint distribution $P^\star$. We assume that $P^\star$ is faithful to a CPDAG $\G^\star$ over the vertex set $V=[p]$, where each vertex $i\in V$ corresponds to the random variable $X_i$. This is formalized next.
\begin{assumption}[Faithfulness]For all disjoint subsets $A,B,C \subset V$:
$$X_A \text{ and }X_B \text{ d-separated by }X_C \text{ in }\mathcal{G}^\star \Leftrightarrow X_A \perp\!\!\!\perp X_B \mid X_C \text{ in }P^\star.$$
\label{ass:faithfulness}
\end{assumption}
We let $\Gc$ denote the collection of all CPDAGs over $V$ and suppose that we observe $n$ independent and identically distributed realizations of $X$ from $P^\star$.

A central goal in causal structure learning is to infer $\G^\star$ from observational data. While existing causal discovery methods typically produce a single estimated CPDAG, such point estimates do not directly quantify the uncertainty associated with the learned graphical structure. To address this limitation, we consider constructing data-dependent collections of CPDAGs that contain the true CPDAG with a prescribed probability, at least asymptotically. Such collections provide a principled way to represent uncertainty in the underlying causal structure.

\begin{definition}[Asymptotically valid confidence region] A collection of CPDAGs $\widehat{\mathfrak{G}}_n$ learned from data of size $n$ is said to be an asymptotically valid $1-\alpha$ confidence region for $\G^\star$ if $\liminf_{n\to \infty}\mathbb{P}(\G^\star\in \widehat{\Gc}_n) \geq 1-\alpha$. 
\end{definition}

\subsection{Two types of confidence regions}\label{sec:region-types}
In this paper, we consider two types of confidence regions, based on balls and intervals, both constructed using the poset structure of CPDAGs.

The first approach constructs a confidence ball around a data-dependent center, typically a point estimate of the true CPDAG. The ball contains all CPDAGs within a specified distance of this center, where the distance quantifies structural differences between graphical models using the underlying poset. The radius determines the size of the region and is chosen to achieve the desired coverage probability. Formally, we define a confidence ball as follows.
\begin{definition}[Ball] A region $\widehat{\Gc}_n^\text{ball} \subseteq \Gc$ is called a ball if \begin{equation}\widehat{\Gc}_n^\text{ball}= \{\G \in \Gc, d(\G,\widehat{\G}_{0,n}) \leq \widehat{r}_n\},
\label{eqn:ball}
\end{equation}
where \(\widehat{\G}_{0,n}\in\Gc\) and \(\widehat r_n\in\mathbb Z_{\geq 0}\) are a data-dependent center and radius, respectively, and $d(\cdot,\cdot)$ is a metric over the space of CPDAGs.
\label{defn:ball}
\end{definition}
In Section~\ref{sec:model-oriented}, we describe a distance metric tailored to the CPDAG poset. A valid confidence region can always be defined by setting $\widehat r_n =
\max_{\G_1,\G_2\in\Gc}
d(\G_1,\G_2)$, which yields $\widehat{\Gc}_n= \Gc$. This construction, however, simply returns the full collection and is therefore uninformative. Our goal is to construct smaller confidence balls that retain the desired coverage guarantee.

The second approach constructs a region using the partial ordering of CPDAGs. Rather than considering graphs within a specified distance of a center, this approach identifies data-dependent lower and upper bounds, and includes all CPDAGs lying between them in the poset. We formalize this notion below.
\begin{definition}[Interval]A region $\widehat{\Gc}_n^\text{interval}\subseteq \Gc$ is called an interval if: $$\widehat{\Gc}_n^\text{interval} = \{\G \in \Gc, \widehat{\G}_{n,L} \preceq \G \preceq \widehat{\G}_{n,U}\},$$ where \(\widehat{\G}_{n,L},\widehat{\G}_{n,U}\in\Gc\) are data-dependent lower and upper bounds satisfying \(\widehat{\G}_{n,L}\preceq\widehat{\G}_{n,U}\), and \(\preceq\) denotes the partial order on CPDAGs.
\label{defn:CI}
\end{definition}

Choosing the least and greatest elements of the poset as bounds yields the trivial region $\widehat{\Gc}_n^\text{interval}= \Gc$, which has coverage probability one. Our goal is instead to construct tighter intervals that retain the desired coverage guarantee, with width measured by
$$\mathrm{width}(\widehat{\mathfrak{G}}_n^\text{interval}) = \mathrm{rank}(\widehat{\G}_{n,U})-\mathrm{rank}(\widehat{\G}_{n,L}) \in [0,p(p-1)/2]$$

The region $\widehat{\mathfrak{G}}_n^\text{interval}$ has a natural interpretation. If it is a valid confidence region, then, with high probability, the conditional dependencies encoded by the true graph contain those encoded by \(\widehat{\G}_{n,L}\) and are contained in those encoded by \(\widehat{\G}_{n,U}\). In particular, with high probability, every edge of $\widehat{\mathcal{G}}_{n,L}$ and every v-structure shared by $\widehat{\mathcal{G}}_{n,L}$ and $\widehat{\mathcal{G}}_{n,U}$ must also be present in the true graph.

\subsection{Model-oriented distance between CPDAGs} 
\label{sec:model-oriented}
\begin{figure}[h]
    \centering
   \includegraphics[width=.5\columnwidth]{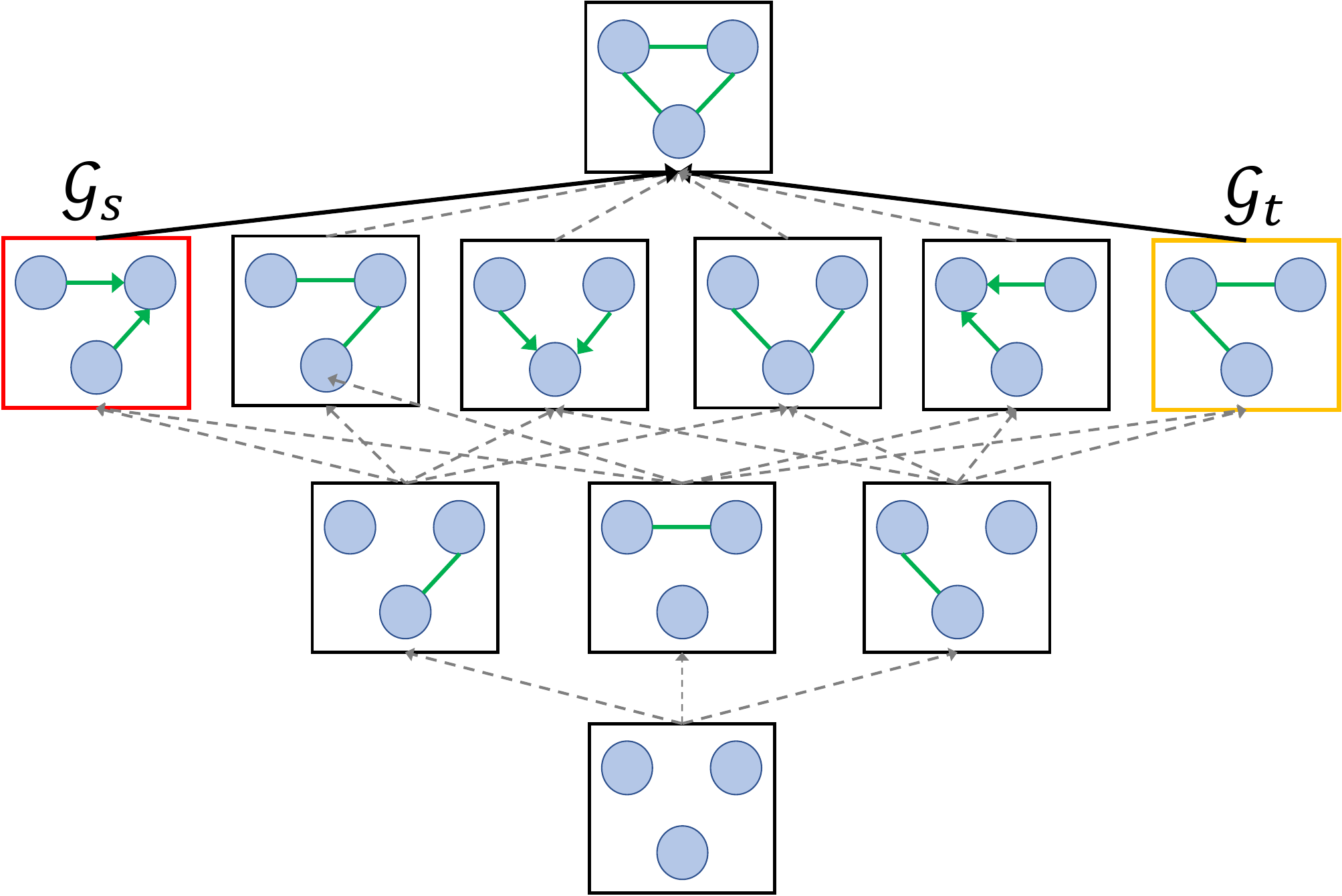}
    \caption{A path in the poset between two CPDAGs $\G_s$ and $\G_t$, shown in black. The path has length two, and no shorter path exists. Thus, $d_\text{model}(\G_s,\G_t)=2$.}
    \label{fig:path}
\end{figure}

For the distance metric in the definition of ball region in Definition~\ref{defn:ball}, we use the model-oriented distance, first proposed in \citep{taeb2025distance} and defined next. 
\begin{definition}[Model-oriented distance]
Let $\mathcal{L}:=(\mathfrak{G},\preceq)$ be the CPDAG poset. Then, for $\G_s, \G_t \in \mathfrak{G}$, the \emph{model-oriented distance} between $(\G_s, \G_t)$ is:
\begin{equation}
d_\text{model}(\G_s,\G_t):= \min\big\{\mathrm{length}(p): \text{$p$ is a path between $\G_s$ and $\G_t$} \text{ in the Hasse diagram of $\mathcal{L}$} \big\},
\end{equation}
{\noindent}where the path can contain dashed arrows in both directions. In other words, a path is a sequence $(\G_0 := \G_s, \G_1, \G_2, \dots,\G_k:=\G_t)$ such that $\G_{i+1}$ and $\G_i$ are neighbors (i.e., $\G_{i+1} \coverby \G_i$ or $\G_i \coverby \G_{i+1}$); see Definition~\ref{defn:covering}.
\label{defn:model_oriented}
\end{definition}
As \(d_{\text{model}}\) is a shortest path distance, it is a metric. Since the poset is organized by model inclusion, the model-oriented distance \(d_{\text{model}}(\G_s,\G_t)\) measures the minimum number of basic operations needed to transform the model of \(\G_s\) into that of \(\G_t\). Figure~\ref{fig:path} provides an illustration. We note that defining distance through the minimal number of basic operations to transform one model to the other is common in other combinatorial settings such as ranking \citep{kendall1938new} and clustering \cite{meila2005comparing}.

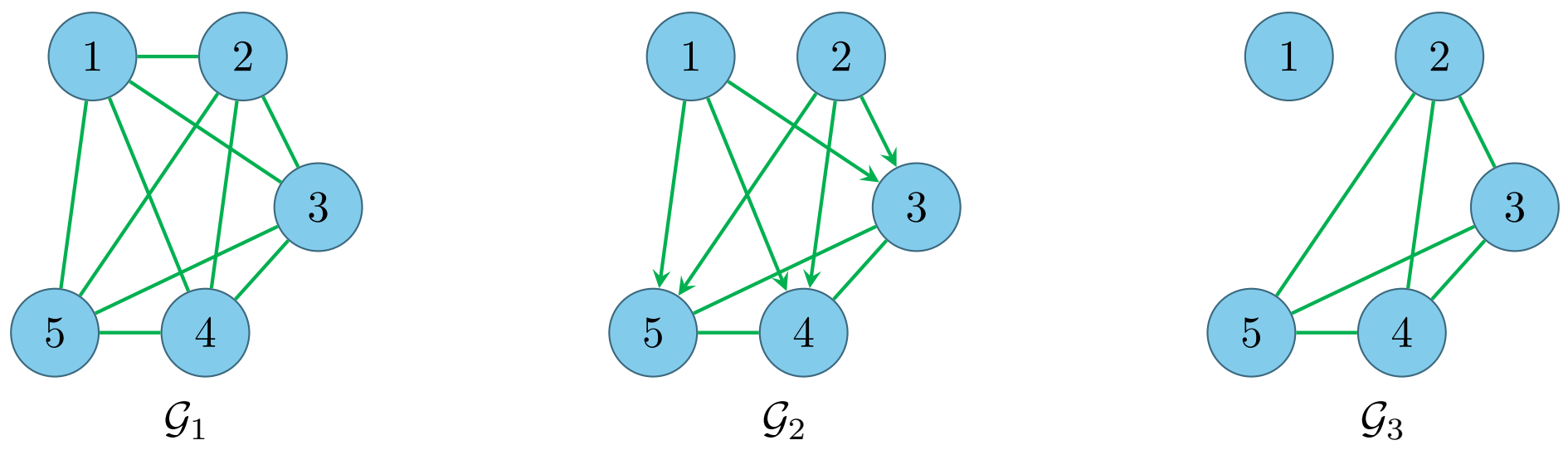
\begin{figure}[t]
\centering
\begin{center}
\begin{tikzpicture}[>=stealth,
vertex/.style={circle,draw=nodeoutline,fill=nodeblue,
minimum size=7mm,inner sep=1pt}]

\hspace*{-0.5in}
\begin{scope}[xshift=0cm]
\node[vertex] (a1) at (0,1.2) {$1$};
\node[vertex] (b1) at (1.2,1.2) {$2$};
\node[vertex] (c1) at (1.8,0) {$3$};
\node[vertex] (d1) at (0.9,-1.0) {$4$};
\node[vertex] (e1) at (-0.3,-1.0) {$5$};

\foreach \i/\j in {a1/b1,a1/c1,a1/d1,a1/e1,
b1/c1,b1/d1,b1/e1,c1/d1,c1/e1,d1/e1}
\draw[thick,edgegreen] (\i)--(\j);

\node at (0.75,-1.7) {$\G_1$};
\end{scope}

\hspace*{0.5in}
\begin{scope}[xshift=3.5cm]
\node[vertex] (a2) at (0,1.2) {$1$};
\node[vertex] (b2) at (1.2,1.2) {$2$};
\node[vertex] (c2) at (1.8,0) {$3$};
\node[vertex] (d2) at (0.9,-1.0) {$4$};
\node[vertex] (e2) at (-0.3,-1.0) {$5$};

\foreach \j in {c2,d2,e2} {
\draw[->,thick,edgegreen] (a2)--(\j);
\draw[->,thick,edgegreen] (b2)--(\j);
}

\draw[thick,edgegreen] (c2)--(d2);
\draw[thick,edgegreen] (c2)--(e2);
\draw[thick,edgegreen] (d2)--(e2);

\node at (0.75,-1.7) {$\G_2$};
\end{scope}

\hspace*{0.5in}

\begin{scope}[xshift=7cm]
\node[vertex] (a1) at (0,1.2) {$1$};
\node[vertex] (b1) at (1.2,1.2) {$2$};
\node[vertex] (c1) at (1.8,0) {$3$};
\node[vertex] (d1) at (0.9,-1.0) {$4$};
\node[vertex] (e1) at (-0.3,-1.0) {$5$};

\foreach \i/\j in {
b1/c1,b1/d1,b1/e1,c1/d1,c1/e1,d1/e1}
\draw[thick,edgegreen] (\i)--(\j);

\node at (0.75,-1.7) {$\G_3$};
\end{scope}

\end{tikzpicture}
\end{center}
\caption{Three CPDAGs associated with Example~\ref{example:first} with $p=5$.}
\label{fig:cpdag-model-comparison}
\end{figure}

The most popular approach for measuring distance between CPDAGs is the structural Hamming distance (SHD) \cite{tsamardinos2006max}, which counts the number of edges that differ between the two graphs. However, as highlighted in \cite{taeb2025distance}, this distance is \emph{model-agnostic} -- it disregards the specific models the graphs represent and is
solely defined on the graphs themselves. SHD treats each edge individually, ignoring that not all edge configurations may give a valid CPDAG. The model-agnostic nature of SHD leads to many undesirable behaviors: as outlined in the following example, a pair of CPDAGs that are very close in terms of the models they represent can have a large SHD distance. 
\begin{example}Let $\G_1$ be a fully connected CPDAG on $p$ nodes. Let $\G_2$ have no edge between $1$ and $2$, with $1 \rightarrow j \leftarrow 2$, $j = 3,4,\dots,p$, and all pairs among $3,\dots,p$ connected by undirected edges. Let $\G_3$ be the complete undirected CPDAG on $2,\dots,p$ with $1$ isolated. These can be seen in Figure~\ref{fig:cpdag-model-comparison}. From this we can say, $\G_3 \preceq \G_2 \preceq \G_1$, $d_\text{model}(\G_1,\G_2)=1$ and $d_{\mathrm{SHD}}(\G_1,\G_2) = 2p-3$, whereas $d_\text{model}(\G_1,\G_3) = d_\text{SHD}(\G_1,\G_3)=p-1$. Since SHD ignores the models that the graphs represent, it regards $\G_3$ as closer to $\G_1$ than $\G_2$, even though $\G_2$ differs from $\G_1$ by only one model change.  
\label{example:first}
\end{example}

The discrepancy is especially important when constructing confidence balls. Under the model-oriented distance, increasing the radius by one corresponds to allowing one additional minimal model change from the center. SHD does not have this interpretation: a single model change can induce many compelled orientations and therefore a large jump in distance. Consequently, for any $p-1\leq r <2p-3$, the SHD ball centered at $\G_1$ contains $\G_3$, which is $p-1$ model changes away, but excludes the immediate neighboring model $\G_2$. Thus, SHD balls can be poorly aligned with the underlying model geometry, including models that are many model changes from the center while excluding models that differ by only one. 

A second related issue is that SHD balls can also have gaps between successive distance levels. For the three-node collider $1 \rightarrow 2 \leftarrow 3$, there is no distinct CPDAG at SHD 1, while valid CPDAGs exist at SHD 2. Hence, the balls of radii 0 and 1 are identical. Thus, increasing the SHD radius by one need not correspond to adding a new layer of models. 

\begin{remark}\label{rem:down-up}
Down-up and up-down paths play an important role in our search for the shortest path in the CPDAG poset. They are easier to compute than unrestricted shortest paths and provide useful bounds and candidate routes for the search. A down-up path from $\G_s$ to $\G_t$ is any path of the form 
$$
\G_0 := \G_s \cover \dots \cover\G_{i^*} \coverby \dots \coverby\G_k:=\G_t
$$
while an up-down path is of the form
$$
\G_0 := \G_s \coverby \dots \coverby \G_{i^*} \cover \dots \cover \G_k:=\G_t
$$
Such paths always exist. A down-up path can pass through the empty CPDAG, while an up-down path can pass through the complete CPDAG. We write $d_{\downarrow\uparrow}(\mathcal{G}_s,\mathcal{G}_t)$ and $d_{\uparrow\downarrow}(\mathcal{G}_s,\mathcal{G}_t)$ for the lengths of a shortest down-up path and a shortest up-down path from $\mathcal{G}_s$ to $\mathcal{G}_t$. Since both are lengths of paths, $d_{\mathrm{model}}(\mathcal{G}_s,\mathcal{G}_t)\le\min\{d_{\downarrow\uparrow}(\mathcal{G}_s,\mathcal{G}_t),d_{\uparrow\downarrow}(\mathcal{G}_s,\mathcal{G}_t)\}$.
\end{remark}

\section{Constructing a confidence region via the perturbed PC algorithm}
\label{sec:perturbed_pc}
In this section, we propose strategies for building confidence regions using the perturbed PC algorithm. The perturbed PC algorithm \citep{chang2026post} replaces a single run of PC with a collection of runs based on perturbed conditional independence statistics, capturing sensitivity to borderline decisions and increasing the chance that one run recovers the true CPDAG. We study a generalized perturbed PC procedure in which the threshold, perturbation scale, and number of runs may vary, with the choices of \citet{chang2026post} as a special case, and characterize how these choices affect coverage. Our default threshold grows with the number of runs to control increasingly extreme perturbations under the null, thereby suppressing spurious edge decisions across the collection. At the same time, it remains small enough for weak nonzero signals to cross the threshold in at least one run. This choice accommodates arbitrarily weak signals without imposing a strong faithfulness condition while producing tighter collections than the tuning of \citet{chang2026post}.

Section~\ref{sec:preliminaries} reviews PC and perturbed PC. Section~\ref{sec:analysis} analyzes coverage and gives tuning parameters that guarantee it. Section~\ref{sec:regions} constructs confidence balls and intervals from the perturbed PC outputs. 

\subsection{Preliminaries: The PC algorithm and its perturbed version}
\label{sec:preliminaries}
We describe both the PC algorithm \citep{spirtes2000causation} and its perturbed version introduced in \cite{chang2026post}. Throughout, we have that $X$ follows a jointly Gaussian distribution satisfying Assumption~\ref{ass:faithfulness}, and that we observe $n$ iid samples from this distribution. The data is denoted by $\mathcal{D}$. 

\vspace{0.1in}
\noindent\textbf{Standard PC algorithm.} The PC algorithm 
starts from a complete undirected graph and removes edges through a sequence of conditional independence tests. A full description, including extensions with tiered background knowledge, is given in Appendix~\ref{app:pc}. In principle, any valid conditional independence test may be used, although we focus here on building test statistics from partial correlation coefficients. 

For a multivariate Gaussian $X$, conditional independence is equivalent to zero partial correlation. For distinct $i, j \in V$ and $S \subseteq V \setminus \{i,j\}$, let $\rho_{ij|S}$ be the partial correlation of $X_i$
and $X_j$ given $X_S$, and $\widehat\rho_{ij|S}$ its sample analogue. Then, $X_i \perp\!\!\!\perp X
_j|X_S$ if and only if $\rho_{ij|S}=0$. The PC algorithm tests this hypothesis using the Fisher statistic $\widehat T_{ij|S} = |\widehat\psi_{ij|S}|/\sigma_{ij|S}$ where 
\[
\psi_{ij|S} := \frac12\log\frac{1 + \rho_{ij|S}}{1 - \rho_{ij|S}}, \qquad
\widehat\psi_{ij|S} := \frac12\log\frac{1 + \widehat\rho_{ij|S}}{1 - \widehat\rho_{ij|S}}, \qquad
\sigma_{ij|S} := (n - |S| - 3)^{-1/2}.
\]
The null is rejected when $\widehat T_{ij|S}$ exceeds a chosen threshold $t$. Let $\mathcal{T}$ be the set of triples $(i,j,S)$ with distinct $i,j\in V, S\subseteq V\setminus\{i,j\}$, and let $L:=|\mathcal{T}|$. Indexing $\mathcal{T}$ by $l$, write:
\[
\widehat T_l := \left|\frac{\psi_l}{\sigma_l} + \widehat U_l\right|, \qquad
\kappa_l := \frac{|\psi_l|}{\sigma_l}, \qquad
\widehat U_l := \frac{\widehat\psi_l - \psi_l}{\sigma_l},
\]
where $\kappa_l$ is the signal strength and $\widehat U_l$ is a standardized sampling error. Let
$\mathcal{T}_\text{null} := \{l \in \mathcal{T} : \psi_l = 0\}$, $\mathcal{T}_\text{non-null}:= \mathcal{T} \setminus \mathcal{T}_\text{null}$, $L_0 := |\mathcal{T}_\text{null}|$, $L_1 := |\mathcal{T}_\text{non-null}|$, and define the minimum signal strength and minimum nonzero partial correlation:
\[
\kappa_{\min,n} := \min_{l \in \mathcal{T}_\text{non-null}} \kappa_l, \qquad
\rho_{\min,n} := \min_{l \in \mathcal{T}_\text{non-null}} |\rho_l|.
\]
Since $|\tfrac12\log\{(1+x)/(1-x)\}| \ge |x|$, we have $\kappa_{\min,n} \ge ({n - k - 3})^{1/2}\rho_{\min,n}$, where $k:=p-2$ is the largest conditioning-set size. Let $z_a := \Phi^{-1}(1 - a)$ where $\Phi$ is the standard normal CDF. For each fixed triple $l$, under multivariate Gaussian sampling, and provided the corresponding population partial correlation stays bounded away from $\pm 1$, $
\widehat U_l
\xrightarrow{d}\mathcal N(0,1)$. Thus, $\limsup_n \mathbb{P}(|\widehat U_l| \ge z_{a/2}) \le a$. Define $$E_n := \Big\{\max_{l \in \mathcal{T}} |\widehat U_l| \le z_{\nu/2L}\Big\}.$$ It is straightforward to see that by a union bound, $\liminf_n \mathbb{P}(E_n) \ge 1 - \nu$.

\vspace{0.1in}
\noindent\textbf{Perturbed PC algorithm.} Standard PC makes a single sequence of conditional independence decisions by comparing noisy test statistics to a fixed threshold. When population signals lie near this threshold, small sampling fluctuations can reverse individual decisions, and these errors may propagate through subsequent edge deletions and orientation steps, producing a CPDAG far from the truth. A single run therefore can easily be inaccurate and gives no indication of the sensitivity of the estimated graph to such borderline decisions.

This motivates replacing a single run of PC with a collection of nearby runs. The perturbed PC algorithm of \cite{chang2026post} repeatedly perturbs the estimated CI statistics and reruns PC. The resulting collection of CPDAGs captures sensitivity to borderline decisions while increasing the chance that at least one run recovers the true CPDAG. 

We describe a generalized version of the procedure in \cite{chang2026post}. In run $m \in [M]$, for each $l \in \mathcal{T}$, replace  $\widehat{\psi}_l$ by an independent draw
$$\widehat\psi^{[m]}_l\sim \mathcal{N}(\widehat\psi_l,\eta^2\sigma_l^2),$$ where $\eta>0$ controls the perturbation scale. Given a threshold $t>0$, the corresponding null hypothesis is rejected when
$$\widehat{T}_l^{[m]}>t\quad \quad \text{ where }\quad \widehat{T}_l^{[m]}:=|\widehat\psi^{[m]}_l|/\sigma_l.$$ Because sampling errors can lead to incompatible edge orientations, not every run produces a valid CPDAG. We only retain runs that return a valid CPDAG. These valid runs are indexed by the elements of $\widehat{\mathcal{K}}_n$ and define $$\widehat{\mathfrak{G}}_n:=\{\widehat{\G}^{[m]}_n:m\in\widehat{\mathcal{K}}_n\}.$$ The full procedure is given in Algorithm~\ref{alg:perturbed PC}. The implementation of  \cite{chang2026post} corresponds to $\eta=1$, and $t = c^\star(\log n/M)^{1/L} z_{\nu/2L}$ for some constant $c^\star >0$. 

\begin{algorithm}[t]
\caption{Perturbed PC algorithm}
\label{alg:perturbed PC}
\KwIn{Data $\mathcal{D}$, threshold $t>0$, perturbation scale $\eta>0$, number of runs $M$. Default choice $\eta = 1$ and $t = \sqrt{2\log M/B}$ for some tuning parameter $B$}
\KwOut{Retained indices $\widehat{\mathcal{K}}_n$ and CPDAG collection $\widehat{\mathfrak{G}}_n$}

Initialize $\widehat{\mathcal{K}}_n \gets \emptyset$ and $\widehat{\mathfrak{G}}_n \gets \emptyset$\;

\For{$m=1,\ldots,M$}{
    Run the PC algorithm with the following modified CI tests:\\
    \Indp
    \ForEach{CI test $X_i \mathrel{\perp\!\!\!\perp} X_j \mid X_S$}{
        Draw $\widehat\psi^{[m]}_{ij|S}
        \sim \mathcal{N}(\widehat\psi_{ij|S},\eta^2\sigma_{ij|S}^2)$\;
        Reject independence if
        $|\widehat\psi^{[m]}_{ij|S}|/\sigma_{ij|S}>t$\;
    }
    \Indm
    Let $\widehat{\mathcal{G}}^{[m]}_n$ denote the resulting graph\;

    \If{$\widehat{\G}^{[m]}_n$ is a valid CPDAG}{
       $\widehat{\mathcal{K}}_n \gets \widehat{\mathcal{K}}_n\cup\{m\}$\;
        $\widehat{\mathfrak{G}}_n
        \gets \widehat{\mathfrak{G}}_n\cup\{\widehat{\G}^{[m]}_n\}$\;
    }
}
\end{algorithm}

\subsection{Analyzing the perturbed PC algorithm}\label{sec:analysis}
We give conditions under which $\widehat{\mathfrak{G}}_n$ contains
$\mathcal{G}^\star$ (coverage) and under which it equals
$\{\mathcal{G}^\star\}$ (collapse). We call a run \emph{correct} if every conditional independence test it encounters is decided correctly. For the analysis, we use the stronger event that all tests $l\in\mathcal T$ are decided correctly, including tests that may not be encountered by a particular run. Since $\psi_l=0$ exactly when $i$ and $j$ are d-separated by $S$ in $\mathcal G^\star$, this event guarantees that the run follows the oracle PC path and returns $\mathcal G^\star$.

\subsubsection{A preliminary lemma}

We begin with a general finite-sample bound, proved in
Appendix~\ref{app:analysis-proofs}.

\begin{lemma}
\label{lemma:main}
Let $z := z_{\nu/2L}$ and, for $t, \eta > 0$,
\[
\pi(t,\eta) :=
\Big\{
\Phi\Big(\frac{t-z}{\eta}\Big)
-
\Phi\Big(\frac{-t-z}{\eta}\Big)
\Big\}^{L_0}
\prod_{l \in \mathcal{T}_\text{non-null}}
\Phi\Big(\frac{\kappa_l-z-t}{\eta}\Big).
\]
On $E_n$ and given data $\mathcal{D}$, each run of
Algorithm~\ref{alg:perturbed PC} is correct with probability at least
$\pi := \pi(t,\eta)$, independently across runs. Hence, for any sample
size $n$,
\[
\mathbb{P}(\mathcal{G}^\star \in \widehat{\mathfrak{G}}_n)
\ge
\{1-(1-\pi)^M\}\mathbb{P}(E_n),
\qquad
\mathbb{P}(\widehat{\mathfrak{G}}_n=\{\mathcal{G}^\star\})
\ge
\pi^M\mathbb{P}(E_n).
\]
\end{lemma}
Lemma~\ref{lemma:main} reduces both coverage and collapse to the probability that a single perturbed PC run is correct. The quantity $\pi$ gives a lower bound on the single-run probability. Each factor in $\pi$ lower bounds, on $E_n$, the probability that one perturbed conditional independence test makes the correct decision. The product arises from the independence of the perturbations across tests. Coverage is guaranteed when at least one of the $M$ runs is correct, yielding the factor $1-(1-\pi)^M$, whereas collapse is guaranteed when all $M$ runs are correct, yielding $\pi^M$.

\subsubsection{Asymptotic guarantees}

Building on Lemma~\ref{lemma:main}, we next give conditions on
$t,\eta$ and $M$ that ensure asymptotic coverage and collapse.
Let $t=t_n$, $\eta=\eta_n$ and $M=M_n$ be sequences that may depend
on $n$, and write $\pi_n=\pi(t_n,\eta_n)$. For $\delta_n\in(0,1)$,
define the \emph{global reach}
\[
\omega_n :=
\eta_n\Phi^{-1}\left(
1/2+(1-\delta_n)^{1/(LM_n)}/2
\right).
\]
For test $l$ in run $m$,
\begin{equation}
\widehat{T}_l^{[m]}
=
\left|
\frac{\psi_l}{\sigma_l}
+
\widehat{U}_l
+
\widehat U_l^{[m]}
\right|,
\quad\text{where}\quad
\widehat U_l^{[m]}
:=
\frac{\widehat\psi_l^{[m]}-\widehat\psi_l}{\sigma_l}.
\label{eqn:test_statistic_decomposition}
\end{equation}
Given the data, the perturbations $\widehat U_l^{[m]}$ are
independent $\mathcal{N}(0,\eta_n^2)$ across tests and runs.
By construction,
\[
\mathbb{P}\left(
\max_{m\le M_n,\,l\le L}
|\widehat U_l^{[m]}|
\le \omega_n
\right)
=
1-\delta_n.
\]
Thus, with probability $1-\delta_n$, every perturbation across
all tests and runs has magnitude at most $\omega_n$. For an integer $s\ge1$, define the \emph{joint favorable reach}
\[
r_n(s)
:=
\eta_n\Phi^{-1}\left(
1-(1-\delta_n^{1/M_n})^{1/s}
\right).
\]
We define $r_n(0)=0$. Then for any specified set $W \subseteq \mathcal{T}_\text{non-null}$ of $s$ nonnull tests, 
\[
\mathbb{P}\left(
    \exists m \le M_n:
    \min_{l \in W}\widehat U_l^{[m]} \ge r_n(s)
\right)=1-\delta_n.
\]
Thus, with probability $1-\delta_n$,
some run has perturbations of at least $r_n(s)$ for all
tests in $W$.

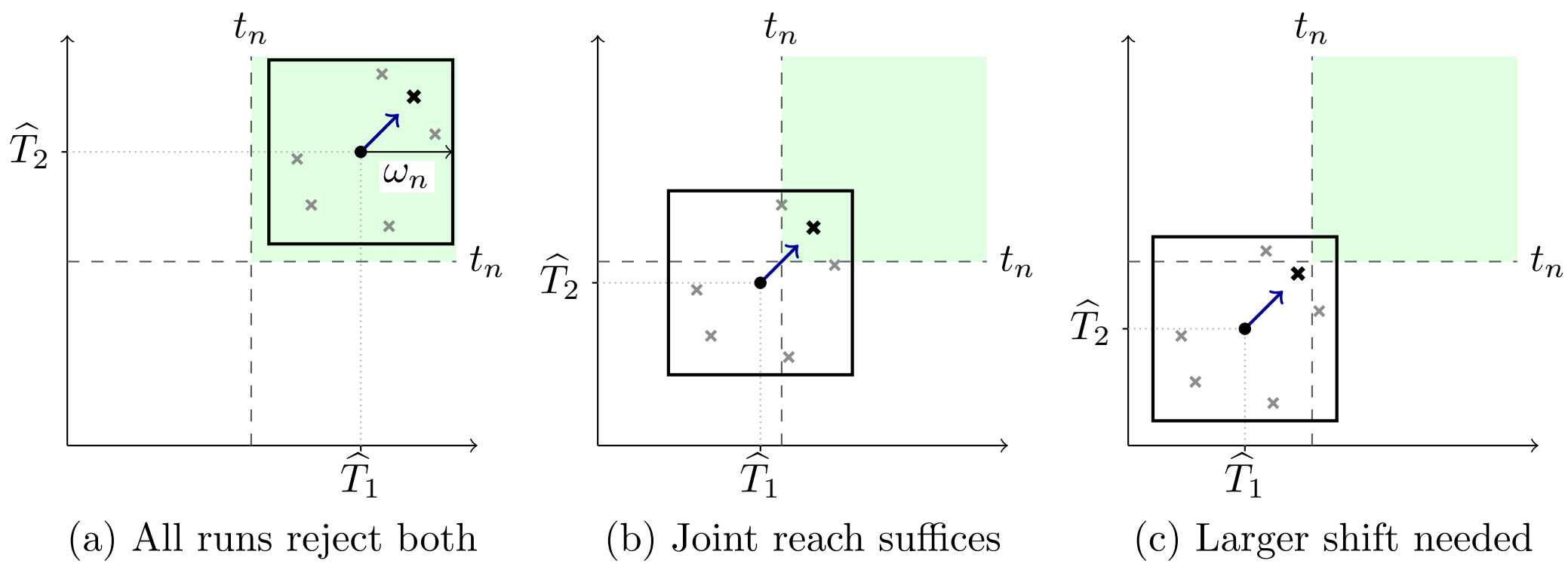
\begin{figure}[t]
\centering
\begin{tikzpicture}[scale=0.62, font=\small]
\def\reach{1.3}
\def\jointreach{0.535311}
\def\threshold{2.6}

\tikzset{
  guideline/.style={thin,densely dotted,black!30},
  threshold line/.style={thin,dashed,black!65},
  observed x/.style={
    anchor=north,inner sep=0pt,yshift=-4pt,
    text height=1.7ex,text depth=0.5ex
  },
  observed y/.style={
    anchor=east,inner sep=0pt,xshift=-4pt,
    text height=1.7ex,text depth=0.5ex
  },
  threshold x/.style={
    anchor=south,inner sep=0pt,yshift=3pt
  },
  threshold y/.style={
    anchor=west,inner sep=0pt,xshift=3pt
  },
  panel label/.style={
    anchor=north,align=center,text width=4.3cm,inner sep=0pt
  }
}

\foreach \x/\y/\lab [count=\i] in {
  4.15/4.15/{(a) All runs reject both},
  2.30/2.30/{(b) Joint reach suffices},
  1.65/1.65/{(c) Larger shift needed}} {
\begin{scope}[shift={({7.5*(\i-1)},0)}]

\fill[green!12!white]
  (\threshold,\threshold) rectangle (5.5,5.5);

\draw[->] (0,0) -- (5.8,0);
\draw[->] (0,0) -- (0,5.8);
\draw[threshold line]
  (\threshold,0) -- (\threshold,5.5);
\draw[threshold line]
  (0,\threshold) -- (5.5,\threshold);

\node[threshold x] at (\threshold,5.5) {$t_n$};
\node[threshold y] at (5.5,\threshold) {$t_n$};

\draw[guideline] (\x,0) -- (\x,\y);
\draw[guideline] (0,\y) -- (\x,\y);
\draw[thin] (\x,0) -- (\x,-0.08);
\draw[thin] (0,\y) -- (-0.08,\y);

\node[observed x] at (\x,0) {$\widehat T_1$};
\node[observed y] at (0,\y) {$\widehat T_2$};

\draw[thick]
  ({\x-\reach},{\y-\reach})
  rectangle
  ({\x+\reach},{\y+\reach});

\foreach \dx/\dy in {
  -0.90/-0.10,
   1.05/0.25,
   0.30/1.10,
   0.40/-1.05,
  -0.70/-0.75} {
  \draw[black!45,thick]
    ({\x+\dx-0.07},{\y+\dy-0.07}) --
    ({\x+\dx+0.07},{\y+\dy+0.07});
  \draw[black!45,thick]
    ({\x+\dx-0.07},{\y+\dy+0.07}) --
    ({\x+\dx+0.07},{\y+\dy-0.07});
}

\draw[very thick]
  ({\x+0.75-0.08},{\y+0.78-0.08}) --
  ({\x+0.75+0.08},{\y+0.78+0.08});
\draw[very thick]
  ({\x+0.75-0.08},{\y+0.78+0.08}) --
  ({\x+0.75+0.08},{\y+0.78-0.08});

\ifnum\i=1
  \draw[->,thin]
    (\x,\y) -- ({\x+\reach},\y);
  \node[below=2pt,fill=white,inner sep=1pt]
    at ({\x+0.65},\y) {$\omega_n$};
\fi

\draw[->,thick,blue!60!black]
  (\x,\y) --
  ({\x+\jointreach},{\y+\jointreach});

\fill (\x,\y) circle (2.5pt);

\node[panel label] at (2.9,-1.05) {\lab};

\end{scope}
}
\end{tikzpicture}

\caption{
Global and joint favorable reach for two nonnull tests.
All displayed signed statistics are positive and hence coincide with
their absolute values. The dot marks the observed pair, crosses
illustrate perturbed runs, and shading indicates joint rejection.
The square has half-width $\omega_n$; the blue arrow has displacement
$(r_n(2),r_n(2))$, with the black cross illustrating a run beyond its
tip in both coordinates. On the respective reach events, each of
probability at least $1-\delta_n$,
(a) every run rejects both tests and (b) at least one run rejects both.
In (c), individual rejections occur in different illustrated runs,
while simultaneous rejection requires an increase greater than
$r_n(2)$ in each coordinate.
}
\label{fig:perturbation}
\end{figure}

The distinction between $\omega_n$ and $r_n(s)$ is important. The global reach $\omega_n$ describes the largest perturbation that can occur anywhere in the entire collection of runs, and is therefore the relevant quantity for collapse: to ensure every run is correct, even the most adverse perturbation must not change any decision.  In contrast, $r_n(s)$ is relevant for coverage: it measures how far a single run can simultaneously move $s$ specified nonnull tests in the favorable direction. If some nonnull statistics do not clear the threshold on their own, such a favorable run can shift them above the threshold simultaneously. See Figure~\ref{fig:perturbation} for an illustration.

For coverage, only the nonnull tests whose signals do not clear the threshold by a diverging margin need favorable perturbations. This motivates separating the nonnull tests according to whether they asymptotically lie above the threshold.
 
\begin{definition}[Strong and weak tests]\label{defn:strong-weak}
A nonnull test $l\in\mathcal{T}_\text{non-null}$ is called 
\emph{strong} if $\kappa_l-t_n\to\infty$ as $n\to\infty$, and
\emph{weak} otherwise.
\end{definition}
 
Let $L_{\text{weak}}$ denote the number of weak tests in $\mathcal{T}_\text{non-null}$. When $\delta_n\le1/2$, we have that $r_n(s)\le\omega_n$.
Moreover, if $M_n\to\infty$,
$\delta_n\to0$ and $\log(1/\delta_n)=o(\log M_n)$, then
\[
\omega_n
=
(1+o(1))\eta_n\sqrt{2\log M_n},
\qquad
r_n(L_{\text{weak}})
=
(1+o(1))\eta_n\sqrt{\frac{2\log M_n}{L_{\text{weak}}}}.
 \]
Thus, coordinating favorable perturbations across all $L_{\text{weak}}>0$ weak tests reduces the available reach by an asymptotic factor $\sqrt{1/L_{\text{weak}}}$ relative to the global reach.

\begin{theorem}[Coverage and collapse]\label{thm:main}
Suppose $\underline{\eta} \leq \eta_n \leq \bar{\eta}$ for constants $\bar{\eta},\underline{\eta} > 0$, fix $0<\nu<1$ and let $\delta_n\to0$.
Then:
\begin{enumerate}
\item[(a)] \textbf{Coverage}.
$\liminf_n
\mathbb{P}(\mathcal{G}^\star\in\widehat{\mathfrak{G}}_n)
\ge1-\nu$
if either of these two conditions holds:
\begin{itemize}
\item[(i)]
$t_n$ is bounded, and
$M_nt_n^{L_0}\to\infty$ as $n\to\infty$;
\item[(ii)]
$t_n\to\infty$ as $n\to\infty$, and every nonnull weak test $l$
satisfies $\kappa_l-t_n+r_n(L_{\text{weak}})\to\infty$.
\end{itemize}
\item[(b)] \textbf{Collapse}.
$\liminf_n
\mathbb{P}(\{\mathcal{G}^\star\}=\widehat{\mathfrak{G}}_n)
\ge1-\nu$
if $t_n-\omega_n\to\infty$ as $n\to\infty$ and
$\kappa_{\min,n}-t_n-\omega_n\to\infty$ as $n\to\infty$.
\end{enumerate}
\end{theorem}

We prove Theorem~\ref{thm:main} in
Appendix~\ref{app:analysis-proofs}.

\subsubsection{Interpreting the theorem}\label{sec:interpreting}
Part (a) gives two routes to coverage. In part (a)(i), the threshold remains bounded. Null statistics can therefore cross the threshold with nonnegligible probability, leading to substantial variability in the estimated graphs across runs. Nevertheless, each run has a nonvanishing probability of returning a true CPDAG, and sufficiently many independent runs make it likely that at least one run is correct. 

In part (a)(ii), the threshold grows with the number of runs. Null statistics are then increasingly unlikely to cross the threshold, reducing variability in the estimated graphs due to null tests. The main difficulty shifts to the weak nonnull signals. Coverage requires these signals to remain close enough to the threshold that favorable perturbations (i.e., $r_n(L_\text{weak})$) can push them across in at least one run. Strong signals already exceed the threshold by a diverging margin and therefore do not rely on favorable perturbations. 

Part (b) gives a stronger requirement for collapse. Every null statistic must lie below the threshold for every run, while every nonnull statistic must remain above it. The threshold must therefore dominate the largest null perturbation (i.e., $\omega_n$), and the weakest nonnull signal must exceed the threshold by more than the largest perturbation. 

The goals of coverage and collapse impose different signal requirements. Perturbations help coverage because a favorable run can reveal a signal, but they make collapse harder because not every run must be correct. Without a quantitative lower bound on the nonzero signals, no tuning can guarantee both coverage and collapse uniformly over faithful distributions. This agrees with the impossibility of uniform consistency for CPDAG estimation \citep{robins2003uniform,zhang2003strong}.

We next consider three threshold regimes.

\subsubsection{Default: reach-scaled threshold}\label{sec:default} We take
$$\eta_n = 1,\quad t_n = \sqrt{\frac{2\log M_n}{B}},$$
where $B > 0$ is a tuning parameter. Since $\omega_n = (1+o(1))\sqrt{2\log M_n}$ for large $M_n$, the threshold remains a fixed fraction $1/\sqrt{B}$ of the global reach as $M_n$ grows. If $L_{\text{weak}}=0$, Theorem~\ref{thm:main}(a)(ii) gives coverage for
every $B>0$. Otherwise, for large $M_n$
$$r_n(L_{\text{weak}}) - t_n = \Big(\frac{1+o(1)}{\sqrt{L_{\text{weak}}}} - \frac{1}{\sqrt{B}}\Big)\sqrt{2\log M_n}.$$
Thus, if $B>L_{\text{weak}}$, this difference diverges, so every weak test satisfies the condition of
Theorem~\ref{thm:main}(a)(ii). Hence, $B$ determines how many weak tests the coverage guarantee can accommodate. In particular, $B > L_1$ requires no lower bound on the nonzero partial correlations. 

For $B>1$, $t_n-\omega_n\to-\infty$, so the collapse condition in part~(b) of Theorem~\ref{thm:main} fails. Null statistics can cross the threshold in some runs, so the
collection need not collapse. Larger $B$ lowers the threshold, which
improves weak-signal recovery at the cost of more false rejections of
conditional independence; smaller $B$ does the reverse. 

\begin{remark}[Choosing $B$ in practice] We use $75\%$ of the data to construct, for each candidate value of $B$, the collection of CPDAGs returned by perturbed PC. For each collection, we compute the largest Gaussian likelihood attained by any graph in the collection. We then choose the smallest $B$ whose best likelihood is within $1\%$ of the largest likelihood achieved over all candidate values of $B$. This favors the least aggressive perturbation level that still recovers a graph with a near-optimal fit.
\end{remark}

\subsubsection{Collapse-oriented threshold}Alternatively, take $t_n = c\sqrt{2\log n}$ and $\eta_n = 1$ for any fixed $c>0$, with a fixed number of runs. Then, $\omega_n = O(1)$ so both coverage and collapse hold whenever $\kappa_{\text{min},n}-c\sqrt{2\log n} \to \infty$. In particular, it is sufficient that $\kappa_{\text{min},n}/\sqrt{\log n} \to \infty$, or equivalently, $\rho_{\text{min},n}\sqrt{n}/\sqrt{\log n} \to \infty$. Thus, the minimum nonzero partial correlation may decay nearly as fast as $\sqrt{\log n/n}$. This is the consistency regime: collapse is possible, but it requires a quantitative signal-separation condition analogous to strong faithfulness. For a fixed number of nodes $p$, this condition is weaker than Assumption (A4) of \cite{kalisch2007estimating}, which requires $O(n^\xi)\leq \rho_{\text{min},n}$ for some $\xi < 1/2$.

\subsubsection{Bounded threshold and the tuning of \cite{chang2026post}}
Part (a)(i) gives a third route to coverage. Suppose $t_n$ is bounded and $\eta_n$ is bounded above and away from zero. Then the nonnull factors in $\pi$ are bounded below by positive constants, while the null factors are of order $t_n$. Hence, $\pi_n \asymp t_n^{L_0}$. Given data \(\mathcal{D}\) for which \(E_n\) holds, the runs are independent, and the probability that \(\mathcal G^\star\) is returned in at least one of the \(M_n\) runs is at least \(1-\exp\{-M_n\pi_n\}\). Therefore, $M_nt_n^{L_0}\to \infty$ is sufficient for coverage without a quantitative lower bound on the nonzero signals. The tuning of \cite{chang2026post} is a special case: $\eta = 1$, $M_n \ge \log n$ and $t_n = c^\star(\log n/M_n)^{1/L}z_{\nu/2L} \le c^\star z_{\nu/2L}$ for some $c^\star>0$ with $c^\star(\log n/M_n)^{1/L} \leq 1$. Since
$x^{L_0/L} \ge x$ for $0 < x \le 1$, $M_nt_n^{L_0} \ge (c^\star z_{\nu/2L})^{L_0}\log n \to \infty$.

\begin{remark}\label{rem:chang-threshold}The threshold of \citet{chang2026post} and our default threshold have fundamentally different scaling. Since $c^\star(\log n/M_n)^{1/L}\leq 1$, their threshold is at most $ z_{\nu/2L}$ and therefore remains bounded in $M_n$. Our default threshold instead grows on the order of $\sqrt{\log M_n}$. Thus, no fixed choice of $c^\star$ can make the two thresholds comparable for large $M_n$.

Both thresholds permit coverage without a lower bound on the nonzero signals. Their different scaling, however, leads to different behavior under the null. The largest perturbation of a null standardized test statistic is of order $\sqrt{\log M_n}$, while the threshold of \citet{chang2026post} remains bounded. Null statistics therefore cross the threshold in some runs with high probability, producing denser graphs and greater variability in the retained collection. See experimental results in Section~\ref{sec:confidence-experiments}.
\end{remark}

\subsection{Constructing our confidence regions}\label{sec:regions}
While Theorem~\ref{thm:main} gives asymptotic coverage for $\widehat{\mathfrak{G}}_n$, two challenges remain. First, at moderate sample sizes, the collection may fail to contain the true CPDAG, especially when the signals are weak. Second, even with adequate coverage, a large collection of CPDAGs can be difficult to interpret. We therefore use $\widehat{\mathfrak{G}}_n$ to construct the confidence balls and intervals of Definitions~\ref{defn:ball} and~\ref{defn:CI}. These regions enlarge the collection to improve the coverage, and they summarize graph uncertainty in a concise form: a center and a radius, or a lower and an upper graph.

To construct the ball, we first estimate a center  $\widehat{\mathcal{G}}_{0,n}$ by applying PC to the unperturbed test statistics with threshold $\sqrt{2\log n}$.
This threshold diverges slowly enough to retain power against nonzero partial correlations while driving false rejections of conditional independence to zero. We then use $\widehat{\mathfrak{G}}_n$ to define the confidence ball as:
\begin{equation}
\widehat{\mathfrak{G}}_n^\text{ball}:= \{\G \in\mathfrak{G}: d_\text{model}(\G,\widehat{\G}_{0,n})\leq \widehat r_n\}~~\text{ with } \widehat r_n:= \max_{m\in\widehat{\mathcal{K}}_n} d_\text{model}(\widehat{\G}^{[m]}_n,\widehat{\mathcal{G}}_{0,n}),
\label{eqn:distance}
\end{equation}
which is the collection of graphs within model-oriented distance $\widehat r_n$ from the center $\widehat{\mathcal{G}}_{0,n}$. Here, $\widehat{r}_n$ is the maximum distance from the center to a graph in the collection $\widehat{\mathfrak{G}}_n$. 

Using $\widehat{\mathfrak{G}}_n$, we construct the confidence interval as:
\begin{eqnarray}
\begin{aligned}
\widehat{\mathfrak{G}}_n^\text{interval}:= \{\G\in\mathfrak{G}: \widehat{\G}_{n,L}\preceq \G \preceq \widehat{\G}_{n,U}\}~~\text{ where }\widehat{\G}_{n,L} \preceq  \widehat{\G}^{[m]}_n ~~  \text{ and }~~ \widehat{\G}^{[m]}_n \preceq \widehat{\G}_{n,U} ~~ \forall m\in \widehat{\mathcal{K}}_n.
\end{aligned}
\label{eqn:upper_lower_cover}
\end{eqnarray}
This region contains all CPDAGs lying in between $\widehat{\G}_{n,L}$ and $\widehat{\G}_{n,U}$ in the partial order, where $\widehat{\G}_{n,L}$ is any lower bound on all graphs in $\widehat{\mathfrak{G}}_n$, and $\widehat{\G}_{n,U}$ is any upper bound. Such bounds always exist: one may take $\widehat{\G}_{n,L}$ to be the empty CPDAG, the least element in the CPDAG poset, and  $\widehat{\G}_{n,U}$ to be the fully connected CPDAG, the greatest element. We therefore seek bounds that make the interval narrow, as measured by $\mathrm{rank}(\widehat{\G}_{n,U})-\mathrm{rank}(\widehat{\G}_{n,L})$. We will describe some strategies in Section~\ref{sec:computation}. 

By construction, \(\widehat{\mathfrak G}_n\subseteq\widehat{\mathfrak G}_n^{\mathrm{ball}}\) and \(\widehat{\mathfrak G}_n\subseteq\widehat{\mathfrak G}_n^{\mathrm{interval}}\). The following corollary therefore follows directly from Theorem 2.
\begin{corollary}\label{cor:regions}
Suppose the conditions for coverage hold from Theorem~\ref{thm:main}. Then, we have that $\widehat{\mathfrak{G}}_n^\text{ball}$ and $\widehat{\mathfrak{G}}_n^\text{interval}$ are asymptotically valid confidence regions, i.e., 
$$\lim\inf_n \mathbb{P}(\G^\star \in \widehat{\mathfrak{G}}_n^\text{ball}) \geq 1-\nu ~~\text{ and }~~\lim\inf_n \mathbb{P}(\G^\star \in \widehat{\mathfrak{G}}_n^\text{interval}) \geq 1-\nu.$$
\end{corollary}

Algorithm~\ref{alg:regions} provides full details for constructing $\widehat{\mathfrak{G}}_{n}^\text{ball}$ and $\widehat{\mathfrak{G}}_{n}^\text{interval}$.

\begin{algorithm}[t]
\label{alg:confidence_region}
\caption{Confidence ball and interval}\label{alg:regions}
\KwIn{$(\widehat{\mathcal{K}}_n,\widehat{\mathfrak{G}}_n)$ from Algorithm~\ref{alg:perturbed PC}}
\KwOut{Ball $\widehat{\mathfrak{G}}^{\mathrm{ball}}_n$ and interval $\widehat{\mathfrak{G}}^{\mathrm{interval}}_n$}

Run PC with threshold $\sqrt{2\log n}$ to get the center $\widehat{\mathcal{G}}_{0,n}$\;
$\widehat r_n \gets \max_{m  \in \widehat{\mathcal{K}}_n} d_{\text{model}}(\widehat{\G}_{0,n},\widehat{\G}_n^{[m]})$\;
$\widehat{\mathfrak{G}}^{\mathrm{ball}}_n \gets \{\mathcal{G} : d_{\mathrm{model}}(\mathcal{G}, \widehat{\mathcal{G}}_{0,n}) \le \widehat r_n\}$\;
$\widehat{\mathcal{G}}_{n,L} \gets$ CPDAG that satisfies $\widehat{\mathcal{G}}_{n,L} \preceq \G$ for all $\G\in \widehat{\mathfrak{G}}_n$\;
$\widehat{\mathcal{G}}_{n,U} \gets$ CPDAG that satisfies $\G \preceq \widehat{\mathcal{G}}_{n,U}$ for all $\G \in \widehat{\mathfrak{G}}_n$\;
$\widehat{\mathfrak{G}}^{\mathrm{interval}}_n \gets \{\mathcal{G} : \widehat{\mathcal{G}}_{n,L} \preceq \mathcal{G} \preceq \widehat{\mathcal{G}}_{n,U}\}$\;
\end{algorithm}

\section{Theoretical results for computation}\label{sec:distance-theory}
 
The confidence interval of Section~\ref{sec:regions} requires common lower and upper graphs of the retained collection. The confidence ball requires model-oriented distances from the center to each retained graph. Finding the narrowest interval or computing exact distances in general requires searching over the CPDAG poset. This section develops bounds to guide these searches. Section~\ref{sec:common-bounds} bounds the rank of the smallest common upper graph and the largest common lower graph. Section~\ref{sec:structured} gives exact distances for structured pairs. Section~\ref{sec:mismatch} provides bounds on the model-oriented distance, which partially uses the results from the previous two subsections. The theoretical results in this section are useful for the algorithms developed for obtaining confidence regions in Section~\ref{sec:computation}. 
 
Throughout, we denote a pair of CPDAGs of interest by $\G_s$ and $\G_t$, and use the shorthand notation $d = d_{\mathrm{model}}$.

 \subsection{Bounds for common lower and upper graphs}\label{sec:common-bounds}
Let ${\widehat{\mathfrak{G}}}_n\subseteq \mathfrak{G}$ be a collection of CPDAGs obtained from Algorithm~\ref{alg:perturbed PC}. As described in Algorithm~\ref{alg:regions}, for building a confidence interval, we seek a lower graph $\widehat{\G}_{n,L} \in \mathfrak{G}$ and upper graph $\widehat{\G}_{n,U} \in \mathfrak{G}$ satisfying $\widehat{\G}_{n,L} \preceq \G \preceq \widehat{\G}_{n,U}$ for all $\G \in {\widehat{\mathfrak{G}}}_n$. The width of this interval is given by $\mathrm{rank}(\widehat{\G}_{n,U})-\mathrm{rank}(\widehat{\G}_{n,L})$. To obtain these two graphs, our algorithm (described in Section~\ref{sec:computation}) constructs these graphs sequentially by finding common lower and upper graphs for pairs of CPDAGs. To keep the final interval narrow, we seek the largest-rank common lower graph and the smallest-rank common upper graph for any pair $\G_s,\G_t\in\mathfrak G$:
\begin{eqnarray*}
\begin{aligned}
\argmax_{\G\in\mathfrak{G}: \G \preceq \G_s,\G \preceq \G_t}\mathrm{rank}(\G) \quad \text{ and }\quad \argmin_{\G\in\mathfrak{G}: \G_s \preceq \G,\G_t\preceq \G}\mathrm{rank}(\G). 
\end{aligned}
\end{eqnarray*}
These optimization problems are well-defined as the empty graph $\G_\text{empty}$ is always a feasible graph in the first optimization and the fully connected graph $\G_\text{full}$ is always feasible in the second optimization problem. To solve these optimization problems, the procedures described in Section~\ref{sec:computation} require bounds on the maximum and minimum objective values of the optimization problem above. We first define:
\begin{align*}
\rho_L(\G_s,\G_t) := \max_{\G\in\mathfrak{G}: \G \preceq \G_s, \G\preceq \G_t} \mathrm{rank}(\G)\quad \text{ and } \quad
\rho_U(\G_s,\G_t) := \min_{\G\in\mathfrak{G}: \G_s \preceq \G, \G_t \preceq \G} \mathrm{rank}(\G).
\end{align*}
Section~\ref{bounds:lower} develops lower and upper bounds for $\rho_L(\G_s,\G_t)$. Section~\ref{bounds:upper} develops bounds for $\rho_U(\G_s,\G_t)$. Each subsection constructs a feasible common graph and bounds the best achievable rank. The constructed graphs provide feasible interval endpoints, while the remaining bounds limit how much their ranks can improve during the search of Section~\ref{sec:common-bounds-computation}. We use the following notation:
\begin{itemize}
\item $\mathcal D_\cap(\G_s,\G_t)$ is the intersection of chosen DAG representatives of $\G_s$ and $\G_t$;
\item $\mathrm{Skel}_\cap(\G_s,\G_t)$ and $\mathrm{Skel}_\cup(\G_s,\G_t)$ are the intersection and union of their skeletons;
\item $\mathcal G_{\mathrm{ch}}(\G_s,\G_t)$ is a chordal completion of the union of the moral graphs of $\G_s$ and $\G_t$.
\end{itemize}

\subsubsection{Bounds for constructing a common lower graph}
\label{bounds:lower}
For the confidence interval, we want a common lower graph with rank as large as possible, since a larger lower endpoint produces a narrower interval. For a pair $\G_s,\G_t$, this means determining how many edges can be retained while remaining below both graphs. We first construct a feasible common lower graph, which gives a lower bound on $\rho_L(\G_s,\G_t)$, and then identify edge conflicts that limit how large any common lower graph can be.

A simple feasible construction comes from intersecting DAG representatives. The DAG $\mathcal D_\cap(\G_s,\G_t)$ is a subgraph of the chosen DAG representatives of $\G_s$ and $\G_t$. Hence, the CPDAG of $\mathcal D_\cap(\G_s,\G_t)$ is a common lower graph, giving
$$|E(\mathcal D_\cap(\G_s,\G_t))| \leq \rho_L(\G_s,\G_t).$$
Here, $E(\cdot)$ denotes the edge set of the input graph. The quality of this bound depends on the chosen representatives. For example, if both CPDAGs consist of the undirected edge $u-v$, opposite orientations of their DAG representatives give an empty intersection, whereas matching orientations retain the edge. Thus, the choice of representatives affects the tightness of the bound, but not its validity. 

To upper bound $\rho_L(\G_s,\G_t)$, we ask which shared edges cannot be retained simultaneously. Every common lower graph has its skeleton inside $\mathrm{Skel}_\cap(\G_s,\G_t)$, but orientation differences can force some of these shared edges to be deleted. The simplest example is
\[
\G_s:\ u\to w\leftarrow v,
\qquad
\G_t:\ u-w-v.
\]
The collider implies marginal independence of $u$ and $v$, whereas the chain implies independence conditional on $w$. Any common lower graph must satisfy both constraints, so it must delete at least one of the two shared edges. The following definition identifies this type of conflict inside larger graphs.

\begin{definition}[Forbidden triple, legs, and shields]\label{def:forbidden-triple}
Let $\operatorname{v-str}(\G)$ denote the set of triples $(u,w,v)$
that form a v-structure at $w$ in $\G$. The set of \emph{forbidden triples} is
\[
\mathcal S(\G_s,\G_t):=
\left\{
(u,w,v)\in V^3:
\begin{array}{l}
\{u,w\},\{v,w\}\in\mathrm{Skel}(\G_s)\cap\mathrm{Skel}(\G_t),\\
\{u,v\}\notin\mathrm{Skel}(\G_s)\cup\mathrm{Skel}(\G_t),\\
(u,w,v)\in\operatorname{v-str}(\G_s)\triangle\operatorname{v-str}(\G_t)
\end{array}
\right\}.
\]
For each forbidden triple $(u,w,v)$, where the order of $u$ and $v$ is irrelevant, $\{u,w\}$ and $\{v,w\}$ are its \emph{legs}. The missing outer pair $\{u,v\}$ is its \emph{shield}.
\end{definition}
 
Each forbidden triple requires at least one of its two legs to be omitted from any common lower graph. Because one omitted edge may resolve several forbidden triples, the edge-deletion requirements can overlap. We capture this overlap with an auxiliary conflict graph.

\begin{definition}[Skeleton conflict graph]\label{def:skeleton-conflict}
The \emph{skeleton conflict graph} $\Gamma_L(\G_s,\G_t)$ is an undirected graph with
vertex set $\mathrm{Skel}_\cap(\G_s,\G_t)$. Two vertices
are adjacent if the corresponding shared skeleton edges are the two legs of a forbidden triple.
\end{definition}
Vertices of $\Gamma_L(\G_s,\G_t)$ therefore represent shared edges that could potentially appear in a common lower graph, while edges of $\Gamma_L(\G_s,\G_t)$ represent pairs that cannot be retained together. The shared edges omitted by any common lower graph must therefore form a vertex cover of $\Gamma_L(\G_s,\G_t)$. If $\tau(\Gamma_L(\G_s,\G_t))$ denotes the minimum vertex-cover size, then
$$\rho_L(\G_s,\G_t) \leq |\mathrm{Skel}_\cap(\G_s,\G_t)|-\tau(\Gamma_L(\G_s,\G_t)).$$
This bound caps the rank of every common lower graph, but computing a minimum vertex cover is NP-hard \citep{karp1972reducibility}. 

For the search algorithm, we therefore also want a weaker bound that can be computed quickly. A maximum matching provides such a bound. A \emph{matching} of $\Gamma_L(\G_s,\G_t)$ is a set of edges with no shared endpoints, and a \emph{maximum matching} of $\Gamma_L(\G_s,\G_t)$ has the largest possible size, denoted by
$$\nu(\Gamma_L(\G_s,\G_t)).$$
Since every matched edge requires a distinct vertex in any vertex cover, we have
$\nu(\Gamma_L(\G_s,\G_t))\leq\tau(\Gamma_L(\G_s,\G_t))$. 

Any maximum matching on a graph with $n$ vertices and $m$ edges takes $O(m\sqrt{n})$ time \citep{micali1980v}. Since $\Gamma_L(\mathcal{G}_s,\mathcal{G}_t)$ has at most $|V|^2$ vertices and one edge per forbidden triple, the forbidden triples can be listed and the matching bound computed in polynomial time. The following proposition summarizes the results, with formal proof in Appendix~\ref{app:bound-proofs}. 
\begin{proposition}[Bounds for common lower graphs]
\label{prop:common-lower}
For every $\G_s,\G_t \in \mathfrak{G}$, we have
\begin{align*}
|E(\mathcal D_\cap(\G_s,\G_t))|
&\leq \rho_L(\G_s,\G_t)
\leq |\mathrm{Skel}_\cap(\G_s,\G_t)|-\tau(\Gamma_L(\G_s,\G_t))\leq |\mathrm{Skel}_\cap(\G_s,\G_t)|-\nu(\Gamma_L(\G_s,\G_t)).
\end{align*}
\end{proposition}
The bounds can sometimes determine the optimal common lower graph without any search. In particular, if  $|E(\mathcal{D}_\cap(\mathcal{G}_s,\mathcal{G}_t))|=|\mathrm{Skel}_\cap(\mathcal{G}_s,\mathcal{G}_t)|-\nu(\Gamma_L(\mathcal{G}_s,\mathcal{G}_t))$, every inequality in Proposition~\ref{prop:common-lower} holds with equality. The CPDAG of $\mathcal{D}_\cap(\mathcal{G}_s,\mathcal{G}_t)$ is then a largest-rank common lower graph, so the lower graph can be obtained directly. 

\subsubsection{Bounds for constructing a common upper graph}
 \label{bounds:upper}
For the confidence interval, we also want a common upper graph with rank as small as possible, since a smaller upper endpoint produces a narrower interval. Every common upper graph must contain $\mathrm{Skel}_\cup(\G_s,\G_t)$, giving the immediate lower bound
$$|\mathrm{Skel}_\cup(\mathcal{G}_s,\mathcal{G}_t)| \leq \rho_{U}(\mathcal{G}_s,\mathcal{G}_t).$$
To obtain an upper bound on $\rho(\G_s,\G_t)$, we construct a feasible common upper graph. We first take the union of the moral
graphs of $\G_s$ and $\G_t$, where each moral graph is obtained by joining the two endpoints of every v-structure. We then add edges until the union is chordal, meaning that every cycle of length at least four has an edge joining two nonconsecutive vertices. This operation is a \emph{chordal completion} \citep{rose1976algorithmic}. We denote the resulting undirected graph by $\mathcal G_{\mathrm{ch}}(\G_s,\G_t)$.

The graph $\mathcal G_{\mathrm{ch}}(\G_s,\G_t)$ is a CPDAG and a common upper graph, and therefore
\[
\rho_U(\G_s,\G_t)
\leq
\mathrm{rank}(\mathcal G_{\mathrm{ch}}(\G_s,\G_t)).
\]
Any chordal completion gives a valid bound, but completions with fewer added edges give tighter bounds. A minimal completion, in which no added edge can be removed without breaking chordality, can be computed in $O(|V|^3)$ time \citep{rose1976algorithmic}. In contrast, finding a completion with the fewest added edges is NP-hard \citep{yannakakis1981computing}. The proof of the following proposition is given in Appendix~\ref{app:bound-proofs}.

\begin{proposition}[Bounds for common upper graphs]
\label{prop:common-upper}
For every $\G_s,\G_t \in \mathfrak{G}$, we have
\[
|\mathrm{Skel}_\cup(\G_s,\G_t)|
\leq \rho_U(\G_s,\G_t)
\leq \mathrm{rank}(\mathcal G_{\mathrm{ch}}(\G_s,\G_t)).
\]
\end{proposition}
 
The constructions and Propositions~\ref{prop:common-lower} and~\ref{prop:common-upper}
extend  directly to any finite collection $\mathcal A$ of CPDAGs. We replace pairwise intersections and unions by intersections and unions over all graphs in $\mathcal A$. A triple is forbidden for $\mathcal A$ if both legs appear in every graph, the shield appears in no graph, and the triple is a v-structure in some but not all graphs in $\mathcal A$.

\subsection{Exact distances for structured pairs}\label{sec:structured}
We show how the model-oriented distance can be computed efficiently when the endpoint graphs $\G_s$ and $\G_t$ have special structure, namely shared disconnected components or polytree skeletons. In Section~\ref{sec:distance-computation}, we use these results to obtain upper bounds on the distance for general CPDAGs by projecting them onto these structured classes.

\subsubsection{Distance decomposition over disconnected components}

When both graphs $\G_s$ and $\G_t$ 
decompose into the same disconnected components, the distance can be computed separately within each component. For $W\subseteq V$, define $\G|_W$ as the CPDAG corresponding to $\mathcal D[W]$,
where $\mathcal D[W]$ is the subgraph induced by $W$ in any DAG representative $\mathcal D$ of $\G$. The restriction does not depend on the representative, since all choices have the same induced skeleton and v-structures. Let $d_W$ denote the model-oriented distance between CPDAGs on the vertex set $W$.

\begin{theorem}[Decomposition]\label{thm:decomposition}
Let $V_1,\ldots,V_K$ partition $V$ so that neither $\G_s$ nor
$\G_t$ has an edge between distinct components. Then
\[
d(\G_s,\G_t)
=\sum_{k=1}^K d_{V_k}(\G_s|_{V_k},\G_t|_{V_k}),
\]
and no graph on any shortest path has an edge between different components.
\end{theorem}
We prove this theorem in Appendix~\ref{app:decomposition-proof}. The theorem shows that when the two endpoint graphs decompose over the same groups, the distance decomposes additively across them. Thus, the full distance problem can be solved separately within each component, potentially leading to substantial computational savings. 

The connected components of
$\mathrm{Skel}(\G_s)\cup\mathrm{Skel}(\G_t)$ give the finest
partition satisfying the conditions of the theorem. 

\subsubsection{Polytrees}
We call a CPDAG a \emph{polytree} when its skeleton has no undirected cycles. For two polytrees, as we establish next, there always exists a shortest path that is both a down-up path and consists only of polytrees.
We use $d_{\downarrow\uparrow}(\mathcal{G}_s,\mathcal{G}_t)$ from Remark~\ref{rem:down-up}.
\begin{theorem}[Polytrees]\label{thm:polytree}
Let $\mathcal{G}_s$ and $\mathcal{G}_t$ be polytrees. Then
\[
d(\G_s,\G_t)
=d_{\downarrow\uparrow}(\G_s,\G_t).
\]
Moreover, every graph on a down-up path from $\mathcal{G}_s$ to $\mathcal{G}_t$ is a polytree. 
\end{theorem}
We prove this theorem in Appendix~\ref{app:polytree-proof}. \citet{taeb2025distance} establish a similar result when the search space is restricted to polytrees. Our theorem removes this restriction: only the two endpoint graphs need to be polytrees, and the minimum is taken over all paths, including paths through arbitrary CPDAGs.

\subsection{Bounds on the model-oriented distance}\label{sec:mismatch}
In this section, we develop upper and lower bounds on the model-oriented distance $d(\G_s,\G_t)$. 

\subsubsection{Upper bounds}
\label{sec:upper_bounds_model}
A down-up path, defined in Remark~\ref{rem:down-up}, can descend from $\G_s$ to a common lower graph $\G_L$
and then ascend to $\G_t$. Since each step changes rank by one,
this path has length
$\mathrm{rank}(\G_s)+\mathrm{rank}(\G_t)-2\mathrm{rank}(\G_L)$.
One can similarly define an up-down path through a common upper graph $\G_U$. Since both of these are just one set of feasible paths, we have
\[
d(\G_s,\G_t)\leq\min\left\{
\begin{aligned}
&\mathrm{rank}(\G_s)+\mathrm{rank}(\G_t)-2\mathrm{rank}(\G_L),\\
&2\mathrm{rank}(\G_U)-\mathrm{rank}(\G_s)-\mathrm{rank}(\G_t)
\end{aligned}
\right\}.
\]
Thus, Propositions~\ref{prop:common-lower} and~\ref{prop:common-upper} give us distance upper bounds without requiring optimal common graphs. 

We also construct stronger upper bounds by projecting $\G_s$ and $\G_t$ onto polytree and disconnected component graph classes for which the distance is easier to compute; see Section~\ref{sec:structured}. For the polytree projection, we retain a depth-first-search forest of each skeleton, delete the remaining edges from a consistent DAG extension, and convert the result back to the CPDAG. The projected graphs are polytrees, so their distance is given exactly by Theorem~\ref{thm:polytree}. For the component projection, we partition the vertex set of the union skeleton into disjoint blocks, delete all edges between blocks from consistent DAG extensions, and convert the results back to CPDAGs. The distance between the projected graphs then decomposes across these blocks and can be computed exactly.

Each projection is obtained from the original graph through a known number of edge deletions. The triangle inequality therefore bounds the original distance by the projected distance plus the deletion counts. We denote the resulting bounds by
$$u_\text{component}(\G_s,\G_t) \text{ and } u_\text{polytree}(\G_s,\G_t).$$
 
\subsubsection{Detour operations}
 
Neighboring CPDAGs have DAG representatives that differ by one edge
addition or deletion \citep{chickering2002optimal}. Hence, every path must change each pair in 
$\mathrm{Skel}(\G_s)\triangle\mathrm{Skel}(\G_t)$ at least once.
Orientation constraints can force additional temporary changes.
 
\begin{definition}[Detour operations]\label{def:detour-operations}
A step of a path from $\G_s$ to $\G_t$ is a \emph{detour} if it changes the status of a vertex pair away from its state in the target skeleton. There are two types of \emph{detour operations}:
\[
\begin{aligned}
\mathrm{ER}(a,b):&\quad\text{remove }\{a,b\}\in\mathrm{Skel}(\G_t),\\
\mathrm{EA}(a,b):&\quad\text{add }\{a,b\}\notin\mathrm{Skel}(\G_t).
\end{aligned}
\]
\end{definition}
 Every detour operation must later be undone. If $q(P)$ counts the detour operations along a path $P$, including repetitions, then
\begin{equation}
\mathrm{length}(P)
=|\mathrm{Skel}(\G_s)\triangle\mathrm{Skel}(\G_t)|+2q(P).
\label{eq:path-detour-operations}
\end{equation}
Thus, every unavoidable detour operation increases the skeleton lower bound by two.

\begin{figure}[t]
\centering
\begin{minipage}[t]{0.46\linewidth}
\vspace{0pt}
\centering
\begin{tikzpicture}[>=stealth,
every node/.style={circle,draw=nodeoutline,fill=nodeblue,
minimum size=7mm,inner sep=1pt}]
\node (a) at (0,1.4) {$a$};
\node (b) at (1.2,0) {$b$};
\node (c) at (2.4,1.4) {$c$};
\draw[->,thick,edgegreen] (a)--(b);
\draw[->,thick,edgegreen] (c)--(b);
\end{tikzpicture}
 
$\G_s$
 
\medskip
 
\begin{tikzpicture}[>=stealth,
every node/.style={circle,draw=nodeoutline,fill=nodeblue,
minimum size=7mm,inner sep=1pt}]
\node (a) at (0,1.4) {$a$};
\node (b) at (1.2,0) {$b$};
\node (c) at (2.4,1.4) {$c$};
\draw[thick,edgegreen] (a)--(b);
\draw[thick,edgegreen] (c)--(b);
\end{tikzpicture}
 
$\G_t$
 
\medskip
{\small Conflict graph $\Gamma_L(\G_s,\G_t)$}
 
\smallskip
\begin{tikzpicture}[every node/.style={draw=nodeoutline,
rounded corners,fill=nodeblue,inner sep=3pt,font=\small}]
\node (e1) at (0,0) {$\{a,b\}$};
\node (e2) at (2.2,0) {$\{b,c\}$};
\draw[thick,edgegreen] (e1)--(e2);
\end{tikzpicture}
 
$\tau(\Gamma_L(\G_s,\G_t))=\nu(\Gamma_L(\G_s,\G_t))=1$
 
\medskip
(a) Forbidden triple
\end{minipage}
\hfill
\begin{minipage}[t]{0.46\linewidth}
\vspace{0pt}
\centering
\begin{tikzpicture}[>=stealth,
every node/.style={circle,draw=nodeoutline,fill=nodeblue,
minimum size=7mm,inner sep=1pt}]
\node (a) at (0,1.5) {$a$};
\node (b) at (0,0) {$b$};
\node (c) at (2,0) {$c$};
\node (d) at (2,1.5) {$d$};
\draw[->,thick,edgegreen] (a)--(b);
\draw[->,thick,edgegreen] (d)--(b);
\draw[->,thick,edgegreen] (b)--(c);
\draw[->,thick,edgegreen] (d)--(c);
\end{tikzpicture}
 
$\G_s$
 
\medskip
 
\begin{tikzpicture}[>=stealth,
every node/.style={circle,draw=nodeoutline,fill=nodeblue,
minimum size=7mm,inner sep=1pt}]
\node (a) at (0,1.5) {$a$};
\node (b) at (0,0) {$b$};
\node (c) at (2,0) {$c$};
\node (d) at (2,1.5) {$d$};
\draw[thick,edgegreen] (a)--(b);
\draw[thick,edgegreen] (b)--(c);
\draw[thick,edgegreen] (c)--(d);
\end{tikzpicture}
 
$\G_t$
 
\medskip
(b) Overlapping configuration
\end{minipage}
 
\caption{\small Two configurations that force detour operations.
(a) The skeletons agree, but only the source has the collider
$a\to b\leftarrow c$. Every path must temporarily delete
$\{a,b\}$ or $\{b,c\}$, or add $\{a,c\}$, and later undo
that change. The distance is two.
(b) The skeletons differ only in $\{b,d\}$. Deleting that edge
directly creates a collider absent from the target. The distance
is three, attained by deleting $\{a,b\}$, deleting $\{b,d\}$,
and restoring $\{a,b\}$ through suitable DAG representatives.}
\label{fig:mismatch-examples}
\end{figure}
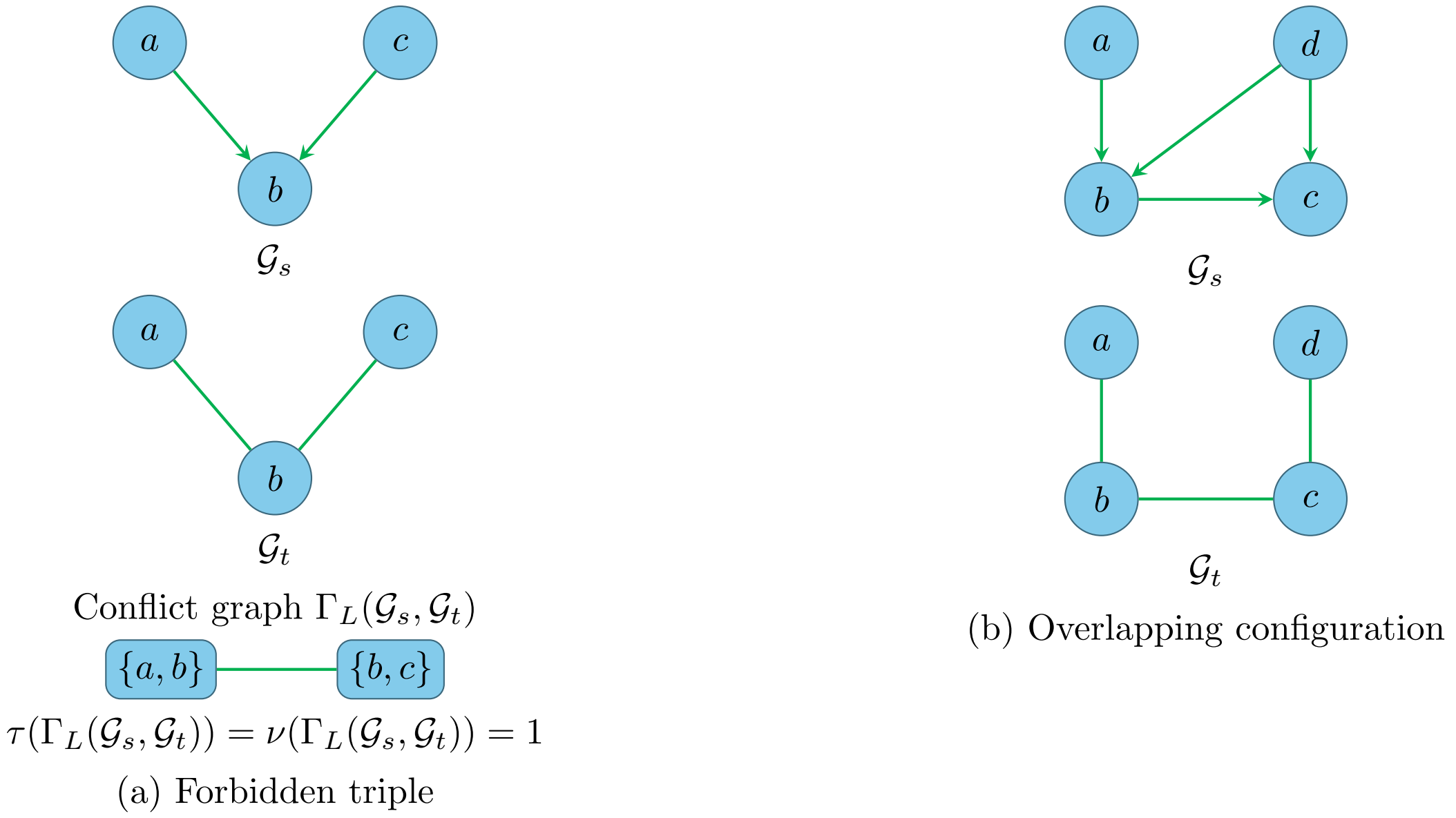

For the collider and chain in Figure~\ref{fig:mismatch-examples}(a),
any path must delete a leg or add the shield before the collider
can change. Each such step is a detour and must later be reversed, so the distance is at least two. Deleting one leg and restoring that leg
through a different DAG representative attains this bound.

More generally, every forbidden triple $(u,w,v)$ (see Definition~\ref{def:forbidden-triple}) requires at least one operation from
\begin{equation}
M_{(a,b,c)}=
\{\mathrm{ER}(a,b),\ \mathrm{ER}(b,c),\ \mathrm{EA}(a,c)\}.
\label{eqn:forbidden}
\end{equation}

A single operation may resolve several forbidden triples, so counting them separately can overcount. The bound of \citet{taeb2025distance} accounts for this overlap. We next extend the same idea to conflicts involving more than three vertices.

\subsubsection{General mismatches and their lower bound}
Forbidden triples do not capture every configuration that forces detour operations. The following example shows a four-node configuration with no forbidden triple, yet the distance is strictly larger than the skeleton difference.
 
\begin{example}[A conflict on four vertices]
\label{ex:overlapping-mismatch}
In Figure~\ref{fig:mismatch-examples}(b), the skeletons of $\G_s$ and $\G_t$ differ only in the edge $\{b,d\}$. However, deleting $\{b,d\}$ from $\G_s$ creates the v-structure $b \rightarrow c \leftarrow d$, which is absent from $\mathcal G_t$. Thus, any path must make an additional change before deleting $\{b,d\}$, and later undo that change. Hence, $d(\G_s,\G_t)\geq 3$. This bound is attained by deleting from $\G_s$ $\{a,b\}$, then $\{b,d\}$, and finally restoring $\{a,b\}$.

\end{example}
This example motivates the following general notion.
 
\begin{definition}[Mismatch]\label{def:mismatch}
A \emph{mismatch} for $(\G_s,\G_t)$ is a set $M$ of detour 
operations such that every path from $\G_s$ to $\G_t$ must perform at least one operation in $M$.
\end{definition}
A forbidden triple is one example of a mismatch. The four-node configuration above gives another. In that example,
\[
\{\mathrm{ER}(a,b),\mathrm{ER}(b,c),\mathrm{ER}(c,d),
\mathrm{EA}(a,c),\mathrm{EA}(a,d),\mathrm{EA}(b,d)\},
\]
is a mismatch for $(\mathcal G_s,\mathcal G_t)$.

Now let $\mathcal M$ be any collection of mismatches. Since every path must hit every $M \in \mathcal{M}$, its detour operations must form a hitting set for $\mathcal{M}$. Define
\[
\widetilde m_{\mathrm{op}}(\mathcal M)
=\min\left\{|H|:
H\subseteq\bigcup_{M\in\mathcal M}M,\quad
H\cap M\neq\varnothing\ \text{for every }M\in\mathcal M
\right\}.
\]
Thus, $\widetilde m_{\mathrm{op}}(\mathcal M)$ is the minimum number of distinct detour operations needed to hit every mismatch. If one operation resolves several mismatches, it is counted only once.
 
\begin{theorem}[Mismatch lower bound]\label{thm:mismatch}
For every collection $\mathcal M$ of mismatches for $(\G_s,\G_t)$,
\[
d(\G_s,\G_t)
\geq|\mathrm{Skel}(\G_s)\triangle\mathrm{Skel}(\G_t)|
+2\widetilde m_{\mathrm{op}}(\mathcal M).
\]
\end{theorem}
We prove this theorem in Appendix~\ref{app:mismatch-proof}. The bound follows because every path's detour operations form a hitting set, and each detour operation must eventually be undone (hence, the factor of two). When $\mathcal{M}$ contains only the mismatches induced by forbidden triples, the theorem recovers the bound of
\citet{taeb2025distance}. Adding mismatches from larger configurations strengthens the bound, as the four-node example above shows. In Figure~\ref{fig:additional-mismatches}, we provide three additional four-node mismatch configurations. 
 
\subsubsection{A quickly computable lower bound}
 
Solving the minimum hitting-set problem can be costly. A simpler bound follows by selecting mismatches that share no detour operation. Each selected mismatch must then contribute a distinct detour operation.

To check disjointness, define the support of a mismatch as the set of vertices appearing in its operations. Since every operation acts on a pair of vertices, two mismatches whose supports share at most one vertex cannot share an operation.

\begin{example}[Disjoint and overlapping mismatches]\label{ex:disjoint-mismatches}
Let $\G_s$ be $u\to w\leftarrow v\to x\leftarrow y$, and let $\G_t$
be the undirected path $u-w-v-x-y$. The two skeletons agree. The
forbidden triples $(u,w,v)$ and $(v,x,y)$ have supports $\{u,w,v\}$
and $\{v,x,y\}$, which share only $v$. Their mismatches are therefore disjoint, so Corollary~\ref{thm:independent-conflicts} below gives
$d(\G_s,\G_t)\geq 2\cdot 2=4$, which is tight. In contrast, consider the four-node example on $\{a,b,c,d\}$ where $\G_s$ has edges $a\to c$, $b\to c$ and $d\to c$, and $\G_t$ is the undirected star centered at $c$. The forbidden triples $(a,c,b)$ and $(a,c,d)$ share the pair $\{a,c\}$. Hence, both corresponding mismatches contain $\mathrm{ER}(a,c)$, so they are not disjoint and at most one can be included in a collection $I$ of pairwise disjoint mismatches.
\end{example}
 
\begin{corollary}[Disjoint mismatches]
\label{thm:independent-conflicts}
Let $I$ be a collection of pairwise disjoint mismatches. Then
\[
d(\G_s,\G_t)
\geq|\mathrm{Skel}(\G_s)\triangle\mathrm{Skel}(\G_t)|+2|I|.
\]
\end{corollary}
 
A collection $I$ can be built greedily
by retaining a mismatch whenever its operations are disjoint from
those already selected; the collection need not have maximum size
for the bound to hold. Scanning all forbidden triples takes
$O(|V|^3)$ time when selected operations are recorded as the
collection grows. Four-vertex configurations may improve the bound
at additional computational cost.

\section{Computational strategies}
\label{sec:computation}
The confidence regions of Section~\ref{sec:regions} need two computations on the retained collection $\widehat{\mathfrak{G}}_n$ from perturbed PC. 
First, we compute distances from the center $\widehat{\mathcal{G}}_{0,n}$ to the retained graphs to obtain the radius $\widehat r_n$, and construct confidence balls. Second, we find common lower and upper bounds of the retained graphs to construct confidence intervals. We compute both with A$^\ast$ search over the Hasse diagram of $\mathcal{L}$, leveraging the bounds from Section~\ref{sec:distance-theory} as heuristics and feasible solutions.

\subsection{A unified A$^\ast$ procedure}
\label{sec:unified-astar}
We perform the required computations using the A$^\ast$ algorithm of \citet{hart1968formal} on the Hasse diagram of $\mathcal{L}$. Unlike breadth-first search, A$^\ast$ prioritizes states using lower bounds on the remaining search cost, allowing shortest paths or extremal feasible elements to be found efficiently. In our setting, the search graph is the undirected Hasse diagram of $\mathcal{L}$, whose edges are covering relations of unit length. Thus, shortest path search yields the model-oriented distance needed for the confidence ball, while constrained searches over the same poset yield common lower and upper graphs needed for the confidence interval. 

We use a common A$^\ast$ framework for these tasks. Each search is specified by the following ingredients:
\begin{itemize}
\item $\G_s$: initial graph where the search begins;
\item $S(\mathcal{G})$: the neighboring CPDAGs that the search may visit from $\mathcal{G}$;
\item $T(\G) \in \{0,1\}$: $T(\G) = 1$ if $\G$ is a terminal state; 
\item $h(\G)$: a lower bound, called a heuristic, on the remaining distance from $\G$ to a terminal state;
\item $C(\G)$: a tuple consisting of a feasible terminal state together with the length of a path from \(\mathcal G\) to that state;
\item $U_0$: initial upper bound for the distance from $\G_s$ to a terminal state.
\end{itemize}
A$^\ast$ also keeps $g(\mathcal{G})$, the length of the shortest path from $\mathcal{G}_s$ to $\mathcal{G}$ found so far, and expands states in increasing order of
$$f(\G) := g(\G)+h(\G).$$
The completion $C(\mathcal{G})$ provides an upper bound on the shortest feasible path length from the start to a terminal state, so states that cannot improve the current best solution can be pruned. Once the smallest remaining \(f\)-value reaches the current upper bound, the best solution is certified optimal. Note that constrained searches use the same algorithm, with \(S\) and \(T\) restricted to the admissible portion of the Hasse diagram. 

Algorithm~\ref{alg:unified-astar} summarizes the unified search. It returns the best feasible terminal state found and certifies optimality whenever the A$^\ast$ lower bound meets the current upper bound. If terminated early, it returns the best feasible solution without certification. Using the returned path lengths as distance upper bounds for the ball, or the returned common graphs as feasible interval endpoints, yields conservative regions and therefore preserves the coverage guarantee.

\begin{algorithm}[t]
\caption{Unified A$^\ast$ search over the CPDAG poset}
\label{alg:unified-astar}
\KwIn{Initial graph $\mathcal{G}_s$; successor operator $S$; terminal condition $T$; heuristic $h$; completion $C$; initial upper bound $U_0$ (default $\infty$)}
\KwOut{Best terminal graph and path length, with optimality flag}

$(\mathcal{G}_{\mathrm{best}},\ell)\gets C(\mathcal{G}_s)$; $U\gets\min\{\ell,U_0\}$\;
$g(\mathcal{G}_s)\gets 0$; $f(\mathcal{G}_s)\gets h(\mathcal{G}_s)$\;
$\mathrm{OPEN}\gets\{\mathcal{G}_s\}$, ordered by $(f,-g)$\;

\While{$\mathrm{OPEN}\neq\emptyset$ and time remains}{
    $\mathcal{G}\gets\mathrm{OPEN.popmin}()$\;

    \If{$f(\mathcal{G})\ge U$}{
        \Return{$\mathcal{G}_{\mathrm{best}},U$, optimal}\;
    }

    \If{$T(\mathcal{G})$}{
        $\mathcal{G}_{\mathrm{best}}\gets\mathcal{G}$;
        $U\gets g(\mathcal{G})$\;
        \Return{$\mathcal{G}_{\mathrm{best}},U$, optimal}\;
    }

    \ForEach{$\mathcal{G}'\in S(\mathcal{G})$, subject to pruning restrictions}{
        $g'\gets g(\mathcal{G})+1$\;

        \If{$\mathcal{G}'$ is unseen or $g'<g(\mathcal{G}')$}{
            $g(\mathcal{G}')\gets g'$\;

            $(\mathcal{G}_c,\ell)\gets C(\mathcal{G}')$\;
            \If{$g'+\ell<U$}{
                $\mathcal{G}_{\mathrm{best}}\gets\mathcal{G}_c$\;
                $U\gets g'+\ell$\;
            }

            $f(\mathcal{G}')\gets g'+h(\mathcal{G}')$\;

            \If{$f(\mathcal{G}')<U$}{
                $\mathrm{OPEN}\gets\mathrm{OPEN}\cup\{\mathcal{G}'\}$\;
            }
        }
    }
}

\If{$\mathrm{OPEN}=\emptyset$}{
    \Return{$\mathcal{G}_{\mathrm{best}},U$, optimal}\;
}
\Return{$\mathcal{G}_{\mathrm{best}},U$, not certified optimal}\;
\end{algorithm}

\subsection{Computing the confidence ball}
\label{sec:distance-computation}
To construct the confidence ball, we need the model-oriented distance from the center $\widehat{\G}_{0,n}$ to each graph in $\widehat{\mathfrak{G}}_n$. We therefore solve the same shortest path problem repeatedly, taking $\G_s =\widehat{\G}_{0,n}$ as the start graph and $\G_t \in \widehat{\mathfrak{G}}_n$ as the target. 

For a generic pair $(\G_s,\G_t)$, let $\mathcal{N}(\G)$ be the neighbors of graph $\G$ in the CPDAG Hasse diagram, i.e., the CPDAGs that either are covered by $\G$ or cover $\G$. To compute the distance from $\mathcal{G}_s$ to $\mathcal{G}_t$, the search starts at $\mathcal{G}_s$, and uses
\[
S_D(\mathcal{G})=\mathcal{N}(\mathcal{G}),
\qquad
T_D(\mathcal{G})=\mathbf{1}\{\mathcal{G}=\mathcal{G}_t\}.
\]
Thus, the search may move along any covering relation and terminate when it reaches $\G_t$.

\subsubsection{Computing the heuristic $h$}
In order to certify optimality of A$^\ast$, the heuristic must be \emph{admissible}, meaning $h(\mathcal G)$ lower bounds the shortest path length from $\mathcal G$ to a terminal
state \citep{hart1968formal}. By Theorem~\ref{thm:mismatch}, every collection of mismatches for
$(\G,\G_t)$ gives an admissible heuristic. A heuristic
is also \emph{consistent} if $h(\mathcal G)=0$ at terminal states and
$h(\mathcal G)\le1+h(\mathcal G')$ for every $\mathcal G'\in S_D(\mathcal G)$. With a consistent heuristic, no state is reopened, so each state is expanded
at most once, improving the efficiency of the search. We next establish consistency for the collection of all mismatches
and for the collection of all forbidden-triple mismatches.

\begin{proposition}[Consistency]\label{prop:consistency}
Fix a target $\G_t$.
For every $\G$, let $\mathcal M(\G,\G_t)$ be the collection
of all mismatches for $(\G,\G_t)$.
Then the heuristic
\[
h(\G)
=|\mathrm{Skel}(\G)\triangle\mathrm{Skel}(\G_t)|
+2\widetilde m_{\mathrm{op}}(\mathcal M(\G,\G_t))
\]
is consistent for the distance search.
The same conclusion holds when $\mathcal M(\G,\G_t)$ is
the collection of all forbidden-triple mismatches for every $\G$.
In particular, the bound of \citet{taeb2025distance} is consistent.
\end{proposition}

We prove both cases of Proposition~\ref{prop:consistency}
in Appendix~\ref{app:consistency-proof}.
 
Computing $h$ requires solving a minimum hitting-set problem,
which is impractical for a high number of nodes. In practice, we use the polynomial-time bound of
Corollary~\ref{thm:independent-conflicts}. For the current
state-target pair $(\G,\G_t)$, we detect the mismatch patterns listed
below and form their \emph{mismatch-conflict graph}, which has one
vertex per detected mismatch and joins two mismatches when they share
an operation. An independent set of this graph is a collection of
pairwise disjoint mismatches. We construct such a set $I(\G,\G_t)$
greedily, by repeatedly selecting a vertex of minimum degree and
deleting it and its neighbors. We then set
$$h_D(\G) = |\mathrm{Skel}(\mathcal{G})\triangle\mathrm{Skel}(\mathcal{G}_t)|
+2|I(\mathcal{G},\mathcal{G}_t)|.$$
Each detected pattern gives a mismatch
(Appendix~\ref{app:mismatch-certification}), so by
Corollary~\ref{thm:independent-conflicts}, $h_D$ is a lower bound on
$d(\G,\G_t)$. Unlike $h$, the heuristic $h_D$ need not be consistent. Since $h_D$ is admissible, the search remains exact, at
the cost of occasionally reopening a state.

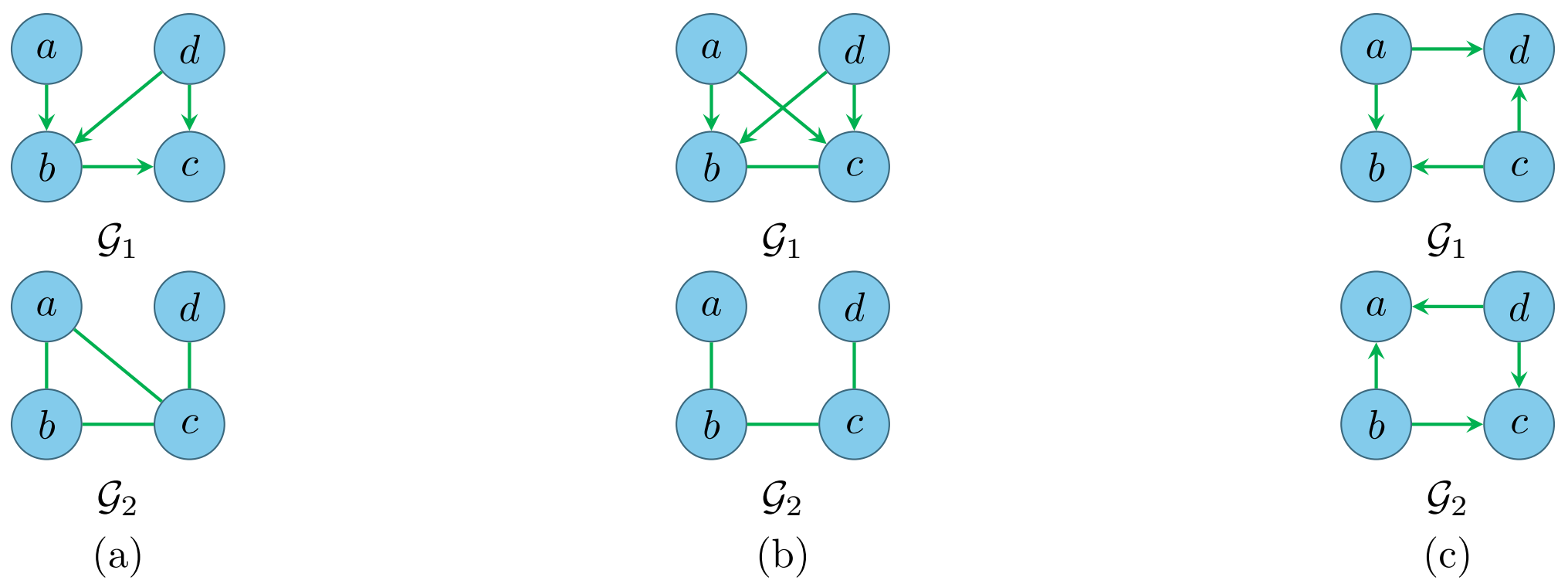
\begin{figure}[t]
\centering
\begin{minipage}{0.31\linewidth}
\centering
\begin{tikzpicture}[>=stealth,scale=.72,
every node/.style={circle,draw=nodeoutline,fill=nodeblue,minimum size=6mm,inner sep=1pt}]
\node (a) at (0,1.4) {$a$}; \node (b) at (0,0) {$b$};
\node (c) at (1.7,0) {$c$}; \node (d) at (1.7,1.4) {$d$};
\draw[->,thick,edgegreen] (a)--(b); \draw[->,thick,edgegreen] (d)--(b);
\draw[->,thick,edgegreen] (b)--(c); \draw[->,thick,edgegreen] (d)--(c);
\end{tikzpicture}

$\mathcal{G}_1$

\smallskip
\begin{tikzpicture}[>=stealth,scale=.72,
every node/.style={circle,draw=nodeoutline,fill=nodeblue,minimum size=6mm,inner sep=1pt}]
\node (a) at (0,1.4) {$a$}; \node (b) at (0,0) {$b$};
\node (c) at (1.7,0) {$c$}; \node (d) at (1.7,1.4) {$d$};
\draw[thick,edgegreen] (a)--(b); \draw[thick,edgegreen] (a)--(c);
\draw[thick,edgegreen] (b)--(c); \draw[thick,edgegreen] (c)--(d);
\end{tikzpicture}

$\mathcal{G}_2$

\smallskip (a)
\end{minipage}
\hfill
\begin{minipage}{0.31\linewidth}
\centering
\begin{tikzpicture}[>=stealth,scale=.72,
every node/.style={circle,draw=nodeoutline,fill=nodeblue,minimum size=6mm,inner sep=1pt}]
\node (a) at (0,1.4) {$a$}; \node (b) at (0,0) {$b$};
\node (c) at (1.7,0) {$c$}; \node (d) at (1.7,1.4) {$d$};
\draw[->,thick,edgegreen] (a)--(b); \draw[->,thick,edgegreen] (d)--(b);
\draw[->,thick,edgegreen] (a)--(c); \draw[->,thick,edgegreen] (d)--(c);
\draw[thick,edgegreen] (b)--(c);
\end{tikzpicture}

$\mathcal{G}_1$

\smallskip
\begin{tikzpicture}[>=stealth,scale=.72,
every node/.style={circle,draw=nodeoutline,fill=nodeblue,minimum size=6mm,inner sep=1pt}]
\node (a) at (0,1.4) {$a$}; \node (b) at (0,0) {$b$};
\node (c) at (1.7,0) {$c$}; \node (d) at (1.7,1.4) {$d$};
\draw[thick,edgegreen] (a)--(b); \draw[thick,edgegreen] (b)--(c);
\draw[thick,edgegreen] (c)--(d);
\end{tikzpicture}

$\mathcal{G}_2$

\smallskip (b)
\end{minipage}
\hfill
\begin{minipage}{0.31\linewidth}
\centering
\begin{tikzpicture}[>=stealth,scale=.72,
every node/.style={circle,draw=nodeoutline,fill=nodeblue,minimum size=6mm,inner sep=1pt}]
\node (a) at (0,1.4) {$a$}; \node (b) at (0,0) {$b$};
\node (c) at (1.7,0) {$c$}; \node (d) at (1.7,1.4) {$d$};
\draw[->,thick,edgegreen] (a)--(b); \draw[->,thick,edgegreen] (c)--(b);
\draw[->,thick,edgegreen] (a)--(d); \draw[->,thick,edgegreen] (c)--(d);
\end{tikzpicture}

$\mathcal{G}_1$

\smallskip
\begin{tikzpicture}[>=stealth,scale=.72,
every node/.style={circle,draw=nodeoutline,fill=nodeblue,minimum size=6mm,inner sep=1pt}]
\node (a) at (0,1.4) {$a$}; \node (b) at (0,0) {$b$};
\node (c) at (1.7,0) {$c$}; \node (d) at (1.7,1.4) {$d$};
\draw[->,thick,edgegreen] (b)--(a); \draw[->,thick,edgegreen] (d)--(a);
\draw[->,thick,edgegreen] (b)--(c); \draw[->,thick,edgegreen] (d)--(c);
\end{tikzpicture}

$\mathcal{G}_2$

\smallskip (c)
\end{minipage}
\caption{Three additional four-vertex mismatches used by the heuristic $h$. Figure~\ref{fig:mismatch-examples} gives the other two mismatches we consider in this paper.}
\label{fig:additional-mismatches}
\end{figure}

For the mismatch patterns, we consider the two mismatch types in Figure~\ref{fig:mismatch-examples}, as well as the three in Figure~\ref{fig:additional-mismatches}. For the four-vertex mismatches, we consider all node relabelings of the patterns. Their corresponding mismatches are:
\[
\begin{aligned}
M_1={}&\{\mathrm{ER}(a,b),\ \mathrm{ER}(b,c),\ \mathrm{EA}(a,c)\},\\
M_2={}&\{\mathrm{ER}(a,b),\mathrm{ER}(b,c),\mathrm{ER}(c,d),
\mathrm{EA}(a,c),\mathrm{EA}(a,d),\mathrm{EA}(b,d)\},\\
M_3={}&\{\mathrm{ER}(a,b),\mathrm{ER}(a,c),\mathrm{ER}(b,c),\mathrm{ER}(c,d),
          \mathrm{EA}(b,d),\mathrm{EA}(a,d)\},\\
M_4={}&\{\mathrm{ER}(a,b),\mathrm{ER}(b,c),\mathrm{ER}(c,d),
          \mathrm{EA}(a,d),\mathrm{EA}(b,d),\mathrm{EA}(a,c)\},\\
M_5={}&\{\mathrm{ER}(a,b),\mathrm{ER}(b,c),\mathrm{ER}(c,d),\mathrm{ER}(a,d),
          \mathrm{EA}(a,c),\mathrm{EA}(b,d)\}.
\end{aligned}
\]
Here, $M_1$ is the mismatch of Figure~\ref{fig:mismatch-examples}(a) and $M_2$ is the mismatch of Figure~\ref{fig:mismatch-examples}(b). The mismatches $M_3$, $M_4$ and $M_5$ correspond to Figure~\ref{fig:additional-mismatches}(a), (b) and (c).

\subsubsection{Computing feasible completions} We use several feasible paths to obtain upper bounds on $d(\G_s,\G_t)$. Here, a feasible completion from a current state $\G$ is any valid path in the CPDAG Hasse diagram from $\G$ to the target $\G_t$. 

The first feasible path comes from common lower and upper bounds: any common lower graph yields a down-up path, and any common upper graph yields an up-down path; see Remark~\ref{rem:down-up} for the definition of these path types. The shortest such paths have lengths
\[
u_{\downarrow\uparrow}(\mathcal{G},\mathcal{G}_t):=\mathrm{rank}(\mathcal{G})+\mathrm{rank}(\mathcal{G}_t)-2\rho_L(\mathcal{G},\mathcal{G}_t),\qquad
u_{\uparrow\downarrow}(\mathcal{G},\mathcal{G}_t):=2\rho_U(\mathcal{G},\mathcal{G}_t)-\mathrm{rank}(\mathcal{G})-\mathrm{rank}(\mathcal{G}_t),
\]
with $\rho_L$ and $\rho_U$ as in Section~5.1. Computing $\rho_L$ and $\rho_U$ requires a search. Propositions~\ref{prop:common-lower} and~\ref{prop:common-upper} instead give the inexpensive polynomial-time upper bounds
\begin{eqnarray*}
\begin{aligned}
\widehat u_{\downarrow\uparrow}(\mathcal{G},\mathcal{G}_t)&:=\mathrm{rank}(\G)+\mathrm{rank}(\G_t)-2|E(\mathcal{D}_\cap(\G,\G_t))|,\\
\widehat u_{\uparrow\downarrow}(\mathcal{G},\mathcal{G}_t)&:= 2\,\mathrm{rank}(\G_\text{ch}(\G,\G_t))-\mathrm{rank}(\G)-\mathrm{rank}(\G_t).
\end{aligned}
\end{eqnarray*}
We then use the completion
$$C_D(\G) = (\G_t, \min\{\widehat u_{\downarrow\uparrow}(\mathcal{G},\mathcal{G}_t),\widehat u_{\uparrow\downarrow}(\mathcal{G},\mathcal{G}_t)\}).$$

We set the initial upper bound $U_0$ as
$$U_0 = \min\{u_\text{component}(\G_s,\G_t),u_\text{polytree}(\G_s,\G_t), u_{\downarrow\uparrow}(\G_s,\G_t),u_{\uparrow\downarrow}(\G_s,\G_t)\},$$
where $u_\text{component}(\G_s,\G_t)$ and $u_\text{polytree}(\G_s,\G_t)$ are described in Section~\ref{sec:upper_bounds_model}. Note that  during the search, for any intermediate graph $\G$, we update the upper bound using the cheaper polynomial-time upper bounds on $u_{\downarrow\uparrow}(\G,\G_t)$ and $u_{\uparrow\downarrow}(\G,\G_t)$.

\subsubsection{Pruning strategies}\label{sec:distance-pruning}
We further reduce the search by excluding covering operations to graphs that cannot occur on a shortest path. First, let $\G_L$ and $\G_U$ be any common lower and upper graphs of $\G_s$ and $\G_t$, respectively. Every graph $\G$ on a shortest path satisfies
\[
\mathrm{rank}(\G_L)
\leq
\mathrm{rank}(\G)
\leq
\mathrm{rank}(\G_U).
\]
Indeed, a path that visits a graph of rank below $\mathrm{rank}(\mathcal{G}_L)$ or above $\mathrm{rank}(\mathcal{G}_U)$ is longer than the corresponding down-up or up-down path. We therefore generate only neighbors within this rank interval. Tighter common bounds give stronger pruning, although the polynomial-time constructions of Propositions~\ref{prop:common-lower} and~\ref{prop:common-upper} already provide valid bounds.

Second, shortest paths never introduce edges that are disjoint from all differences between the start and target graphs. Let $V_\Delta$ be the set of vertices that are endpoints of a pair in $\mathrm{Skel}(\G_s)\triangle\mathrm{Skel}(\G_t)$ or belong to a triple that is a v-structure in exactly one of the two graphs. A \emph{disjoint stray edge} is an edge absent from both skeletons whose endpoints both lie outside $V_\Delta$.

\begin{proposition}[Disjoint stray edges]\label{prop:stray}
No graph on a shortest path from $\G_s$ to $\G_t$ contains a disjoint stray edge.
\end{proposition}
See Appendix~\ref{app:disjoint-stray} for a proof. We therefore exclude such edge additions when generating neighbors.

\subsection{Computing the confidence interval}
\label{sec:common-bounds-computation}
The goal is to construct a common lower graph and a common upper graph for all CPDAGs in $\widehat{\mathfrak{G}}_n$, which determine the endpoints of the confidence interval. We do this sequentially: starting from two CPDAGs, we compute a common lower and upper graph, then combine each resulting graph with the next CPDAG in the collection, and continue until all graphs have been incorporated. As described next, each common-bound problem is solved using our A$^\ast$ framework. 

For a pair $(\G_1,\G_2)$, the common-lower search starts from the graph of smaller rank, say $\G_1$, and moves only downward in the Hasse diagram. Thus, $S_L(\G)$ consists of CPDAGs covered by $\G$, and a state $\G$ is terminal if it is a lower bound of both graphs:
$$S_L(\G) = \{\G'\in\mathfrak{G}: \G \text{ covers }\G'\}\quad\text{ and }\quad 
T_L(\G):= \mathbf{1}\{\G \preceq \G_1 \text{ and }\G \preceq \G_2\}.$$
Because each downward move decreases rank by one, the shortest terminal path ends at a common lower bound of the largest rank. At an intermediate state \(\mathcal G\), any terminal state reachable from \(\mathcal G\) must be a common lower bound of \(\mathcal G,\mathcal G_2\), so Proposition~\ref{prop:common-lower} provides both the heuristic \(h_L(\mathcal G)\) and a feasible completion \(C_L(\mathcal G)\). Let $\mathcal{G}_\cap(\mathcal{G},\mathcal{G}_2)$ denote the CPDAG of $\mathcal{D}_\cap(\mathcal{G},\mathcal{G}_2)$:
\begin{eqnarray*}
\begin{aligned}
h_L(\G) &= \mathrm{rank}(\mathcal{G})
-
\bigl(
|\mathrm{Skel}_{\cap}(\G,\G_2)|
-
\nu(\Gamma_L(\G,\G_2))
\bigr),\\
C_L(\mathcal{G}) &= \bigl(\mathcal{G}_\cap(\mathcal{G},\mathcal{G}_2),\ \mathrm{rank}(\mathcal{G})-|E(\mathcal{D}_\cap(\mathcal{G},\mathcal{G}_2))|\bigr).
\end{aligned}
\end{eqnarray*}
The common-upper search is symmetric. We relabel the pair so that $\mathrm{rank}(\mathcal{G}_1)\ge\mathrm{rank}(\mathcal{G}_2)$. The search starts from $\mathcal{G}_1$ and moves only upward:
\[
S_U(\mathcal{G})=\{\mathcal{G}'\in\mathfrak{G}:\mathcal{G}'\text{ covers }\mathcal{G}\},\qquad
T_U(\mathcal{G})=\mathbf{1}\{\mathcal{G}_1\preceq\mathcal{G}\text{ and }\mathcal{G}_2\preceq\mathcal{G}\}.
\]
The shortest terminal path therefore ends at a common upper graph of smallest rank. At a state $\mathcal{G}$, every terminal state is a common upper graph of $\mathcal{G}$ and $\mathcal{G}_2$. Proposition~\ref{prop:common-upper} applied to $(\mathcal{G},\mathcal{G}_2)$ gives the heuristic and the completion:
\begin{eqnarray*}
\begin{aligned}
h_U(\G) &=|\mathrm{Skel}_{\cup}(\G,\G_2)| -\mathrm{rank}(\mathcal{G}),
\\
C_U(\G) &= \bigl(
\mathcal{G}_{\mathrm{ch}}(\G,\G_2),\
\mathrm{rank}(\mathcal{G}_{\mathrm{ch}}(\G,\G_2))
-
\mathrm{rank}(\mathcal{G})\bigr).
\end{aligned}
\end{eqnarray*}

\section{Simulations on confidence regions}
\label{sec:confidence-experiments}
In this section, we evaluate the confidence regions constructed by Algorithm~\ref{alg:regions} and assess their coverage and how their radius or width changes with sample size, signal strength, and graph structure. As benchmarks, we compare against the perturbed PC collection of \cite{chang2026post} and the standard PC point estimate. All comparisons use the same population models, datasets, resamples, and random seeds.

Unless otherwise noted, confidence balls and intervals are constructed from the collection returned by Algorithm~\ref{alg:perturbed PC} using its default parameters. We also consider parameter choices in Algorithm~\ref{alg:perturbed PC} that reproduce the perturbed PC collection of \cite{chang2026post}.

\subsection{Setup}\label{sec:confidence-setup} For a given DAG, we generate data from a linear Gaussian structural equation model (SEM):
\[
X_j=\sum_{k\in\mathrm{Pa}(j)}\beta_{kj}X_k+\varepsilon_j,
\qquad
\varepsilon_j\overset{\mathrm{ind}}{\sim}\mathcal N(0,1).
\]
To sample a DAG, we first draw a uniformly random ordering of the $p$ vertices. Each pair of vertices is then connected independently with the edge probability of the regime (Table~\ref{tab:confidence-regimes}), with the edge directed from the earlier to the later vertex in the sampled ordering. Edge coefficients $\beta_{kj}$ are drawn uniformly from $[2,3]$.

For regimes with temporal information, we instead sample a topological ordering consistent with the prescribed tiers: vertices are randomly ordered within each tier, while all vertices in an earlier tier precede those in a later tier. Edges are then sampled and oriented according to this ordering in the same way. Thus, no edge can point from a later tier to an earlier tier, while orientations within a tier remain unrestricted by the temporal information.

We consider five regimes, summarized in Table~\ref{tab:confidence-regimes}. The three sparse $p=10$ regimes differ only in the minimum nonzero partial correlation. We obtain each regime by rejection sampling until the minimum nonzero partial correlation falls in the target interval. We also consider two regimes with temporal background information, which assign variables to ordered tiers and forbid edges from later to earlier tiers. Following \citet{chang2026post}, the dense $p=10$ regime uses the tier assignment $O(V)=(1,1,1,2,2,2,2,2,3,3)$ (Appendix~\ref{app:tiered-pc}), while the sparse $p=15$ regime uses three tiers of five variables.

\begin{center}
\small
\begin{tabular}{p{0.23\textwidth} c c p{0.33\textwidth}}
\hline
Regime & $p$ & Edge probability  & Signal or temporal structure \\
\hline
Dense temporal & 10 & $4/9$ &  Three temporal tiers of sizes $3,5,2$. \\
Sparse, weak signal (non-temporal) & 10 & $2/9$  & Minimum nonzero partial correlation in $[0.005,0.05]$. \\
Sparse, moderate signal (non-temporal) & 10 & $2/9$  & Minimum nonzero partial correlation in $[0.05,0.10]$. \\
Sparse, strong signal (non-temporal) & 10 & $2/9$  & Minimum nonzero partial correlation in $[0.10,0.20]$. \\
Sparse temporal & 15 & $2/14$  & Three temporal tiers of five variables each. \\
\hline
\end{tabular}
\captionof{table}{\small Simulation regimes for the confidence region experiments. The three non-temporal sparse regimes range from weak to strong signals as measured through the minimum nonzero partial correlation coefficient.}
\label{tab:confidence-regimes}
\end{center}
For each regime, we sample 20 population SEMs, each consisting of a DAG and its edge coefficients. For each population model and sample size
\[
n\in\{100,500,1000,10000,30000\},
\]
we generate $30$ independent datasets. We fix the number of perturbed PC resamples at $M = 1000$ and use a nominal confidence level $1-\nu = 0.95$. Each distance search and each common lower or upper graph search has a time limit of $0.5$ seconds. If the timeout is reached, we use the best feasible solution found, which preserves coverage guarantees. 

Following \citet{chang2026post}, their perturbed PC procedure uses the threshold 
$$t_{\text{chang}}=c^\ast\left(\frac{\log n}{M}\right)^{1/L}z_{\nu/(2L)}.$$
Here, in place of $L=|\mathcal{T}|$, we use $L=\binom{p}{2}(d_{\max}+1)$, where $d_{\max}$ is the expected degree of each node. Furthermore, $z_q = \Phi^{-1}(1-q)$ denotes the upper-tail Gaussian quantile. The quantity $c^\star$ is a tuning parameter. As recommended by \citet{chang2026post}, we choose this parameter to be small while controlling the number of retained CPDAGs. Our proposed threshold (see Section~\ref{sec:default}) is
\[
t_\text{ours}=\sqrt{\frac{2\log M}{B}},
\qquad B\in\{3,6,12,24\}.
\]
For each population model, $B$ is selected once using a dataset of size $10000$ from one Monte Carlo run. We split this dataset into $7500$ training observations and $2500$ holdout observations. For each candidate \(B\), we generate $M=10000$ perturbed PC runs using the training data and retain the distinct valid CPDAGs. For each retained CPDAG, we choose a consistent DAG extension, fit the corresponding Gaussian DAG model on the training data, and evaluate its average log-likelihood on the holdout data. We then select the smallest \(B\) whose best holdout score is within \(1\%\) of the best score across candidates. The selected \(B\) is held fixed across all evaluation sample sizes and Monte Carlo trials for that population model.

\subsection{Results}
\label{sec:confidence-ball-experiments}
We evaluate both confidence balls and model confidence intervals on the dense temporal $p=10$ and sparse temporal $p=15$ regimes in Table~\ref{tab:confidence-regimes}. Appendix~\ref{app:sparse_conf} presents the results for the three sparse non-temporal $p=10$ regimes (rows 2--4 of Table~\ref{tab:confidence-regimes}). For each dataset, we compare the PC point estimate, the perturbed PC collection of \citet{chang2026post}, and confidence regions constructed from perturbed PC collections. We construct both balls and intervals from the \citet{chang2026post} collection and from the collection obtained using our proposed threshold $t_{\mathrm{ours}}$. Coverage is the fraction of datasets whose region contains the true CPDAG. Region size is the radius for confidence balls and the width for model confidence intervals. The width is the rank difference between the upper and lower endpoints and is at most $p(p-1)/2$, which equals 45 for $p=10$ and 105 for $p=15$.

\begin{figure}[h!]
\centering
\makebox[\textwidth][c]{%
\begin{minipage}[t]{0.45\textwidth}
\centering
\includegraphics[width=\linewidth,draft=false]{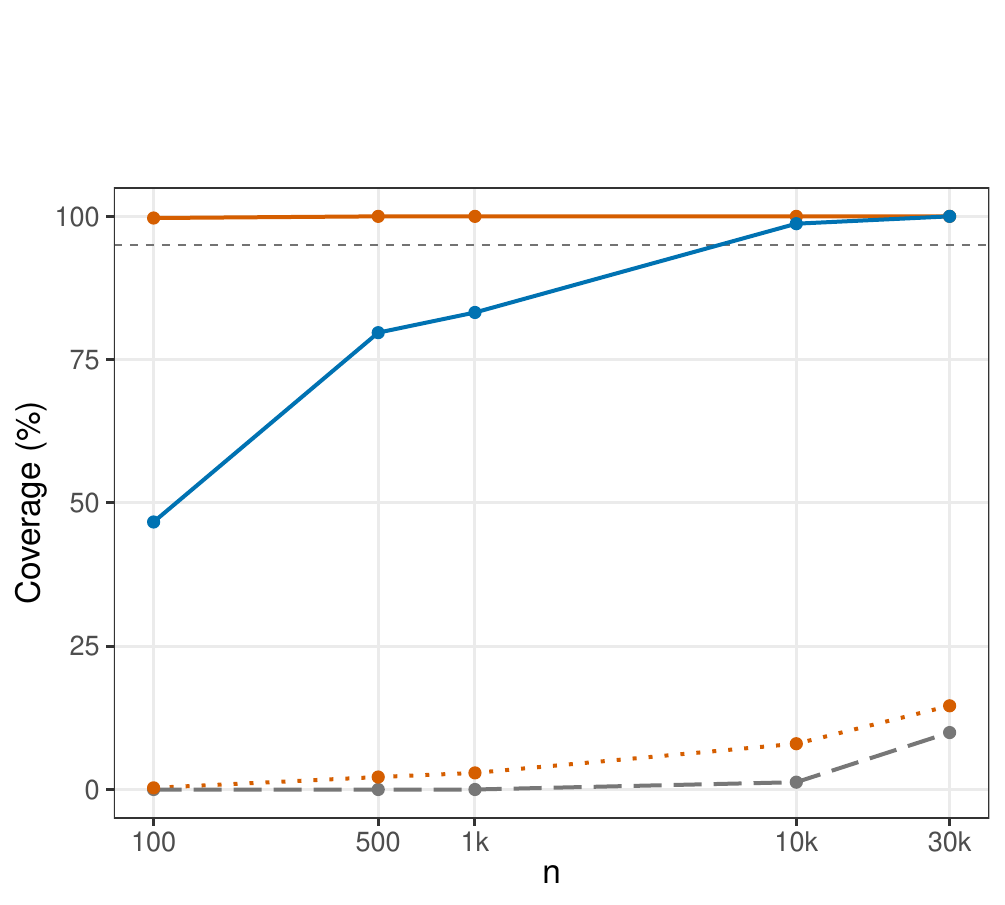}
\end{minipage}
\hspace{0.015\textwidth}
\begin{minipage}[t]{0.45\textwidth}
\centering
\includegraphics[width=\linewidth,draft=false]{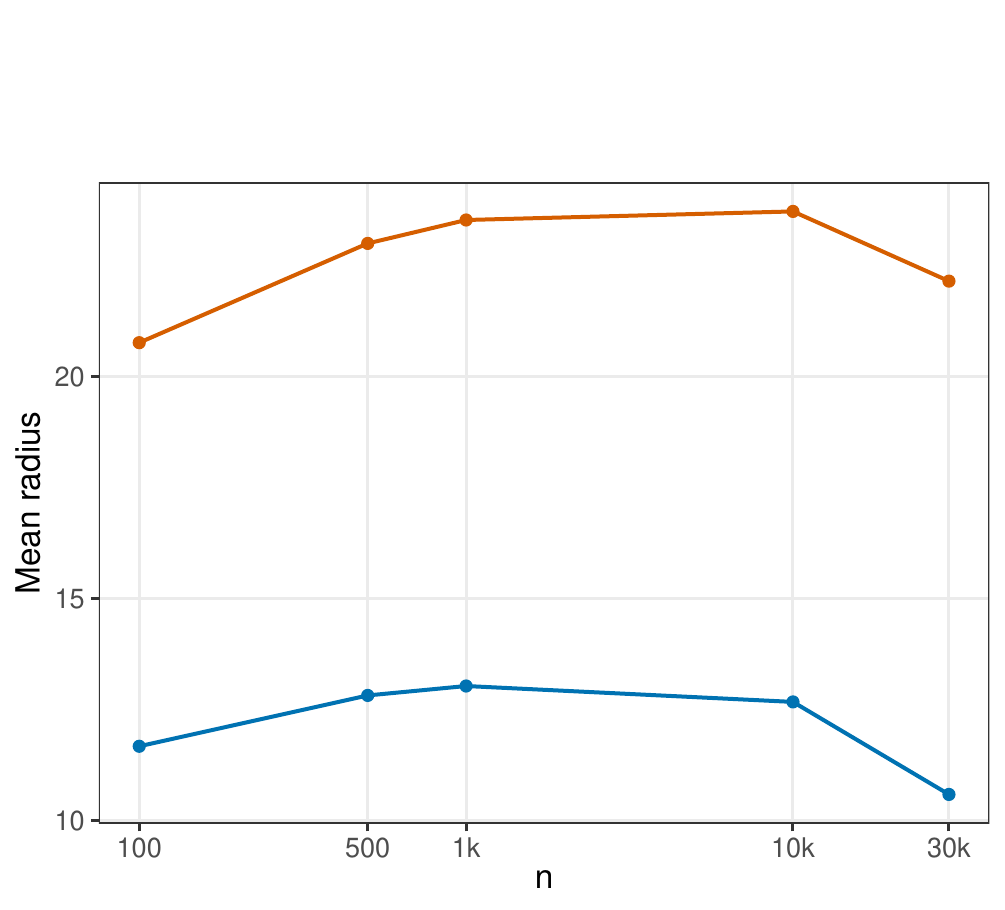}
\end{minipage}
}

\includegraphics[width=0.95\textwidth,draft=false]{plots/conf7_ball_shared_legend_final.pdf}
\caption{Confidence ball results for the dense temporal $p=10$ regime. Left: empirical coverage over 600 instances at each $n$. Right: mean radius over the same instances. }
\label{fig:confidence-ball-dense-temporal}
\end{figure}

\begin{figure}[h!]
\centering
\makebox[\textwidth][c]{%
\begin{minipage}[t]{0.45\textwidth}
\centering
\includegraphics[width=\linewidth,draft=false]{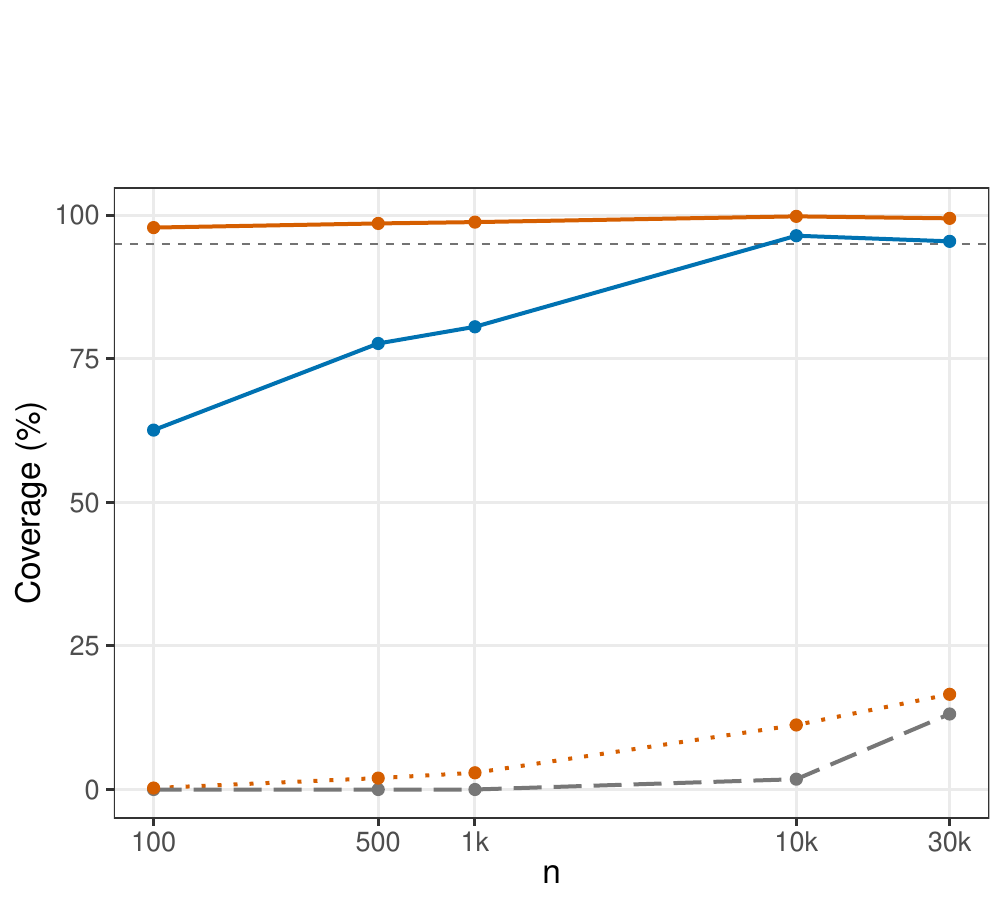}
\end{minipage}
\hspace{0.015\textwidth}
\begin{minipage}[t]{0.45\textwidth}
\centering
\includegraphics[width=\linewidth,draft=false]{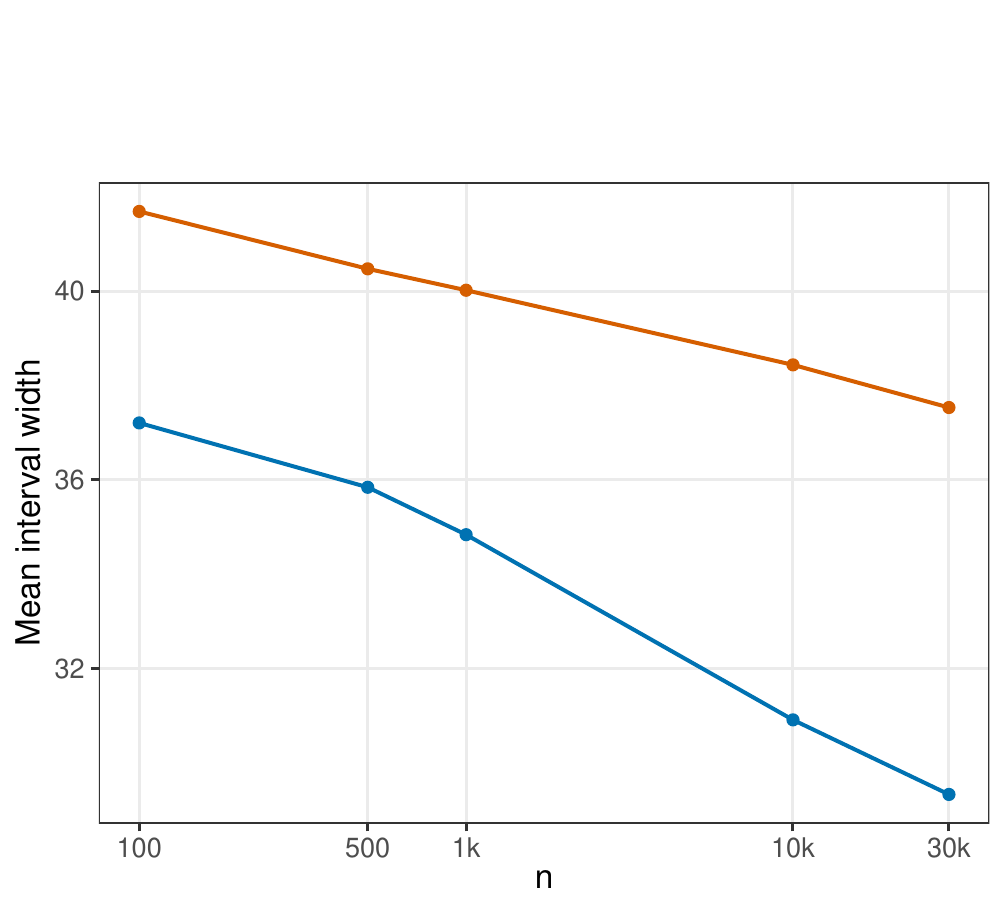}
\end{minipage}
}

\includegraphics[width=0.95\textwidth,draft=false]{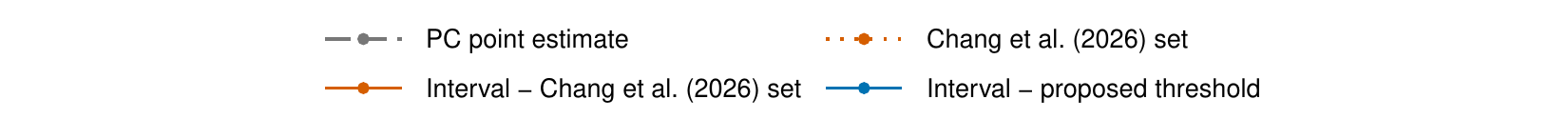}

\caption{Confidence interval results for the dense temporal $p=10$ regime. Left: empirical coverage over 600 instances at each $n$. Right: mean interval width over the same instances.}
\label{fig:confidence-interval-temporal}
\end{figure}

Consider first the dense temporal $p=10$ setting. Figures~\ref{fig:confidence-ball-dense-temporal} and~\ref{fig:confidence-interval-temporal} show the ball and interval results, respectively. The PC point estimate has poor coverage, as expected when structural uncertainty is ignored. The perturbed PC collection of \citet{chang2026post} can also miscover; although this could be due to the choice of its tuning parameter, which we discuss later. Converting this collection into either a confidence ball or a model confidence interval substantially improves coverage. The proposed threshold maintains comparable coverage while producing substantially more compact regions, with smaller ball radii and narrower intervals.


\begin{figure}[h!]
\centering
\makebox[\textwidth][c]{\begin{minipage}[t]{0.45\textwidth}\centering\includegraphics[width=\linewidth,draft=false]{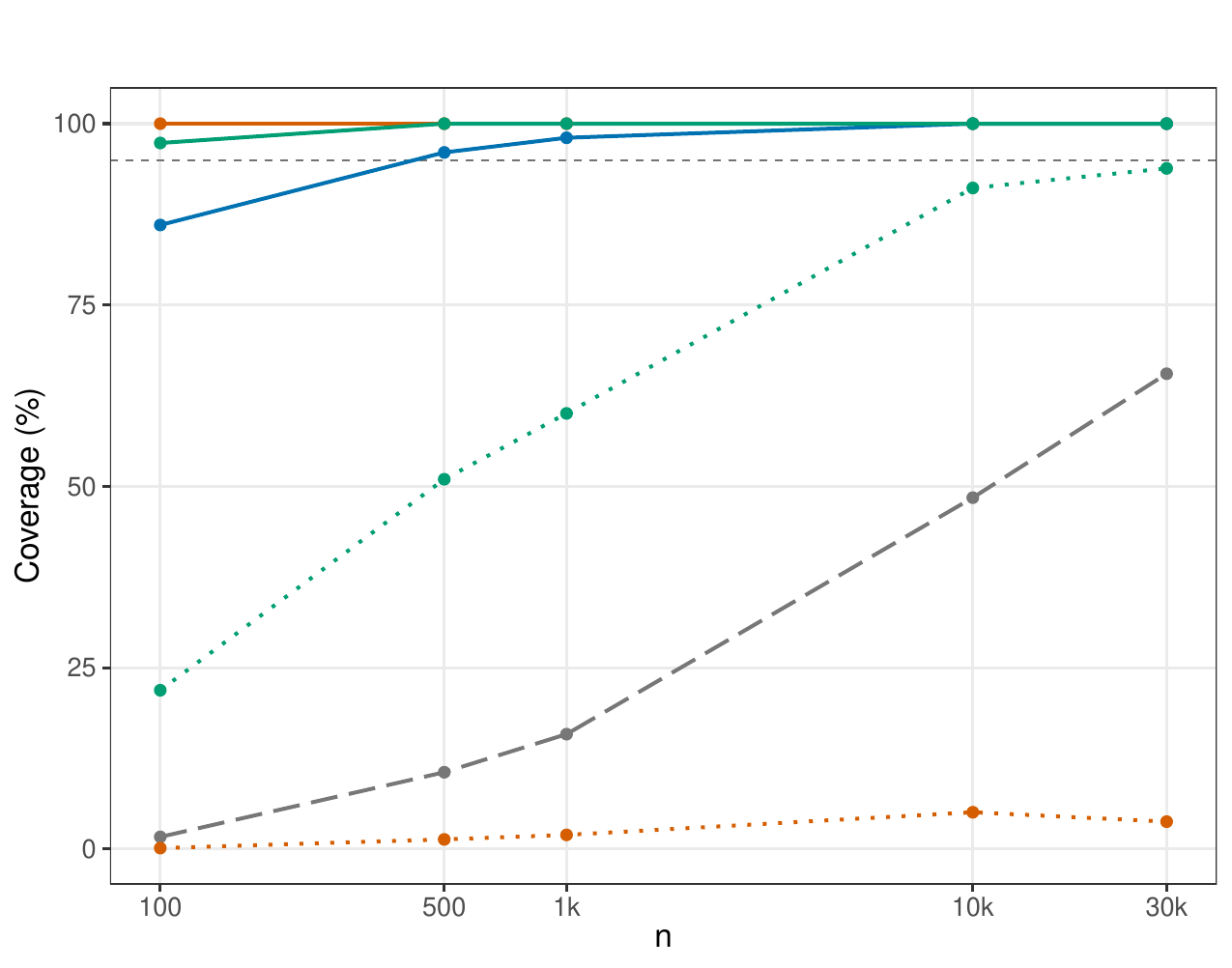}\end{minipage}\hspace{0.015\textwidth}\begin{minipage}[t]{0.45\textwidth}\centering\includegraphics[width=\linewidth,draft=false]{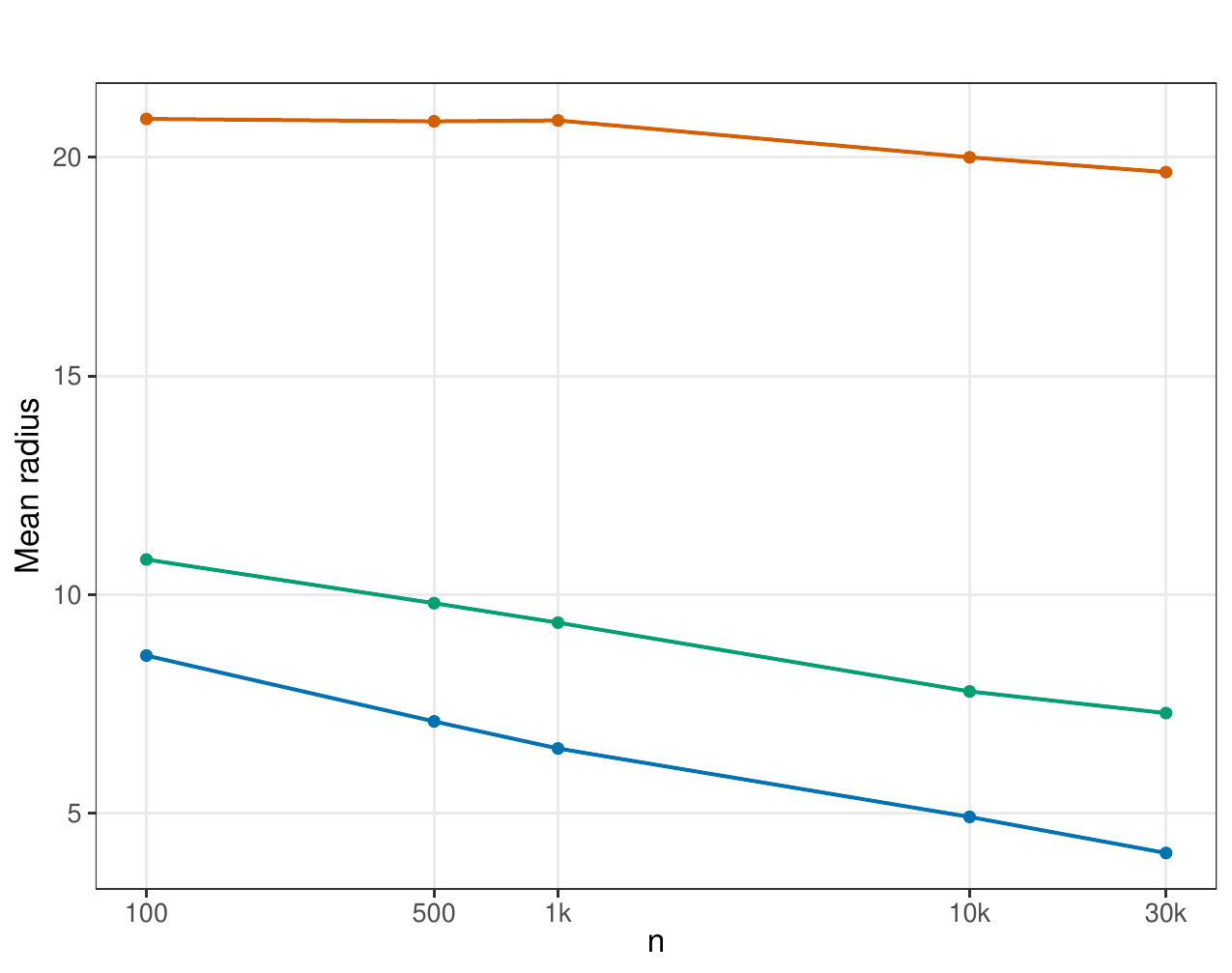}\end{minipage}}
\includegraphics[width=\textwidth,draft=false]{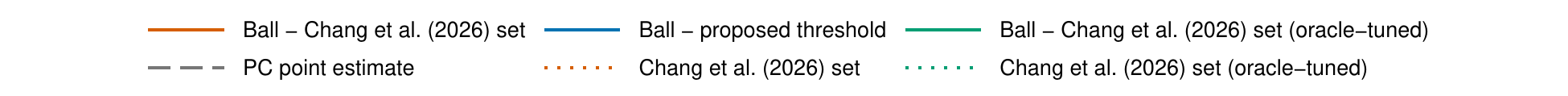}
\caption{Confidence ball results for the sparse temporal $p=15$ regime. Left: empirical coverage over 600 instances at each $n$. Right: mean radius over the same instances. For \citet{chang2026post}, we report results under both their recommended tuning strategy and an oracle choice of $c^\star$ that maximizes coverage.}
\label{fig:confidence-ball-sparse-temporal-p15}
\end{figure}

\begin{figure}[h!]
\centering
\makebox[\textwidth][c]{\begin{minipage}[t]{0.45\textwidth}\centering\includegraphics[width=\linewidth,draft=false]{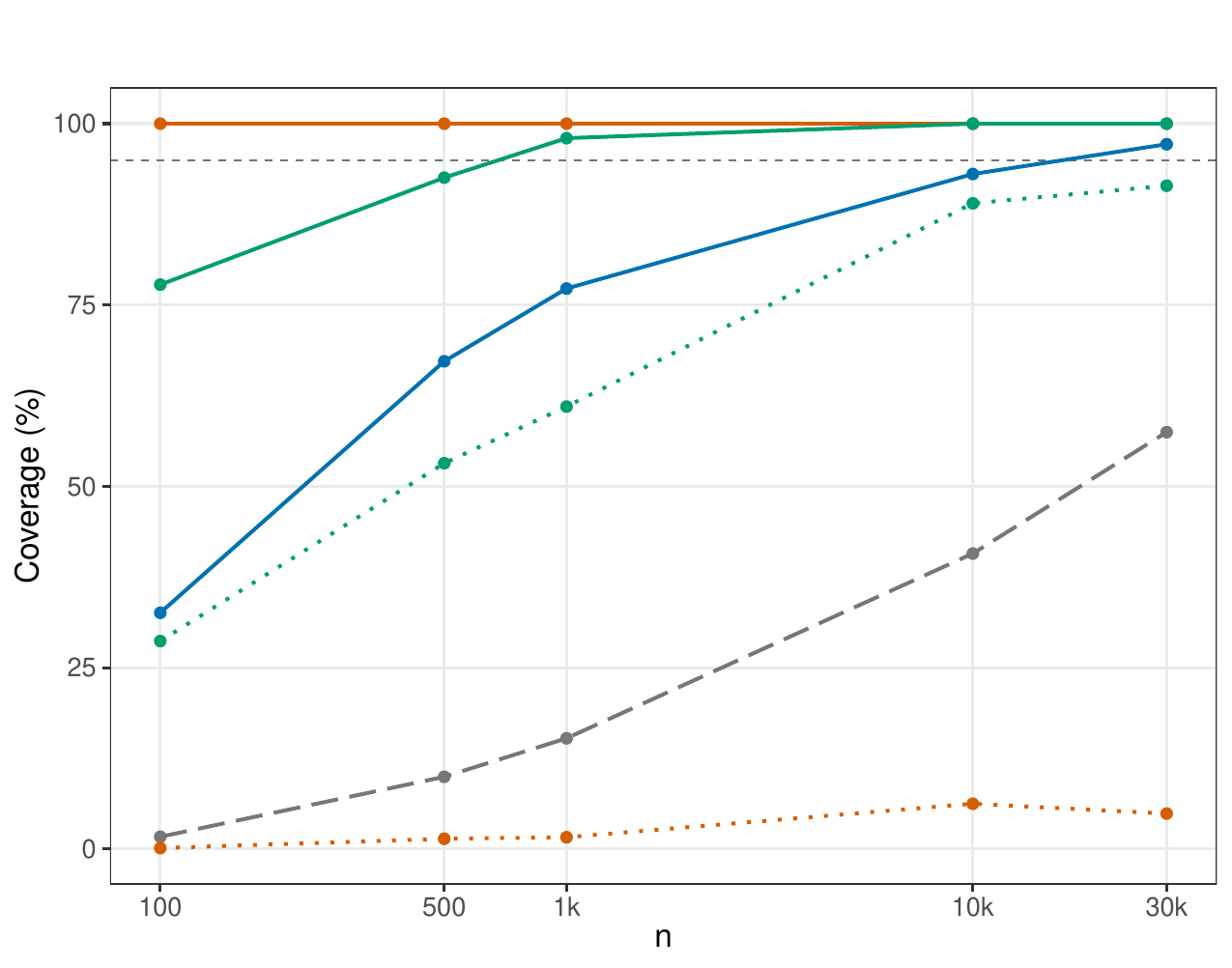}\end{minipage}\hspace{0.015\textwidth}\begin{minipage}[t]{0.45\textwidth}\centering\includegraphics[width=\linewidth,draft=false]{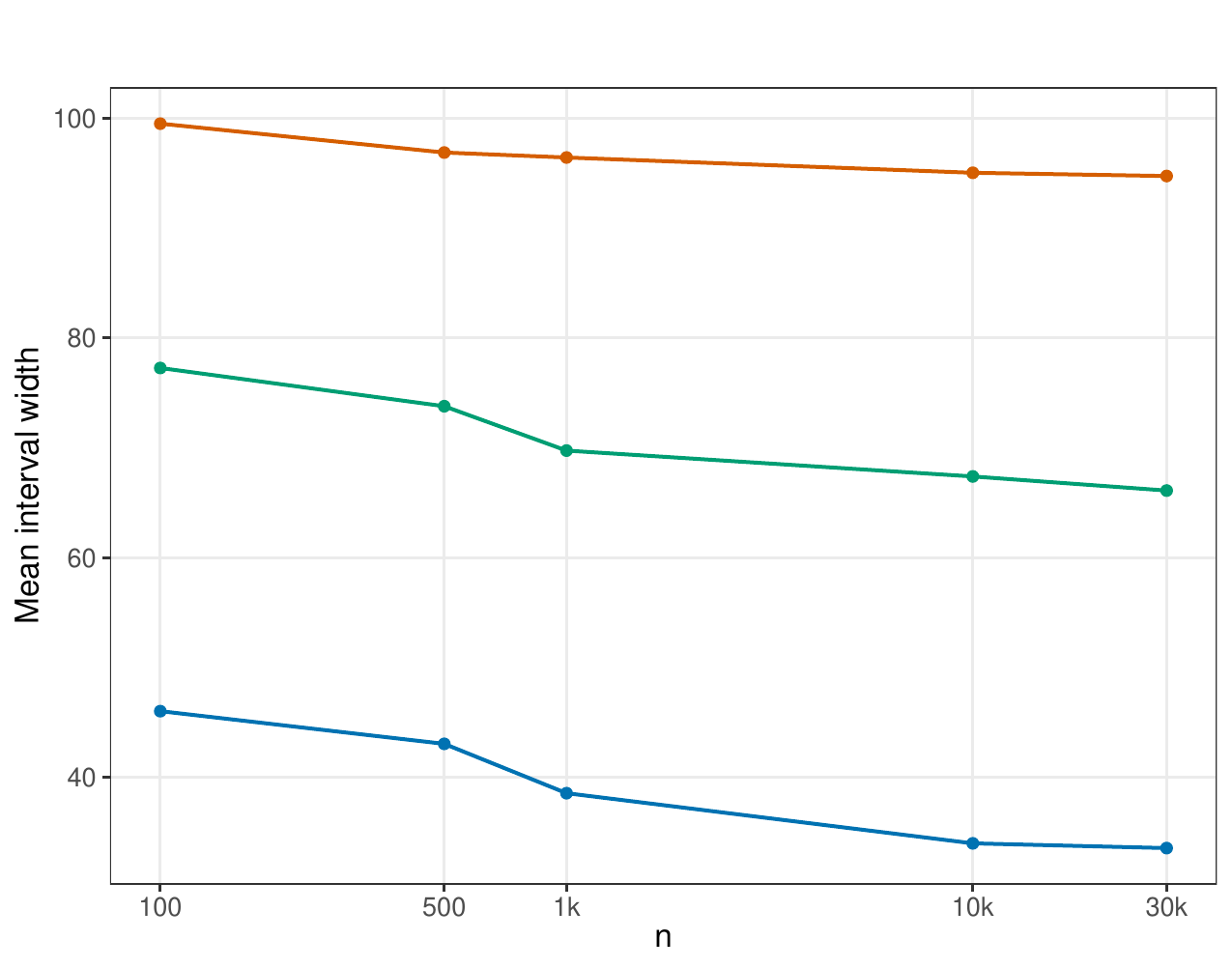}\end{minipage}}
\includegraphics[width=\textwidth,draft=false]{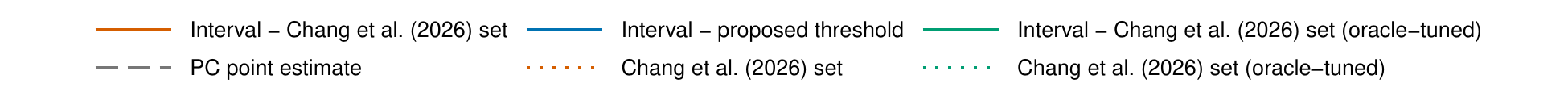}
\caption{Confidence interval results for the sparse temporal $p=15$ regime. Left: empirical coverage over 600 instances at each $n$. Right: mean interval width over the same instances. For \citet{chang2026post}, we report results under both their recommended tuning strategy and an oracle choice of $c^\star$ that maximizes coverage.}
\label{fig:confidence-interval-sparse-temporal-p15}
\end{figure}

Consider next the sparse temporal $p=15$ setting. In this setting, the recommended strategy of \citet{chang2026post} for choosing $c^\star$ leads to poor coverage. We therefore also report an oracle choice of $c^\star$ that maximizes coverage for the perturbed PC collection, providing a more favorable benchmark. We then compare confidence balls and model confidence intervals constructed from this oracle-tuned collection with those obtained using our proposed threshold. Figures~\ref{fig:confidence-ball-sparse-temporal-p15} and~\ref{fig:confidence-interval-sparse-temporal-p15} show the ball and interval results, respectively.  The oracle choice substantially improves coverage of the perturbed PC collection and also reduces the resulting ball radius and interval width. Even under this favorable tuning, however, the confidence regions remain substantially larger than those obtained using our proposed threshold. Our threshold achieves comparable coverage with markedly smaller radii and narrower intervals.

The gain in compactness that we see in all these experiments follows from the threshold choice (see Remark~\ref{rem:chang-threshold}). The threshold of \citet{chang2026post} allows more null statistics to cross in some runs, producing more variable CPDAG collections. Our proposed threshold suppresses these crossings, yielding smaller confidence balls and narrower intervals.\\

\noindent\textbf{Stable edges from confidence intervals.} The lower and upper graphs of the model confidence interval also provide an edge-level summary of structural uncertainty. Any edge that appears in both endpoints is present throughout the interval and is therefore supported across all CPDAGs represented by the interval. Table~\ref{tab:shared-edges} reports the number of such shared edges across sample sizes. As $n$ increases, the number of shared edges grows, reflecting increasing agreement between the lower and upper graphs and greater certainty about the underlying structure.

\begin{table}[htbp]
\centering
\caption{Average number of edges shared by the lower and upper graphs of the model confidence interval across sample sizes. Results are averaged over the $600$ instances.}
\label{tab:shared-edges}
\begin{tabular}{lccccc}
\toprule
& $n=100$ & $n=500$ & $n=1000$ & $n=10000$ & $n=30000$ \\
\midrule
dense $p=10$ setting  & 3.61 & 5.43 & 6.04 & 8.24 & 9.68 \\
\midrule
sparse $p=15$ setting  & 5.17 & 7.93 & 9.05 & 11.54 & 12.19 \\
\bottomrule
\end{tabular}
\end{table}

\section{Simulations on computational performance}
\label{sec:computational_performance}

We explore the extent to which the theoretical results of Section~\ref{sec:distance-theory} improve our computations. We first evaluate the computation of model-oriented distances needed for confidence balls in three regimes: nearly decomposable graphs, graphs close to being polytrees, and graph pairs with many local mismatch configurations. We then evaluate the computation of lower graphs needed for confidence intervals across different graph sizes and densities.

\subsection{Setup}
\label{sec:computational-experiments}

Given two CPDAGs, we record a timeout if the requested lower graph or model-oriented distance is not certified within $60$ seconds. To assess performance under smaller computational budgets, we also consider
\[
  b\in\{0.001,0.003,0.01,0.03,0.1,0.3,1,3,10,30,60\}\ \text{seconds}.
\]
A run is counted as a timeout at budget $b$ if its completion time exceeds $b$. We use the same graph pairs across all budgets and report one-standard-deviation bands around the timeout proportions.

We compare our methods with those of \citet{taeb2025distance}. For model-oriented distance, their procedure uses exact down-up and up-down paths as upper bounds and the exact forbidden-triple mismatch bound as its lower bound.  For lower-graph computation, their method uses the skeleton intersection as its heuristic, completes through the least CPDAG, and has no finite initial upper bound. Section~\ref{sec:computation} describes our corresponding search strategies.

\subsection{Model-oriented distance computation under structural proximity}
\label{sec:exp-components}
\subsubsection{Disconnected components}\
We next study the extent to which Theorem~\ref{thm:decomposition} improves computation of model-oriented distance for graphs that are nearly decomposable. We consider
\[
  q_{\mathrm{in}}=0.7,\qquad
  m\in\{0,1,3,7\},
\]
with $p\in\{10,15,20,30,100\}$ and block sizes three and five. Vertices are partitioned as evenly as possible among the blocks. The start and target graphs share this partition. Conditional on a random topological ordering, within-block edges are included independently with probability $q_\mathrm{in}$, and exactly $m$ cross-block edges are added uniformly at random and oriented according to the same ordering. Start and target DAGs are generated independently and converted to CPDAGs. For each setting, we generate $100$ pairs.

\begin{figure}[htbp]
\centering
\includegraphics[width=\textwidth,draft=false]{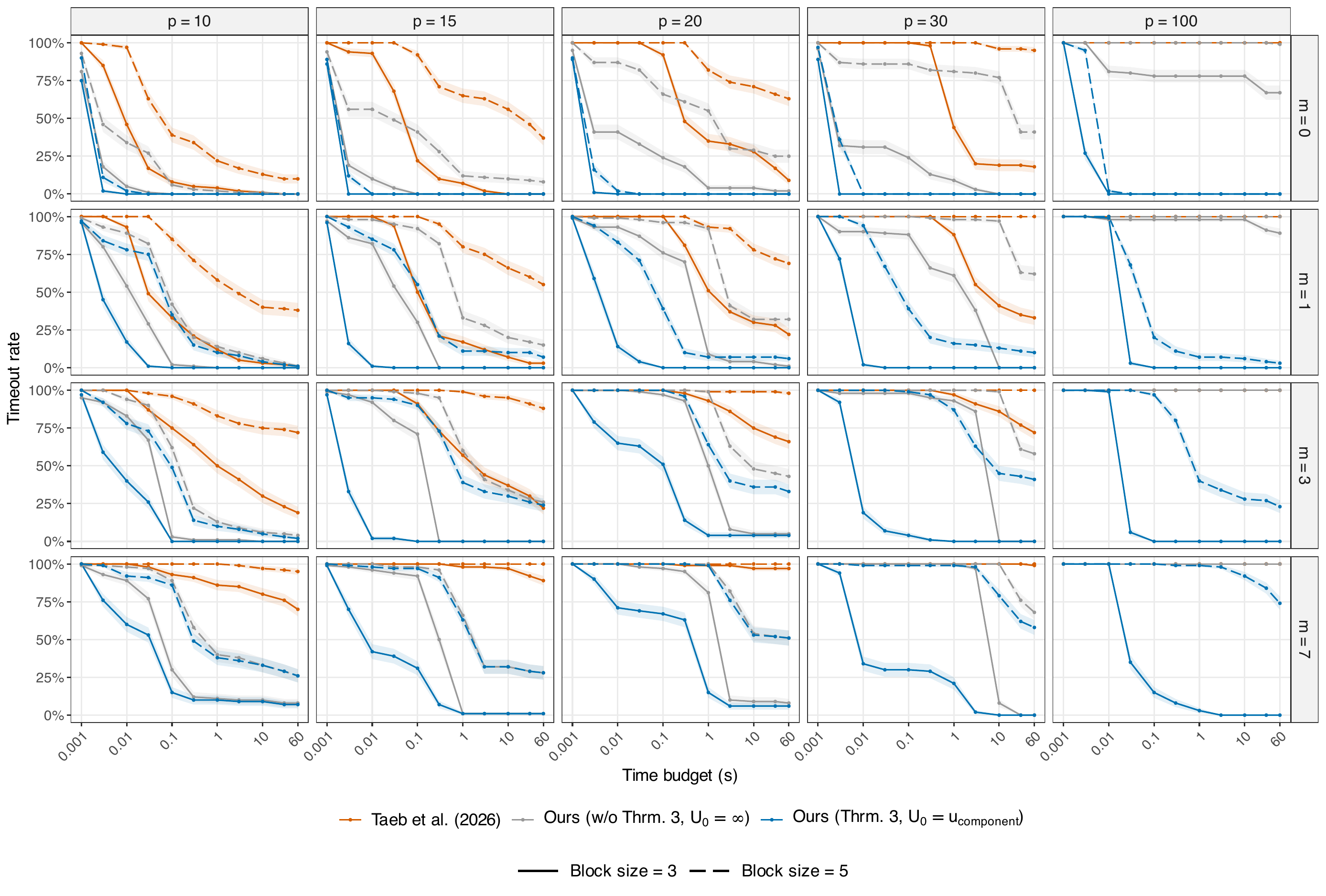}
\caption{Timeout rates for model-oriented distance searches between CPDAG pairs in the disconnected-component regime, with $q_{\mathrm{in}}=0.7$ and $m\in\{0,1,3,7\}$. Columns correspond to $p=10,15,20,30,$ and $100$; rows correspond to $m$. Solid and dashed curves use block sizes three and five, respectively. Each point is the proportion of the 100 pairs not solved within the indicated time budget; shaded regions show one standard deviation around that proportion. The curves compare the model-oriented-distance A$^\ast$ procedure of \citet{taeb2025distance} with our procedure without the disconnected-component upper bound and with our procedure using the disconnected-component upper bound.}
\label{fig:components-mpositive-timeout}
\end{figure}

We compute model-oriented distance with the A$^\ast$ procedure of \citet{taeb2025distance}, our search without the disconnected-component upper bound, and the same search with the upper bound enabled. All other search components are held fixed. Figure~\ref{fig:components-mpositive-timeout} reports timeout rates across computational budgets. When $m=0$, the component upper bound is exact, so our procedure returns it immediately.

The search using the disconnected-component upper bound has the lowest timeout rate in nearly every setting. For $m=0$, the exact bound makes its timeout rate fall to zero almost immediately at every graph size. The advantage of the bound remains pronounced for $m>0$, particularly at $p=30$ and $p=100$: with block size three, it solves nearly all pairs within the budget grid even as $m$ increases, while the procedures without the bound retain substantial timeout rates. For block size five, its timeout rate rises with $m$, reflecting the weaker decomposition, but it remains below the competing procedures. The procedure of \citet{taeb2025distance} times out on most pairs at the larger graph sizes. We include timeout rates for $q_{\text{in}}=0.5$ in Appendix~\ref{app:structural_prox}.

\subsubsection{Polytrees}
\label{sec:exp-polytrees}
We next study the extent to which Theorem~\ref{thm:polytree} improves computation of the model-oriented distance for graphs near the polytree class. We use
\[
  p\in\{10,15,20,30,100\},\qquad
  q_{\mathrm{tree}}=0.8,\qquad
  c\in\{0,1,3,7\}.
\]
For each graph, we sample a uniformly random labeled tree using Wilson's algorithm \citet{Wilson1996GeneratingRS}, and retain each edge independently with probability $q_{\mathrm{tree}}$. This produces a forest. We orient the retained edges according to a random topological ordering and then add $c$ edges between previously non-adjacent vertices in the same connected component, orienting them according to the ordering. Thus, $c$ controls the departure from polytree structure. Start and target graphs are generated independently and converted to CPDAGs. For each setting, we generate $100$ unique pairs.

\begin{figure}[htbp]
\centering
\includegraphics[width=\textwidth,draft=false]{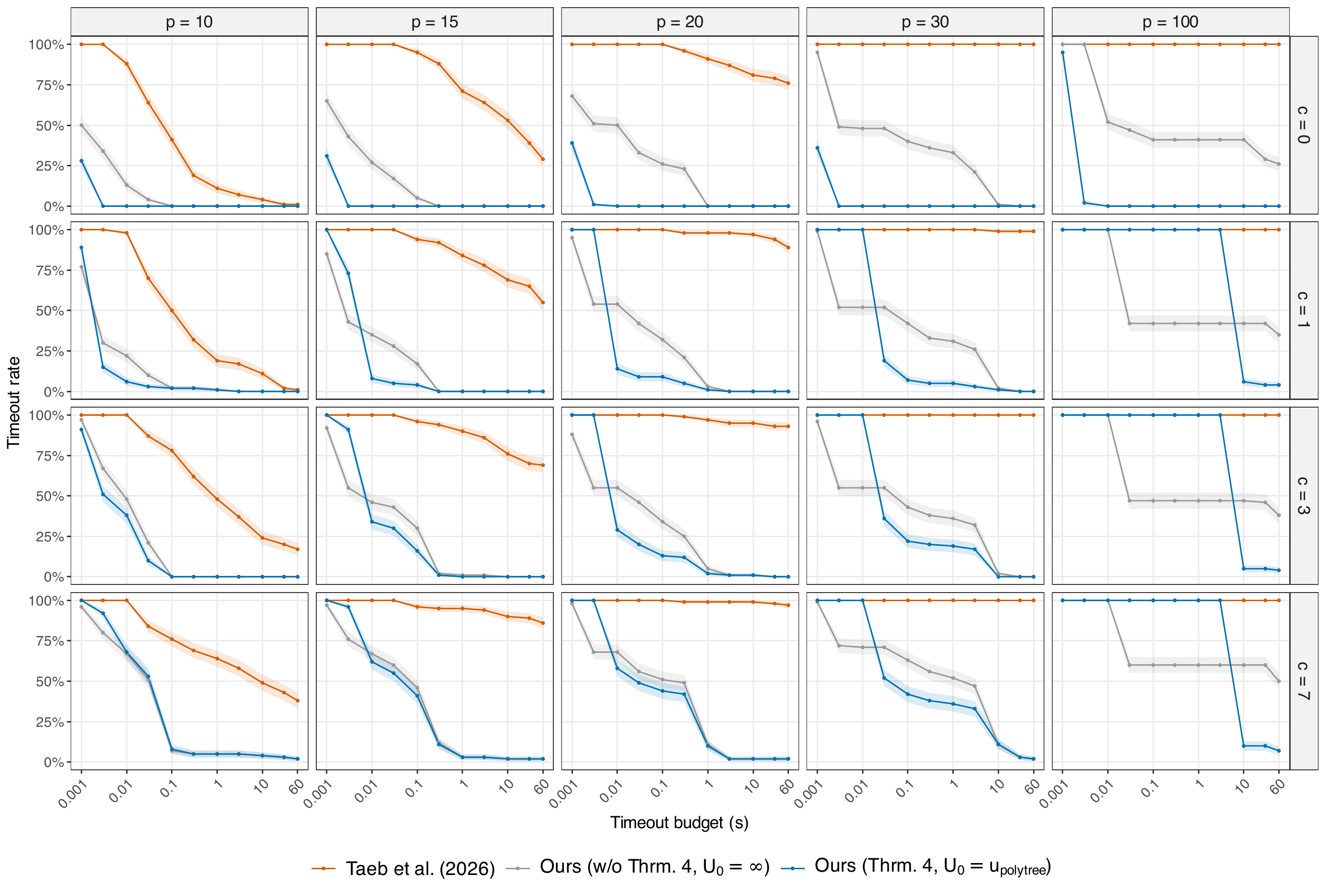}
\caption{Timeout rates for model-oriented distance searches between CPDAG pairs near the polytree class, with $q_{\mathrm{tree}}=0.8$ and $c\in\{0,1,3,7\}$. Columns correspond to $p=10,15,20,30,$ and $100$; rows correspond to $c$. Each point is the proportion of the 100 pairs not solved within the indicated time budget; shaded regions show one standard deviation around that proportion. The curves compare the model-oriented-distance A$^\ast$ procedure of \citet{taeb2025distance} with our procedure without the polytree upper bound and with our procedure using the polytree upper bound.}
\label{fig:polytree-cpositive-timeout}
\end{figure}

We compute model-oriented distance with the A$^\ast$ procedure of \citet{taeb2025distance}, our search without the polytree upper bound, and the same search with the upper bound enabled. All other search components are held fixed. Figure~\ref{fig:polytree-cpositive-timeout} reports timeout rates across computational budgets. When $c=0$, the polytree upper bound is exact, so our procedure returns it immediately.

The search using the polytree upper bound has the lowest timeout rate in nearly every setting. For $c=0$, the exact bound makes its timeout rate fall to zero almost immediately at every graph size. For $c>0$, its advantage over the search without the bound grows with $p$: at $p=30$ and $p=100$, it substantially reduces timeouts at moderate and large budgets, while the search without the bound remains well above zero. The procedure of \citet{taeb2025distance} times out on most pairs at the larger graph sizes, including at the largest budget. We include timeout rates for $q_{\text{tree}}=0.5$ and $q_{\text{tree}}=1$ in Appendix~\ref{app:structural_prox}.

\subsection{Lower bounds from mismatch structure}
\label{sec:exp-mismatches}

We next study the utility of the mismatch lower bound of Theorem~\ref{thm:mismatch} and its independent-set approximation from Corollary~\ref{thm:independent-conflicts} for computing the model-oriented distance. We construct connected chains of overlapping four-node mismatch configurations with
\[
  p\in\{7,10,22,28\},
  \qquad p=3K+1.
\]
Here, $K$ represents the number of four-node blocks. Successive blocks share one vertex. We construct a start and a target DAG from overlapping four-node configurations. For each block, we independently sample one of the four mismatch types in Figures~\ref{fig:mismatch-examples}(b) and~\ref{fig:additional-mismatches}(a)--(c), and place the corresponding paired configurations in the start and target DAGs, respectively. We convert both DAGs to CPDAGs. For each setting, we generate $100$ unique pairs of CPDAGs.

\begin{figure}[htbp]
\centering
\includegraphics[width=\textwidth,draft=false]{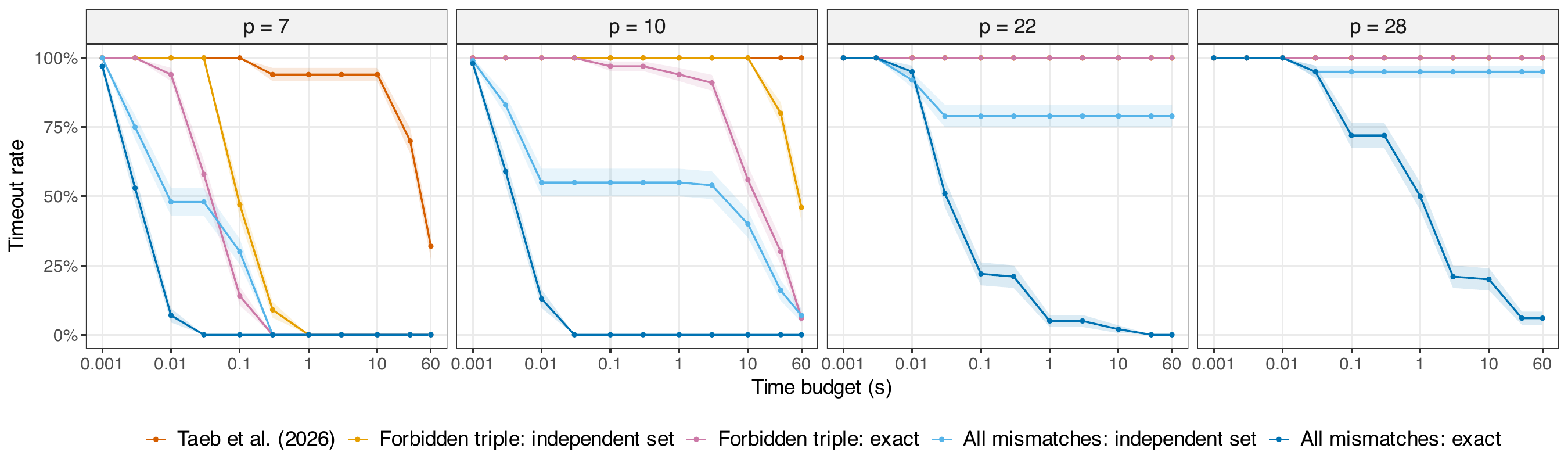}
\caption{Timeout rates for model-oriented distance searches between CPDAG pairs constructed from connected chains of overlapping four-node mismatch blocks. The panels correspond to $p=7,10,22,$ and $28$. Each point is the proportion of the 100 pairs not solved within the indicated time budget, with shaded regions showing one standard deviation. The curves compare the A$^\ast$ procedure of \citet{taeb2025distance} with four variants of our procedure: exact and independent-set lower-bound computations using forbidden triples alone or all five mismatch types.}
\label{fig:mismatch-timeout}
\end{figure}

We compare the model-oriented-distance A$^{\ast}$ procedure of \citet{taeb2025distance} with four variants of our procedure: exact and independent-set computations of the lower bound using only forbidden triples, illustrated in Figure~\ref{fig:mismatch-examples}(a), and using all five mismatch types (dubbed ``all-mismatch"). All other search components are held fixed. Figure~\ref{fig:mismatch-timeout} reports timeout rates across computational budgets.

The search using the exact all-mismatch bound has the lowest timeout rate at each graph size. For $p=7$ and $p=10$, the timeout rate reaches zero at small budgets. The search using the forbidden-triple bounds requires more time, while the procedure of \citet{taeb2025distance} remains substantially slower. The differences become more pronounced for $p=22$ and $p=28$. At these large graph sizes, the search using the exact all-mismatch bound improves substantially with additional time.  In contrast, the search using the all-mismatch independent-set bound plateaus near $80\%$ and $95\%$, respectively. The searches using the forbidden-triple bounds and the procedure of \citet{taeb2025distance} remain at or near $100\%$ timeouts.

\subsection{Common lower graphs}
\label{sec:exp-common-graphs}
Finally, we ask whether the construction of Proposition~\ref{prop:common-lower} improves lower-graph computation relative to the procedure of \citet{taeb2025distance}.

\begin{figure}[htbp]
\centering
\includegraphics[width=\textwidth,draft=false]{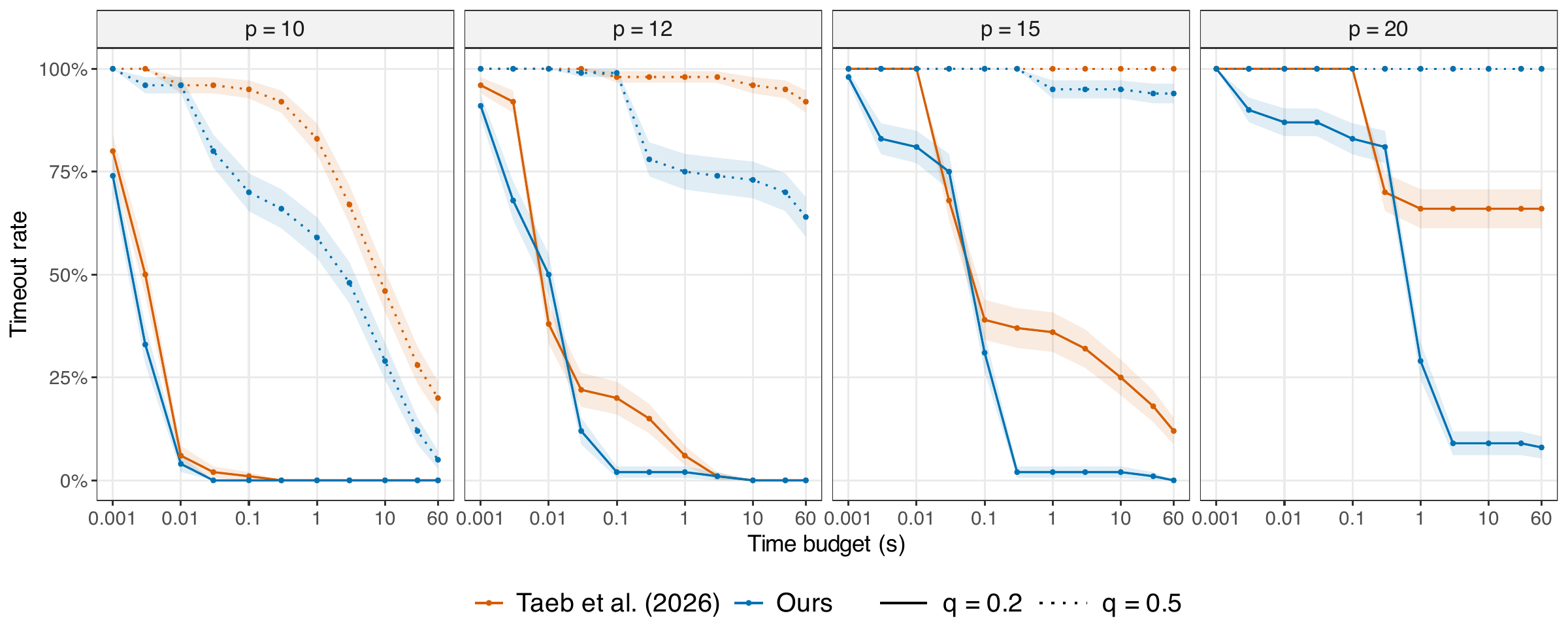}
\caption{Timeout rates for computation of common lower graphs. Panels correspond to $p=10,12,15,$ and $20$. Solid and dotted curves correspond to $q=0.2$ and $q=0.5$, respectively. Each point is the proportion of the 100 pairs not solved within the indicated time budget; shaded regions show one standard deviation. The curves compare the procedure of \citet{taeb2025distance} with our procedure.}
\label{fig:common-graphs-timeout}
\end{figure}

We generate random CPDAGs by first sampling a uniformly random ordering of $p$ labeled vertices. For each pair of vertices, we then include an edge independently with probability $q$ and orient it from the earlier to the later vertex in the sampled ordering. We convert the resulting DAG to its CPDAG. We consider
\[
  p\in\{10,12,15,20\},
  \qquad q\in\{0.2,0.5\}.
\]
For each $(p,q)$, we generate $100$ independent CPDAG pairs, resampling when the two graphs are identical.  We then compute common lower graphs using the procedure of \citet{taeb2025distance} and our procedure. Figure~\ref{fig:common-graphs-timeout} reports timeout rates over the full budget grid.

Our procedure has lower timeout rates in nearly every setting. The improvement is most pronounced for the sparse graphs at $p=15$ and $p=20$, where our timeout rate approaches zero while that of \citet{taeb2025distance} remains substantially higher. The dense instances are more difficult for both procedures, although our procedure continues to improve with additional time at $p=10$ and $p=12$.

\section{Discussion}
We first analyzed the perturbed PC algorithm and proposed a threshold that tightened the collection of plausible CPDAGs while preserving asymptotic coverage. The threshold filtered out most null statistics across runs while allowing weak signals to survive in at least one run, without requiring strong faithfulness. We then used the CPDAG poset under model inclusion to summarize this collection through confidence balls and model confidence intervals. The confidence ball measured structural uncertainty through a shortest path distance from the PC estimate, while the confidence interval bracketed the true CPDAG between common lower and upper graphs. Finally, we developed new computational theory for constructing both regions. Sharper distance bounds, exact decomposition across disconnected components, exact characterization for polytrees, and new bounds for common lower and upper graphs substantially reduced the search space. Simulations showed that the resulting regions attained good coverage while being substantially smaller than regions constructed from the original perturbed PC collections.

Our work opens several directions for future research. First, our guarantees treat the number of variables $p$ as fixed. Extending them to $p \to \infty$ would require uniform control of sample partial correlations and of the growing number of conditional independence tests. Second, kernel conditional independence tests could relax the linear Gaussian assumptions. Their null distributions depend on unknown features of the joint distribution, which complicates calibration of the perturbations. Third, our model confidence intervals construct the lower and upper models sequentially. This procedure guarantees a valid interval containing the perturbed PC collection, but it need not produce the smallest possible interval. A natural extension is to optimize the two endpoints jointly to obtain a minimum width valid interval. Finally, exact computation remains costly for dense graphs that do not decompose into simpler structures. Approximate computations can still yield valid regions because upper bounds on the radius and feasible interval endpoints only enlarge the regions. A natural direction is therefore to relate computational suboptimality to statistical conservativeness, for example by bounding how much an early-stopped search can inflate the ball radius or interval width.

\section{Acknowledgments}
We gratefully acknowledge the Washington Experimental Mathematics Lab at the University of Washington for bringing undergraduates Jian Kang and Arnav Mazumder together with Armeen Taeb through this project. Simulations for this work were completed on Hyak, the University of Washington's high performance computing cluster. This resource was funded by the UW student technology fee.

\bibliographystyle{agsm}
\bibliography{main.bib}

\appendix
\section{The PC and Tiered PC algorithms}
\label{app:pc}

\subsection{The standard PC algorithm}
\label{app:standard-pc}
We use the PC algorithm of \citet{spirtes2000causation} with the statistic $\widehat T_{ij|S}$ of Section~\ref{sec:preliminaries}. A test rejects conditional independence when $\widehat T_{ij|S}>t$. Otherwise, PC deletes the edge $i-j$ and records $\mathrm{sepset}(i,j)=\mathrm{sepset}(j,i)=S$. The algorithm has three stages. The skeleton stage starts from the complete graph. For $\ell=0,1,\ldots$, it tests each remaining adjacent pair $(i,j)$ given the subsets $S$ of size $\ell$ of the current neighbors of $i$, examining pairs and subsets in a fixed order. The v-structure stage orients each unshielded triple $i-k-j$ as $i\to k\leftarrow j$ if and only if $k\notin\mathrm{sepset}(i,j)$. The propagation stage applies the rules $R_1$--$R_4$ of \citet{meek1995causal} until no further edge can be oriented. In run $m$ of Algorithm~\ref{alg:perturbed PC}, $\widehat T_{ij|S}$ is replaced by $\widehat T^{[m]}_{ij|S}$, and all graphical steps are unchanged.

A run is \emph{invalid} if two v-structures require opposite orientations of an edge, if propagation produces a conflict, or if the output is not a CPDAG. Finite-sample errors can produce each of these outcomes, and an acyclic output need not be a CPDAG. Algorithm~\ref{alg:perturbed PC} discards invalid runs.

When every test decision is correct, PC returns $\mathcal{G}^\star$ under Assumption~\ref{ass:faithfulness} \citep{spirtes2000causation,kalisch2007estimating}. Section~\ref{sec:analysis} uses only this property, so the analysis applies to both the order-dependent version of PC and PC-stable \citep{ColomboMaathuis2014}.

\subsection{The tiered PC algorithm}
\label{app:tiered-pc}

A tier assignment
$O:V\to\{1,\ldots,K\}$, written as $O(V)=(O(1),\ldots,O(p))$,
places the variables in ordered temporal tiers. If $O(i)<O(j)$,
then $X_i$ precedes $X_j$, so $j\to i$ is forbidden. No ordering
is imposed within a tier. We assume that the true underlying DAG
respects the tier assignment.

Following \citet{chang2026post}, we use the PC-stable search of
\citet{ColomboMaathuis2014} and incorporate the tiers into both
skeleton estimation and edge orientation. The test statistics
$\hat{T}_{ij|S}$ and threshold $t>0$ are unchanged.

\paragraph{\emph{Skeleton estimation.}}
Start from $\mathcal{G}_{\mathrm{full}}$. At each conditioning-set size
$\ell$, store the current adjacency sets
$\widetilde{\mathrm{adj}}_\ell(i):=\mathrm{adj}(\mathcal{G},i)$ and define
\begin{equation}
\label{eq:tiered-conditioning}
A_{\ell,ij}
:=\bigl\{k\in\widetilde{\mathrm{adj}}_\ell(i)\setminus\{j\}:
O(k)\leq\max\{O(i),O(j)\}\bigr\}.
\end{equation}
Thus, conditioning variables cannot lie in a tier later than both
endpoints. For each ordered adjacent pair $(i,j)$, test subsets
$S\subseteq A_{\ell,ij}$ of size $\ell$. At the first set satisfying
$\hat{T}_{ij|S}\leq t$, delete $i-j$ and record
$\mathrm{sepset}(i,j)=\mathrm{sepset}(j,i)\leftarrow S$.
Pairs and conditioning sets are examined in a fixed order.
Deletions take effect immediately, but the stored adjacency sets
remain fixed throughout level $\ell$. Stop when $|A_{\ell,ij}|<\ell$
for every remaining ordered adjacent pair.
When $O(i)\equiv1$, this reduces to the unrestricted PC-stable search.
\\
\paragraph{\emph{Edge orientation.}}
Orient each edge between distinct tiers from the earlier to the
later tier. For each unshielded triple $(i,k,j)$ in the estimated
skeleton, impose $i\to k\leftarrow j$ if and only if
$k\notin\mathrm{sepset}(i,j)$. Then apply Meek's rules
$R_1$--$R_4$ \citep{meek1995causal} until no
further edge can be oriented. All orientations must be compatible
with the tier assignment and the recorded separating sets.
\\
\paragraph{\emph{CPDAG output.}}
With tier restrictions, a valid output is generally a maximally
oriented partially directed acyclic graph (MPDAG)
\citep{BangDidelez2023}. We convert this graph to its observational
CPDAG so that our confidence regions remain subsets of $\mathfrak{G}$.
Specifically, choose a consistent DAG extension: a DAG with the same
skeleton that preserves all directed edges and introduces no new
v-structures. Return the CPDAG of its Markov equivalence class.
All consistent extensions have the same skeleton and v-structures,
so the choice of extension does not affect the returned CPDAG.
If the orientation requirements conflict or no consistent DAG
extension exists, the run is invalid.

Algorithm~\ref{alg:tiered-pc} summarizes the procedure.
For run $m$ of perturbed PC, replace $\hat{T}_{ij|S}$ by
$\hat{T}_{ij|S}^{[m]}$; all graphical steps remain unchanged.

\begin{algorithm}[t]
\caption{Tiered PC algorithm with CPDAG output}
\label{alg:tiered-pc}
\KwIn{Data $\mathcal{D}$, threshold $t>0$, and tier assignment $O(V)$.}
\KwOut{A CPDAG estimate $\hat{\mathcal{G}}_n\in\mathfrak{G}$,
        or $\mathrm{INVALID}$.}

Initialize $\mathcal{G}\leftarrow\mathcal{G}_{\mathrm{full}}$
and all separating sets as undefined\;

\For{$\ell=0,\ldots,p-2$}{
  Store $\widetilde{\mathrm{adj}}_\ell(i)
    \leftarrow\mathrm{adj}(\mathcal{G},i)$ for all $i\in V$\;
  Form $A_{\ell,ij}$ as in~\eqref{eq:tiered-conditioning}\;

  \If{$|A_{\ell,ij}|<\ell$ for every ordered adjacent pair $(i,j)$}{
    \textbf{break}\;
  }

  \ForEach{ordered adjacent pair $(i,j)$ with $|A_{\ell,ij}|\geq\ell$}{
    Test sets $S\subseteq A_{\ell,ij}$ of size $\ell$ in a fixed order,
    stopping at the first with $\hat{T}_{ij|S}\leq t$\;
    \If{such a set $S$ is found}{
      Delete $i-j$ from $\mathcal{G}$\;
      $\mathrm{sepset}(i,j)=\mathrm{sepset}(j,i)\leftarrow S$\;
    }
  }
}

Orient cross-tier edges and identify v-structures,
respecting $O(V)$\;
Apply $R_1$--$R_4$ until no further edge can be oriented\;

\If{the orientation requirements conflict
     or no consistent DAG extension exists}{
  \KwRet{$\mathrm{INVALID}$}\;
}

Convert $\mathcal{G}$ to its observational CPDAG
$\hat{\mathcal{G}}_n$\;
\KwRet{$\hat{\mathcal{G}}_n$}\;
\end{algorithm}

\section{Proofs for Section~\ref{sec:analysis}}
\label{app:analysis-proofs}

Throughout, write 
$\bar\Phi:=1-\Phi$. Given the data $\mathcal D$, the $\widehat U_l^{[m]}$
are independent $\mathcal N(0,\eta_n^2)$ over $l\in\mathcal T$
and $m\le M_n$. Run $m$ decides a test in $\mathcal{T}_\text{null}$
correctly if $\widehat{T}_l^{[m]}\le t_n$, and a test in
$\mathcal T\setminus\mathcal{T}_\text{null}$ correctly if
$\widehat{T}_l^{[m]}>t_n$. We interpret empty products as one.

\begin{proof}[Proof of Lemma~\ref{lemma:main}]
Fix the data and suppose that $E_n$ holds. Fix $l\in\mathcal{T}_\text{null}$. Then $\psi_l=0$, so $\widehat{T}_l^{[m]}=|\widehat U_l+\widehat U_l^{[m]}|$. Conditional on the observed data:
$$\mathbb{P}(\widehat{T}_l^{[m]} \leq t | \mathcal{D}) = \mathbb{P}(-t \leq \widehat U_l+\widehat U_l^{[m]} \leq t | \mathcal{D}).$$
Because $\widehat U_l$ is fixed given the data, this equals:
$$\mathbb{P}(-t-\widehat U_l \leq \widehat U_l^{[m]} \leq t - \widehat{U}_l | \mathcal{D}) = \Phi\left(\frac{t-\widehat{U}_l}{\eta}\right)-\Phi\left(\frac{-t-\widehat{U}_l}{\eta}\right).$$
Note that the function $g(u) = \Phi((t-u)/\eta)-\Phi((-t-u)/\eta)$ is even and is nonincreasing as $|u|$ increases. Since $|\widehat U_l|\le z$ on $E_n$, we have that:
$$\mathbb{P}(\widehat{T}_l^{[m]} \leq t~|~ \mathcal{D}) \geq \Phi\left(\frac{t-z}{\eta}\right)-\Phi\left(\frac{-t-z}{\eta}\right). $$

For $l\in\mathcal T\setminus\mathcal{T}_\text{null}$, let $h_l = \text{sign}(\psi_l)$. Note that:
$$\widehat{T}_l^{[m]} = \left|h_l\left(\frac{\psi_l}{\sigma_l}+\widehat U_l + \widehat U_l^{[m]}\right)\right| = \left|\kappa_l + h_l\widehat U_l + h_l\widehat U_l^{[m]}\right|.$$
Therefore, we have:
\[
\begin{aligned}
\mathbb P(\widehat{T}_l^{[m]}>t\mid\mathcal D)
&\ge
\mathbb P(\kappa_l+h_l\widehat U_l+h_l\widehat U_l^{[m]}>t\mid\mathcal D)\\
&=
\Phi\left(\frac{\kappa_l+h_l\widehat U_l-t}{\eta}\right)\\
&\ge
\Phi\left(\frac{\kappa_l-z-t}{\eta}\right).
\end{aligned}
\]
The last inequality uses $h_l\widehat U_l\ge-z$ on $E_n$. The perturbations are independent across tests, so the event that all tests in $\mathcal{T}$ are decided correctly has conditional probability at least $\pi$. The event is sufficient for the run to be correct.
The perturbations are also independent across runs, so some run is correct
with conditional probability at least $1-(1-\pi)^M$, and every
run is correct with conditional probability at least $\pi^M$.
Taking expectations over $E_n$ gives both bounds.
\end{proof}

\begin{proof}[Proof of Theorem~\ref{thm:main}]
By Lemma~\ref{lemma:main} and $1-x\le e^{-x}$,
\begin{equation}
\mathbb P\left(
\mathcal G^\star\in\widehat{\mathfrak G}_n
\right)
\ge
(1-e^{-M_n\pi_n})\mathbb P(E_n),
\label{eqn:temp}
\end{equation}
and $\liminf_n\mathbb P(E_n)\ge1-\nu$. Recall that:
\[
\pi_n :=
\Big\{
\Phi\Big(\frac{t_n-z}{\eta_n}\Big)
-
\Phi\Big(\frac{-t_n-z}{\eta_n}\Big)
\Big\}^{L_0}
\prod_{l \in \mathcal{T}_\text{non-null}}
\Phi\Big(\frac{\kappa_l-z-t_n}{\eta_n}\Big).
\]

\medskip
\noindent\textbf{(a)(i).} Since $t_n$ is bounded,  let $t_n\le\bar t$. Recall, $\underline\eta\le\eta_n\le\bar\eta$, with $\underline\eta>0$. For each null test, the corresponding factor of $\pi_n$ is 
$$
\Phi\left(\frac{t_n-z}{\eta_n}\right)-\Phi\left(\frac{-t_n-z}{\eta_n}\right).$$
This is the probability that a standard normal random variable lies in the interval $[(-t_n-z)/\eta_n, (t_n-z)/\eta_n]$. The interval has length $2t_n/\eta_n\ge2t_n/\bar\eta$ and is contained in
\[
\left[
-\frac{\bar t+z}{\underline\eta},
\frac{\bar t+z}{\underline\eta}
\right].
\]
The probability that a standard normal lies in this interval is bounded below by
\[
\frac{2t_n}{\bar\eta}
\phi\left(\frac{\bar t+z}{\underline\eta}\right).
\]
For each non-null test, the corresponding factor in $\pi_n$ is bounded below by the positive constant
$\Phi(-(\bar t+z)/\underline\eta)$. Hence, for some constant $C>0$, $\pi_n\ge Ct_n^{L_0}$. Since $M_nt_n^{L_0} \to \infty$, we obtain $M_n\pi_n \to \infty$. Plugging this into \eqref{eqn:temp} gives the desired result.

\medskip
\noindent\textbf{(a)(ii).}
For a null test, the corresponding factor of $\pi_n$ is
$$
\Phi\left(\frac{t_n-z}{\eta_n}\right)-\Phi\left(\frac{-t_n-z}{\eta_n}\right).$$
Since $z$ is a constant and $\eta_n$ is bounded above by a constant $\bar{\eta}$, and $t_n \to \infty$, we have $t_n-z>0$ for all large $n$. For such $n$:
\begin{eqnarray*}
\begin{aligned}
\frac{t_n-z}{\eta_n} \geq \frac{t_n-z}{\bar{\eta}} &\to \infty,\\
\frac{-t_n-z}{\eta_n} \leq \frac{-t_n-z}{\bar{\eta}}&\to -\infty.
\end{aligned}
\end{eqnarray*}
Hence, every null factor tends to one. Now split the nonnull tests into strong and weak tests. Let:
$$\mathcal{T}_\text{strong} = \{l \in \mathcal{T}_{\text{non-null}}, \kappa_l-t_n \to \infty\},$$
and let $W$ denote the set of weak nonnull tests.  For each strong nonnull test $l$, we have $\kappa_l-t_n-z>0$ for all large $n$. For such $n$:
\[
\frac{\kappa_l-t_n-z}{\eta_n} \geq \frac{\kappa_l-t_n-z}{\bar{\eta}}  \to\infty,
\]
so its factor
$$\Phi\left(\frac{\kappa_l-z-t_n}{\eta_n}\right) \to 1.$$
Define:
$$\mu_n := \left(\Phi\left(\frac{t_n-z}{\eta_n}\right)-\Phi\left(\frac{-t_n-z}{\eta_n}\right)\right)^{L_0}\prod_{l \in\mathcal{T}_\text{strong} }\Phi\Big(\frac{\kappa_l-z-t_n}{\eta_n}\Big). $$
Then, $\mu_n \to 1$. Fix a realization $\mathcal{D} \in E_n$. By the same bounds used to construct $\pi_n$, $$\mathbb{P}(\text{all null tests and strong nonnull tests are correct in run }m|\mathcal{D}) \geq \mu_n.$$
If $W = \emptyset$, then all tests are either null or strong nonnull, so $\pi_n = \mu_n \to 1$. On $E_n$, the first run is correct with conditional probability tending to one, and Lemma~\ref{lemma:main} gives coverage.

Now suppose $W \neq \emptyset$. For each run $m \leq M_n$, define:
\[
A_m
:=
\left\{\tilde U_l^{[m]}\ge r_n(L_{\text{weak}})
\text{ for all }l\in W
\right\},
\]
where $\tilde U_l^{[m]} = \text{sign}(\psi_l)\widehat U_l^{[m]}$. By definition of $r_n(L_{\text{weak}})$,
\[
\mathbb P\left(
\bigcup_{m\le M_n}A_m
\,\middle|\,
\mathcal D
\right)=1-\delta_n \to 1.
\]
On the event $\bigcup_{m\le M_n}A_m$, let $m^\star$ be the first run for which $A_m$ occurs. Notice that:
$$\mathbb{P}(m^\star \text{ exists} | \mathcal{D}) = 1-\delta_n \to 1.$$
Now fix $m$ and condition on $m^\star = m$. Since $m^\star = m$ implies $A_m$, for every $l \in W$, 
\[
\widehat T_l^{[m^\star]}
\ge
\kappa_l-z+r_n(L_{\text{weak}}).
\]
Because
$$
\kappa_l - t_n+r_n(L_{\text{weak}}) \to \infty,$$
we have, for sufficiently large $n$
$$\widehat T_l^{[m^\star]} > t_n.$$
Thus, on $E_n \cap \{m^\star = m\}$, all weak tests are decided correctly for all sufficiently large $n$. 

\begin{eqnarray*}
\begin{aligned}
&~~~~\mathbb{P}(\text{all tests correct in run }m |\mathcal{D}, m^\star=m),\\
&= \mathbb{P}(\text{all null tests and strong nonnull  tests are correct in run }m |\mathcal{D}, m^\star=m),\\
&=\mathbb{P}(\text{all null tests and strong nonnull tests are correct in run }m |\mathcal{D}),\\
&\geq \mu_n,
\end{aligned}
\end{eqnarray*}
for all sufficiently large $n$. Here, the second equality follows from the fact that, conditional on $\mathcal{D}$, the event $\{m^\star=m\}$ depends only on the perturbations of the weak tests, which are independent of the perturbations of the null and strong nonnull tests. 

Let $C_m := \{\text{all tests are correct in run } m\}$. We then have for every $\mathcal{D} \in E_n$ and $n$ sufficiently large:
\begin{eqnarray*}
\begin{aligned}
&\mathbb{P}(\text{there exists a run in which all tests are correct} | \mathcal{D}),\\
&\geq \sum_{m=1}^{M_n} \mathbb{P}(C_m, m^\star = m |\mathcal{D}), \\
&= \sum_{m=1}^{M_n}\mathbb{P}(m^\star = m|\mathcal{D})\mathbb{P}(C_m|\mathcal{D}, m^\star = m), \\
&\geq \mu_n(1-\delta_n) \to 1.
\end{aligned}
\end{eqnarray*}

The first inequality follows because
$$
\bigcup_{m=1}^{M_n}\{C_m,m^\star=m\}
\subseteq
\{\text{there exists a run in which all tests are correct}\}.
$$
The event on the left requires the specially selected run $m^\star$ to be correct and is therefore sufficient, but not necessary, for the existence of a correct run.

\medskip
\noindent\textbf{(b).}
Let
\[
F_n
:=
\left\{
\max_{m\le M_n,\,l\in\mathcal T}
|\widehat U_l^{[m]}|
\le\omega_n
\right\},
\]
so that $\mathbb P(F_n\mid\mathcal D)=1-\delta_n$.
On $E_n\cap F_n$, every null statistic is at most
$z+\omega_n<t_n$, and every nonnull statistic is at least
$\kappa_{\min,n}-z-\omega_n>t_n$, for all large $n$.
Every run is then correct, so
$\widehat{\mathfrak G}_n=\{\mathcal G^\star\}$
with probability at least $(1-\delta_n)\mathbb P(E_n)$.

\end{proof}

\section{Proofs for Section~\ref{sec:distance-theory}}\label{app:distance-proofs}
 
Let $\Delta = \mathrm{Skel}(\mathcal{G}_s) \triangle \mathrm{Skel}(\mathcal{G}_t)$ for brevity. We use three standard facts.
 
\emph{Fact 1.} For CPDAGs $\mathcal{G}$ and $\mathcal{G}'$, $\mathcal{G} \lessdot \mathcal{G}'$ exactly when there are DAGs $\mathcal{D}$ and
$\mathcal{D}' = \mathcal{D} + e$ in their classes \citep{chickering2002optimal}. Here $\mathcal D+e$ is $\mathcal D$ with one added edge on the pair $e$.
 
\emph{Fact 2.} For CPDAGs $\mathcal{G}$ and $\mathcal{H}$, $\mathcal{H} \preceq \mathcal{G}$
exactly when every conditional independence encoded by $\mathcal{G}$ is encoded by $\mathcal{H}$
(Proposition~\ref{prop:background}). In particular, $\mathrm{Skel}(\mathcal{H}) \subseteq \mathrm{Skel}(\mathcal{G})$. Indeed, if $\{u,v\}\notin\mathrm{Skel}(\mathcal G)$, then $u$ and $v$ are d-separated in $\mathcal G$ by some set. The CPDAG $\mathcal H$ encodes the same independence, so $u$ and $v$ are nonadjacent in $\mathcal H$.
 
\emph{Fact 3.} A subgraph of a DAG encodes every independence of the DAG. By Fact~2, the class of the subgraph therefore precedes the class of the DAG. To see the first claim, let $\mathcal D'$ be a subgraph of a DAG $\mathcal D$. Every path of $\mathcal D'$ is a path of $\mathcal D$ with the same colliders. Every descendant of a vertex in $\mathcal D'$ is a descendant of that vertex in $\mathcal D$. Hence a path that d-connects two vertices given a set $S$ in $\mathcal D'$ also d-connects them given $S$ in $\mathcal D$.
 
We also pass from paths in $\mathcal L$ to sequences of DAGs. Lemma~\ref{lem:forbidden-change}, Theorem~\ref{thm:decomposition} and Proposition~\ref{prop:stray} use the following definition and lemma.
 
\begin{definition}[Valid DAG operations]\label{def:valid-dag-ops}
Let $\mathcal{D} = (V, E)$ be a DAG. A \emph{valid operation} on $\mathcal{D}$ yields a new DAG $\mathcal{D}'$ via one of
three mechanisms:
\begin{enumerate}
\item[(1)] edge addition, where $\mathcal{D}' = (V, E \cup \{h\})$, $h \notin E$, and $\mathcal{D}'$ remains
acyclic;
\item[(2)] edge deletion, where $\mathcal{D}' = (V, E \setminus \{h\})$ and $h \in E$;
\item[(3)] covered edge reversal,
where $\mathcal{D}'$ is obtained by reversing a covered edge $h = u \to v$ in $\mathcal{D}$, that is, an edge with
\[
\mathrm{Pa}_{\mathcal{D}}(v) = \mathrm{Pa}_{\mathcal{D}}(u) \cup \{u\}.
\]
\end{enumerate}
A sequence of DAGs is valid if each graph is
obtained from its predecessor by a valid operation.
\end{definition}
 
\begin{lemma}[DAG--CPDAG correspondence]\label{lem:dag-cpdag}
Let $\mathcal{G}_0, \dots, \mathcal{G}_k$ be a path in $\mathcal{L}$. There exists a valid sequence of DAGs
$\mathcal{D}_0, \dots, \mathcal{D}_m$ (where $m \ge k$) such that $\mathcal{D}_0 \in [\mathcal{G}_0]$,
$\mathcal{D}_m \in [\mathcal{G}_k]$, and every transition $\mathcal{D}_j \to \mathcal{D}_{j+1}$ is either a single edge
addition, a single edge deletion, or a covered edge reversal. The sequence can be chosen so that exactly $k$ transitions are edge additions or deletions, and so that the classes of $\mathcal{D}_0, \dots, \mathcal{D}_m$, with consecutive repetitions removed, are $\mathcal{G}_0, \dots, \mathcal{G}_k$ in this order. Such a sequence \emph{represents} the path. Conversely, let $\mathcal{D}_0, \dots, \mathcal{D}_m$ be any valid sequence of DAGs with exactly $r$ edge additions and deletions. Then the classes of $\mathcal{D}_0, \dots, \mathcal{D}_m$, with consecutive repetitions removed, form a path in $\mathcal{L}$ with $r$ steps.
\end{lemma}
 
\begin{proof}
We construct the sequence inductively, starting from any DAG $\mathcal{D}_0 \in [\mathcal{G}_0]$. Let $\mathcal{D}_{\mathrm{curr}}$ be the current DAG representative of
$\mathcal{G}_i$. Consider a step $\mathcal{G}_i \to \mathcal{G}_{i+1}$ in the path. By Fact~1, there exist representatives $\mathcal{D} \in [\mathcal{G}_i]$ and
$\mathcal{D}' \in [\mathcal{G}_{i+1}]$ such that $\mathcal{D}'$ is obtained from $\mathcal{D}$ by adding or deleting a
single edge \citep{chickering2002optimal}. Since $\mathcal{D}_{\mathrm{curr}}$ and $\mathcal{D}$ belong to the same
equivalence class $[\mathcal{G}_i]$, there exists a sequence of covered edge reversals transforming
$\mathcal{D}_{\mathrm{curr}}$ to $\mathcal{D}$ \citep{chickering1995transformational}. We append this sequence to the
DAG sequence, followed by the single operation transforming $\mathcal{D}$ to $\mathcal{D}'$. We repeat this construction for all steps in
the CPDAG path. Every DAG reached by the covered edge reversals stays in $[\mathcal{G}_i]$ \citep{chickering1995transformational}. The step $\mathcal{G}_i \to \mathcal{G}_{i+1}$ therefore contributes exactly one edge addition or deletion, and this operation leads from $[\mathcal{G}_i]$ to $[\mathcal{G}_{i+1}]$. Hence the constructed sequence represents the path.
 
For the converse, a covered edge reversal preserves the class \citep{chickering1995transformational}. By Fact~1, an edge addition or deletion moves to a neighboring class. The classes of the sequence, with consecutive repetitions removed, therefore form a path with one step per edge addition or deletion.
\end{proof}
 
\subsection{Bounds for common lower and upper graphs} \label{app:bound-proofs}
 
We prove Propositions~\ref{prop:common-lower} and~\ref{prop:common-upper} for a finite collection $\mathcal{A}$ of
CPDAGs, as extended at the end of Section~\ref{sec:common-bounds}. The pair $(\mathcal{G}_s,\mathcal{G}_t)$ is the
case $\mathcal{A}=\{\mathcal{G}_s,\mathcal{G}_t\}$.
 
\begin{proof}[Proof of Proposition~\ref{prop:common-lower}]
\emph{Lower bound.} The DAG $\mathcal{D}_\cap(\mathcal{A})$ is a subgraph of the chosen DAG representative $\mathcal{D}_\mathcal{G}$ of every
$\mathcal{G}\in\mathcal{A}$. By Fact~3, the CPDAG of $\mathcal{D}_\cap(\mathcal{A})$ is a
common lower graph, of rank $|E(\mathcal{D}_\cap(\mathcal{A}))|$.
 
\emph{Upper bound.} Let $\mathcal{H}$ be any common lower graph. By Fact~2, $\mathrm{Skel}(\mathcal H)\subseteq\mathrm{Skel}(\mathcal G)$ for every $\mathcal G\in\mathcal A$, so the skeleton of $\mathcal H$ lies in $\mathrm{Skel}_\cap(\mathcal{A})$. Let $\{u,w\}$ and $\{v,w\}$ be adjacent in $\Gamma_L(\mathcal{A})$. Then $(u,w,v)$ is a forbidden triple for $\mathcal A$,
with $u \to w \leftarrow v$ a v-structure in $\mathcal{G}_1 \in \mathcal{A}$ but not in $\mathcal{G}_2 \in \mathcal{A}$.
Let $\mathcal D_1$ be a DAG in the class of $\mathcal{G}_1$. In $\mathcal D_1$, the vertices $u$ and $v$ are nonadjacent and $w$ is a child of both. One of $u$ and $v$ is not a descendant of the other; say $v$ is not a
descendant of $u$. In
$\mathcal{G}_2$, the triple is unshielded and not a v-structure. Hence
\[
\begin{aligned}
&u\text{ and }v\text{ are d-separated in }\mathcal G_1\text{ by }\mathrm{Pa}_{\mathcal D_1}(u)\text{, a set that does not contain }w;\\
&u\text{ and }v\text{ are d-separated in }\mathcal G_2\text{ by some set, and every such set contains }w.
\end{aligned}
\]
By Fact~2, both independences are encoded by $\mathcal{H}$. The pair $\{u,v\}$ is absent from $\mathcal{H}$, since $\{u,v\}$ lies in no skeleton of $\mathcal A$. Suppose $\mathcal{H}$
kept both legs. If the triple were not a v-structure of $\mathcal{H}$, then $\mathcal H$ would violate the first independence. If the triple were a v-structure of $\mathcal H$, then $\mathcal H$ would violate the
second independence. Hence $\mathcal H$ omits $\{u,w\}$ or $\{v,w\}$.
 
\emph{Vertex cover.} The edge of $\Gamma_L(\mathcal A)$ between $\{u,w\}$ and $\{v,w\}$ was arbitrary. The edges of $\mathrm{Skel}_\cap(\mathcal{A})$ missing from $\mathcal{H}$ therefore form a vertex cover of
$\Gamma_L(\mathcal{A})$. Since $\mathrm{rank}(\mathcal H)=|\mathrm{Skel}(\mathcal H)|$,
\[
\mathrm{rank}(\mathcal{H}) \le |\mathrm{Skel}_\cap(\mathcal{A})| -
\tau(\Gamma_L(\mathcal{A})).
\]
Finally, $\tau(\Gamma_L(\mathcal{A})) \ge \nu(\Gamma_L(\mathcal{A}))$, since a vertex cover
contains an endpoint of each edge of a matching, and these endpoints are distinct.
\end{proof}
 
\begin{proof}[Proof of Proposition~\ref{prop:common-upper}]
By Fact~2, a common upper graph contains every skeleton.
Its rank is therefore at least $|\mathrm{Skel}_\cup(\mathcal A)|$.

Write $\mathcal H:=\mathcal G_{\mathrm{ch}}(\mathcal A)$.
Since $\mathcal H$ is chordal, $\mathcal H$ is a CPDAG.
The conditional independences encoded by $\mathcal H$ are exactly
its graph separations \citep{lauritzen1996graphical}.

Let $\mathcal G\in\mathcal A$.
By construction, $\mathrm{moral}(\mathcal G)$ is a subgraph of
$\mathcal H$.
Every separation in $\mathcal H$ is therefore a separation in
$\mathrm{moral}(\mathcal G)$, hence a d-separation in $\mathcal G$
\citep{lauritzen1996graphical}.
By Fact~2, $\mathcal G\preceq\mathcal H$.
Thus $\mathcal H$ is a common upper graph, giving
\[
\rho_U(\mathcal A)
\le \mathrm{rank}(\mathcal H)
=|E(\mathcal H)|.
\]
\end{proof}

\subsection{Lower bounds from mismatches}\label{app:mismatch-proof}
 
\begin{proof}[Proof of Theorem~\ref{thm:mismatch}]
Let $P$ be a shortest path from $\mathcal{G}_s$ to $\mathcal{G}_t$, so that $d(\mathcal{G}_s,\mathcal{G}_t)$ is the
number of steps of $P$. By Fact~1, each step adds or deletes the edge on one pair of vertices. A step that
deletes a pair of $\mathrm{Skel}(\mathcal{G}_t)$ performs $\mathrm{ER}$ on that pair. A step that adds a pair outside
$\mathrm{Skel}(\mathcal{G}_t)$ performs $\mathrm{EA}$ on that pair. These are the detour steps of $P$. Let $F$ be the set
of distinct detour operations performed along $P$.

By Definition~\ref{def:mismatch}, $P$ performs an operation of every $M\in\mathcal{M}$. So
$F\cap\bigcup_{M\in\mathcal{M}}M$ meets every member of $\mathcal{M}$, and is a hitting set for $\mathcal M$. By the definition of
$\widetilde m_{\mathrm{op}}$,
\[
|F|\ge\Bigl|F\cap\bigcup_{M\in\mathcal{M}}M\Bigr|\ge\widetilde m_{\mathrm{op}}(\mathcal{M}).
\]
 
We now count the steps of $P$. Fix a pair $e$ and follow the state of $e$ along $P$. Each step on $e$ switches the
state. The steps on $e$ therefore alternate between moving $e$ away from its state in $\mathcal{G}_t$ and moving $e$ back. If
$e\notin\Delta$, the state of $e$ starts and ends in its target state. The steps on $e$ then come in pairs: a detour step,
then a step that undoes the detour step. If $e\in\Delta$, the first step on $e$ moves $e$ to its target state, and the remaining steps
on $e$ come in such pairs. Hence
\[
(\text{number of steps on }e)=\mathbf{1}\{e\in\Delta\}+2\,(\text{number of detour steps on }e).
\]
Summing over all pairs gives Equation~\eqref{eq:path-detour-operations} for $P$,
where $q(P)$ is the number of detour steps of $P$, counted with repetition. The count uses no property of $P$ other than its endpoints, so Equation~\eqref{eq:path-detour-operations} holds for every path from $\mathcal{G}_s$ to $\mathcal{G}_t$. Each operation in $F$ is performed at least
once, so $q(P)\ge|F|$. Combining these facts with the hitting-set bound gives
\[
d(\mathcal{G}_s,\mathcal{G}_t)=|\Delta|+2q(P)\ge|\Delta|+2|F|\ge|\Delta|+2\widetilde m_{\mathrm{op}}(\mathcal{M}).
\]
\end{proof}
 
\begin{proof}[Proof of Corollary~\ref{thm:independent-conflicts}]
Let $H$ be a hitting set for $I$. Then $H$ contains an operation from each mismatch in $I$. The mismatches in $I$ are pairwise disjoint, so these operations are distinct. Hence $|H|\ge|I|$, and $\widetilde m_{\mathrm{op}}(I)\ge|I|$. Theorem~\ref{thm:mismatch} with $\mathcal{M}=I$ gives $d(\mathcal{G}_s,\mathcal{G}_t)\ge|\Delta|+2|I|$.
\end{proof}
 
The next two lemmas concern a single forbidden triple. The proof of Theorem~\ref{thm:polytree} and Appendix~\ref{app:mismatch-certification} use Lemma~\ref{lem:forbidden-mismatch}.
 
\begin{lemma}\label{lem:forbidden-change}
Let $T = \{u, v, w\}$ be a triple with $\{u,w\}, \{v,w\} \in \mathrm{Skel}(\mathcal{G}_s) \cap \mathrm{Skel}(\mathcal{G}_t)$
and $\{u,v\} \notin \mathrm{Skel}(\mathcal{G}_s) \cup \mathrm{Skel}(\mathcal{G}_t)$. If $T$ forms a v-structure
$u \to w \leftarrow v$ in $\mathcal{G}_s$ but not in $\mathcal{G}_t$, then any path from $\mathcal{G}_s$ to
$\mathcal{G}_t$ in $\mathcal{L}$ contains a graph whose skeleton on $T$ differs from that of $\mathcal{G}_s$ and
$\mathcal{G}_t$.
\end{lemma}
 
\begin{proof}
Assume for contradiction that a path exists on which the skeleton on $T$ remains exactly $\{\{u,w\}, \{v,w\}\}$. By
Lemma~\ref{lem:dag-cpdag}, a valid sequence of DAGs represents this path, from a DAG $\mathcal{D}_s$ in the
class of $\mathcal{G}_s$ to a DAG $\mathcal{D}_t$ in the class of $\mathcal{G}_t$. Every DAG of the sequence lies in the class of a graph on the path. Since the skeleton on $T$ is constant along the path,
no edge addition or deletion occurs within $T$. The edges between $u$ and $w$ and between $v$ and $w$ can therefore change
orientation only by covered edge reversals.
 
Let $\mathcal D$ be a DAG of the sequence that contains the v-structure $u \to w \leftarrow v$. The edge $u \to w$ is reversible only if $u\to w$ is
covered, that is,
\[
\mathrm{Pa}_{\mathcal D}(w) = \mathrm{Pa}_{\mathcal D}(u) \cup \{u\}.
\]
However, $v \in \mathrm{Pa}_{\mathcal D}(w)$, so the covering condition
requires $v \in \mathrm{Pa}_{\mathcal D}(u) \cup \{u\}$. Since $u \neq v$, the covering condition requires $v \in \mathrm{Pa}_{\mathcal D}(u)$. Thus reversing $u \to w$
requires the edge $v \to u$. Similarly, reversing $v \to w$ requires the edge $u \to v$.
 
In both cases, the reversal is valid only if the shielding edge between $u$ and $v$ exists. However, $\{u,v\}$ is in
neither skeleton, and by assumption the skeleton on $T$ is constant along the path. So $\{u,v\}$ is absent throughout.
Neither leg of the v-structure is ever covered. The sequence therefore keeps $u \to w \leftarrow v$ and cannot reach the
non-v-structure of $\mathcal{G}_t$, a contradiction.
\end{proof}
 
\begin{lemma}[Forbidden-triple mismatches]\label{lem:forbidden-mismatch}
Let $(u,w,v)\in\mathcal{S}(\mathcal{G}_s,\mathcal{G}_t)$. Then the set $M_{(u,w,v)}$ of \eqref{eqn:forbidden} is a mismatch for $(\mathcal{G}_s,\mathcal{G}_t)$.
\end{lemma}
 
\begin{proof}
Write $T=\{u,v,w\}$. By the definition of $\mathcal{S}(\mathcal{G}_s,\mathcal{G}_t)$, the legs $\{u,w\}$ and $\{v,w\}$ lie in both skeletons, the shield $\{u,v\}$ lies in neither skeleton, and $u\to w\leftarrow v$ is a v-structure in exactly one of $\mathcal{G}_s$ and $\mathcal{G}_t$. Let $P$ be a path from $\mathcal{G}_s$ to $\mathcal{G}_t$.
 
If $u\to w\leftarrow v$ is a v-structure in $\mathcal{G}_s$, apply Lemma~\ref{lem:forbidden-change} to $P$. Otherwise $u\to w\leftarrow v$ is a v-structure in $\mathcal{G}_t$. The reverse of $P$ is a path from $\mathcal{G}_t$ to $\mathcal{G}_s$, and Lemma~\ref{lem:forbidden-change} applies to the reverse of $P$ with the roles of $\mathcal{G}_s$ and $\mathcal{G}_t$ exchanged. In both cases, some graph on $P$ has a skeleton on $T$ other than $\{\{u,w\},\{v,w\}\}$.
 
 The skeleton of $\mathcal{G}_s$ on $T$ is $\{\{u,w\},\{v,w\}\}$. Consider the first step of $P$ that changes the skeleton on $T$. This step deletes a leg or adds the shield. Both legs lie in $\mathrm{Skel}(\mathcal{G}_t)$, and the shield does not. The step therefore performs $\mathrm{ER}(u,w)$, $\mathrm{ER}(w,v)$ or $\mathrm{EA}(u,v)$, which is an operation of $M_{(u,w,v)}$. Since $P$ was arbitrary, $M_{(u,w,v)}$ is a mismatch for $(\mathcal{G}_s,\mathcal{G}_t)$.
\end{proof}
 
\subsection{Decomposition over disconnected blocks}\label{app:decomposition-proof}
The proof uses Definition~\ref{def:valid-dag-ops} and Lemma~\ref{lem:dag-cpdag}, stated at the start of Appendix~\ref{app:distance-proofs}.
 
\begin{proof}[Proof of Theorem~\ref{thm:decomposition}]
Call an edge \emph{cross} if its vertices lie in different blocks. We first show that no graph on a shortest path has a
cross edge. Suppose that some graph on a shortest path $P$ from $\mathcal{G}_s$ to $\mathcal{G}_t$ has a cross edge. By
Lemma~\ref{lem:dag-cpdag}, there exists a valid sequence of DAGs $\mathcal{D}_0, \dots, \mathcal{D}_m$ representing $P$.
We construct a new sequence $\widetilde{\mathcal{D}}_0, \dots, \widetilde{\mathcal{D}}_m$, where
\[
\widetilde{\mathcal{D}}_j:=\mathcal{D}_j\text{ with its cross edges removed}.
\]
Operations on cross pairs are omitted, and every other operation remains valid:
\begin{itemize}
\item A deleted edge inside a block is an edge of $\widetilde{\mathcal{D}}_j$.
\item An edge that can be added to $\mathcal{D}_j$ without creating a cycle can be added to its subgraph $\widetilde{\mathcal{D}}_j$.
\item Let $a \to b$ be a covered edge of $\mathcal{D}_j$ with
$a,b\in V_i$. Then
$\mathrm{Pa}_{\mathcal{D}_j}(b)
=\mathrm{Pa}_{\mathcal{D}_j}(a)\cup\{a\}$.
Removing cross edges intersects both parent sets with $V_i$.
Since $a\in V_i$, we obtain
\[
\mathrm{Pa}_{\widetilde{\mathcal{D}}_j}(b)
=\mathrm{Pa}_{\widetilde{\mathcal{D}}_j}(a)\cup\{a\}.
\]
Hence $a \to b$ remains covered in $\widetilde{\mathcal{D}}_j$.
\end{itemize}
 
Since $\mathcal{G}_s$ and $\mathcal{G}_t$ have no cross edges, $\widetilde{\mathcal{D}}_0 = \mathcal{D}_0$ and
$\widetilde{\mathcal{D}}_m = \mathcal{D}_m$. After repeated DAGs are dropped, the new sequence is valid. By the converse part of Lemma~\ref{lem:dag-cpdag}, the classes of the new sequence form a path from $\mathcal{G}_s$ to
$\mathcal{G}_t$ with one step per edge addition or deletion. The sequence $\mathcal{D}_0, \dots, \mathcal{D}_m$ has one edge addition or deletion per step of $P$. The path $P$ adds a cross edge. The cross edge is in neither skeleton, so $P$ later removes the cross edge. Both operations
are omitted, so the new path is shorter than $P$, a contradiction.
 
We now compute the distance. Write $d_i = d_{V_i}(\mathcal{G}_s|_{V_i}, \mathcal{G}_t|_{V_i})$. A CPDAG without
cross edges is the class of a disjoint union of DAGs over the blocks. This class depends only on the classes of the
parts, since its skeleton and v-structures are those of the parts. By Fact~1, a step between two such CPDAGs is realized by DAGs
$\mathcal{D}$ and $\mathcal{D} + e$ with $e$ inside a single block $V_i$. The step therefore moves the restriction to $V_i$ to a
neighbor over $V_i$ and leaves the other restrictions unchanged. By the first part, no graph on a shortest path has a cross edge. Along a shortest path, the restrictions to $V_i$, with
repetitions removed, therefore form a path from $\mathcal{G}_s|_{V_i}$ to $\mathcal{G}_t|_{V_i}$ whose steps are the
steps inside $V_i$. This path has at least $d_i$ steps, so
\[
d(\mathcal{G}_s, \mathcal{G}_t)=\sum_i(\text{number of steps inside }V_i)\ge\sum_i d_i.
\]
Conversely, a step between neighbors over
$V_i$, realized by adding an edge to a DAG over $V_i$, remains a step between neighbors over $V$ when the DAGs over the
other blocks are kept fixed. Concatenating shortest paths over the blocks therefore gives a path of length $\sum_i d_i$. Hence
$d(\mathcal{G}_s, \mathcal{G}_t) = \sum_i d_i$. Since each block needs at least $d_i$ steps, a shortest path makes exactly $d_i$ steps inside each $V_i$.
\end{proof}
 
\subsection{Polytree shortest path}\label{app:polytree-proof}
 
We first fix notation for Lemma~\ref{lem:polytree-er}. We identify the triples $(u,w,v)$ and $(v,w,u)$. The three
vertices of a triple are distinct, since its legs $\{u,w\}$ and $\{w,v\}$ and its shield $\{u,v\}$ are pairs of distinct
vertices. For a forbidden triple $T=(u,w,v)$, write $M_T:=M_{(u,w,v)}$. By \eqref{eqn:forbidden},
\[
M_T=\{\mathrm{ER}(u,w),\ \mathrm{ER}(w,v),\ \mathrm{EA}(u,v)\}.
\]
Thus $M_T$ contains exactly two edge removals, one on each leg of $T$, and exactly one edge addition, on the shield of
$T$. As in the definition of $\widetilde m_{\mathrm{op}}$ before Theorem~\ref{thm:mismatch}, a \emph{hitting set} for a
collection $\mathcal{M}$ of mismatches is a set $H$ such that
\[
H\subseteq\bigcup_{M\in\mathcal{M}}M
\quad\text{and}\quad
H\cap M\neq\varnothing\ \text{ for every }M\in\mathcal{M}.
\]
A hitting set with $\widetilde m_{\mathrm{op}}(\mathcal{M})$ elements is \emph{minimum}.
 
\begin{lemma}\label{lem:polytree-er}
Let $\mathcal{G}_s$ and $\mathcal{G}_t$ be CPDAGs where at least one of them is a polytree. Let
$\mathcal{S}=\mathcal{S}(\mathcal{G}_s, \mathcal{G}_t)$ be the set of forbidden triples, and let
$\mathcal{M}_{\mathcal{S}}=\{M_{(u,w,v)}:(u,w,v)\in\mathcal{S}\}$ be the collection of forbidden-triple mismatches. Then some
minimum hitting set for $\mathcal{M}_{\mathcal{S}}$, of size $\widetilde m_{\mathrm{op}}(\mathcal{M}_{\mathcal{S}})$, consists
entirely of edge removal ($\mathrm{ER}$) operations.
\end{lemma}
 
\begin{proof}
Fix a pair $\{u,v\}$. By the form of $M_T$, the
operation $\mathrm{EA}(u,v)$ lies in $M_T$ if and only if $\{u,v\}$ is the shield of $T$, that is, $T=(u,x,v)$ for some
vertex $x$. We show that at most one triple in $\mathcal{S}$ has shield $\{u,v\}$. Suppose that $T_1=(u,w_1,v)$ and
$T_2=(u,w_2,v)$ are distinct triples in $\mathcal{S}$, so that $w_1\neq w_2$. By the definition of $\mathcal{S}$, the legs
of $T_1$ and $T_2$ lie in both skeletons:
\[
\{u,w_1\},\ \{w_1,v\},\ \{v,w_2\},\ \{w_2,u\}\in\mathrm{Skel}(\mathcal{G}_s)\cap\mathrm{Skel}(\mathcal{G}_t).
\]
The vertices $u$, $w_1$, $v$ and $w_2$ are distinct, because $w_1\neq w_2$ and the vertices of each triple are distinct.
Hence $u-w_1-v-w_2-u$ is a cycle of length four in $\mathrm{Skel}(\mathcal{G}_s)$ and in $\mathrm{Skel}(\mathcal{G}_t)$.
The cycle contradicts the assumption that the skeleton of $\mathcal{G}_s$ or of $\mathcal{G}_t$ has no cycles. Therefore at most
one triple in $\mathcal{S}$ has shield $\{u,v\}$, and
\[
\bigl|\{M\in\mathcal{M}_{\mathcal{S}}:\mathrm{EA}(u,v)\in M\}\bigr|\le1.
\]
The pair $\{u,v\}$ was arbitrary, so this bound holds for every pair.
 
Now let $H$ be a minimum hitting set for $\mathcal{M}_{\mathcal{S}}$ that
contains an edge addition $\mathrm{EA}(u,v)$. Since $H\subseteq\bigcup_{M\in\mathcal{M}_{\mathcal{S}}}M$, the operation
$\mathrm{EA}(u,v)$ lies in at least one mismatch of $\mathcal{M}_{\mathcal{S}}$. By the bound on the number of mismatches containing $\mathrm{EA}(u,v)$, the operation $\mathrm{EA}(u,v)$ lies in
exactly one mismatch, say $M_T$ with $T=(u,w,v)\in\mathcal{S}$. Define
\[
H':=\bigl(H\setminus\{\mathrm{EA}(u,v)\}\bigr)\cup\{\mathrm{ER}(u,w)\}.
\]
We check that $H'$ is a hitting set. First, $\mathrm{ER}(u,w)\in M_T$, so
$H'\subseteq\bigcup_{M\in\mathcal{M}_{\mathcal{S}}}M$. Next, let $M\in\mathcal{M}_{\mathcal{S}}$.
\begin{itemize}
\item If $M=M_T$, then $\mathrm{ER}(u,w)\in M\cap H'$.
\item If $M\neq M_T$, then $\mathrm{EA}(u,v)\notin M$ by the bound on the number of mismatches containing $\mathrm{EA}(u,v)$. Hence
$M\cap H\subseteq H\setminus\{\mathrm{EA}(u,v)\}\subseteq H'$. Since $H$ is a hitting set,
$M\cap H'\supseteq M\cap H\neq\varnothing$.
\end{itemize}
Thus $H'$ is a hitting set. By construction, $|H'|\le|H|=\widetilde m_{\mathrm{op}}(\mathcal{M}_{\mathcal{S}})$. No
hitting set has fewer than $\widetilde m_{\mathrm{op}}(\mathcal{M}_{\mathcal{S}})$ elements, so $H'$ is also a minimum
hitting set. The edge additions in $H'$ are exactly the edge additions in $H$ other than $\mathrm{EA}(u,v)$, because
$\mathrm{ER}(u,w)$ is an edge removal. Hence $H'$ contains one fewer edge addition than $H$.
 
The collection $\mathcal{M}_{\mathcal{S}}$ is finite, so
minimum hitting sets exist. Among them, choose $H^\star$ with the fewest edge additions. Suppose $H^\star$ contained an edge
addition. The exchange of the previous paragraph, applied with $H=H^\star$, would then produce a minimum hitting set with fewer edge additions than $H^\star$, which
contradicts the choice of $H^\star$. Hence $H^\star$ contains no edge addition. By Definition~\ref{def:detour-operations}, every detour
operation is an edge removal or an edge addition, so $H^\star$ consists entirely of edge removals.
\end{proof}
 
For a DAG $K$, write $\mathcal{I}(K)$ for the set of conditional independences encoded by $K$. For two DAGs $K$ and $L$, the inclusion
$\mathcal{I}(K) \subseteq \mathcal{I}(L)$, equivalently
$[L] \preceq [K]$ by Fact~2, implies the following \emph{basic conditions}
\citep{Kocka2001OnCI}:
\begin{enumerate}
\item[(a)] $\mathrm{Skel}(L) \subseteq \mathrm{Skel}(K)$;
\item[(b)] every v-structure of $L$ that is unshielded in $K$ is also
a v-structure of $K$;
\item[(c)] every v-structure of $K$ whose legs both lie in $\mathrm{Skel}(L)$ is also a v-structure of $L$.
\end{enumerate}
The next lemma shows that these conditions are also sufficient
for polytrees.
 
\begin{lemma}\label{lem:tree-inclusion}
Let $K$ and $L$ be DAGs such that their skeletons, $\mathrm{Skel}(K)$ and $\mathrm{Skel}(L)$, are polytrees.
Then the basic conditions (a), (b), and (c) hold if and only if $\mathcal{I}(K) \subseteq \mathcal{I}(L)$.
\end{lemma}
 
\begin{proof}
Necessity follows from the result of \citet{Kocka2001OnCI} stated before the lemma.
 
For sufficiency, suppose that (a), (b), and (c) hold. Construct a DAG $J$ by deleting from $K$ every edge whose underlying pair is absent from $\mathrm{Skel}(L)$, retaining the orientations of all remaining edges. By (a),
\[
\mathrm{Skel}(J)=\mathrm{Skel}(L).
\]
We show that $J$ and $L$ also have the same v-structures.
 
Suppose first that $u \to w \leftarrow v$ is a v-structure of $J$. Both arrows belong to $K$. Since $K$ is a polytree, $u$ and $v$ cannot be adjacent in $K$, as an edge between $u$ and $v$ would create a triangle in the skeleton of $K$. Thus $u \to w \leftarrow v$ is also a v-structure of $K$. Both legs lie in $\mathrm{Skel}(J)=\mathrm{Skel}(L)$, so condition (c) implies that $u \to w \leftarrow v$ is a v-structure of $L$.
 
Conversely, suppose that $u \to w \leftarrow v$ is a v-structure of $L$. By (a), both legs belong to $\mathrm{Skel}(K)$. Since $K$ is a polytree, $u$ and $v$ are nonadjacent in $K$. Condition (b) therefore implies that $u \to w \leftarrow v$ is a v-structure of $K$. Both arrows are retained in $J$, so $u \to w \leftarrow v$ is also a v-structure of $J$.
 
Consequently, $J$ and $L$ have the same skeleton and the same v-structures, and hence are Markov equivalent. Since $J$ is obtained from $K$ by edge deletions, Fact~3 shows that every conditional independence encoded by $K$ is also encoded by $J$. Therefore,
\[
\mathcal{I}(K) \subseteq \mathcal{I}(J)=\mathcal{I}(L),
\]
which proves sufficiency.
\end{proof}
 
\begin{proof}[Proof of Theorem~\ref{thm:polytree}]
Let $\mathcal{M}_{\mathcal{S}}$ be the collection of forbidden-triple mismatches defined before
Theorem~\ref{thm:polytree}. The theorem makes three claims. First, some shortest path is a down-up path, so that
$d(\mathcal{G}_s,\mathcal{G}_t)=d_{\downarrow\uparrow}(\mathcal{G}_s,\mathcal{G}_t)$. Second,
$d(\mathcal{G}_s,\mathcal{G}_t)=|\Delta|+2\widetilde m_{\mathrm{op}}(\mathcal{M}_{\mathcal{S}})$. Third, the shortest
down-up path traverses through polytrees. We first construct a down-up path $P^\star$ of length
$|\Delta|+2\widetilde m_{\mathrm{op}}(\mathcal{M}_{\mathcal{S}})$, and then derive the three claims.
 
By Lemma~\ref{lem:polytree-er}, there is a minimum hitting set $H^\star$ for
$\mathcal{M}_{\mathcal{S}}$ that consists entirely of edge removals. Let
\[
E_{\mathrm{conflict}}:=\bigl\{\{a,b\}:\mathrm{ER}(a,b)\in H^\star\bigr\}
\]
be the set of pairs removed by the operations in $H^\star$. Distinct edge removals act on distinct pairs, so
$|E_{\mathrm{conflict}}|=|H^\star|=\widetilde m_{\mathrm{op}}(\mathcal{M}_{\mathcal{S}})$. Each operation in $H^\star$ lies
in some mismatch $M_{(u,w,v)}$, and the edge removals in $M_{(u,w,v)}$ act on the legs $\{u,w\}$ and $\{w,v\}$. Hence
each pair in $E_{\mathrm{conflict}}$ is a leg of a forbidden triple, and
\[
E_{\mathrm{conflict}}\subseteq\mathrm{Skel}(\mathcal{G}_s)\cap\mathrm{Skel}(\mathcal{G}_t).
\]
Let $\mathcal{D}_s$ be a DAG in the equivalence class $[\mathcal{G}_s]$. Let $\mathcal{D}_M$ be the subgraph of $\mathcal{D}_s$ obtained by
removing the edges on the pairs in
\[
E_{\mathrm{conflict}}\cup\bigl(\mathrm{Skel}(\mathcal{G}_s) \setminus \mathrm{Skel}(\mathcal{G}_t)\bigr).
\]
Since a subgraph of a DAG is
a DAG, let $\mathcal{G}_M$ be the CPDAG representing the equivalence class of $\mathcal{D}_M$.
 
We show that $\mathcal{G}_M$ is a common lower bound of $\mathcal{G}_s$ and $\mathcal{G}_t$:
\begin{itemize}
\item $\mathcal{G}_M \preceq \mathcal{G}_s$: Since $\mathcal{D}_M$ is a subgraph of $\mathcal{D}_s$, Fact~3 gives
$\mathcal{G}_M \preceq \mathcal{G}_s$.
\item $\mathcal{G}_M \preceq \mathcal{G}_t$: Apply Lemma~\ref{lem:tree-inclusion} with $K$ a DAG in the class of $\mathcal G_t$ and $L=\mathcal D_M$. The skeleton $\mathrm{Skel}(\mathcal{G}_t)$ is a forest, and $\mathrm{Skel}(\mathcal D_M)\subseteq\mathrm{Skel}(\mathcal G_s)$ has no cycles. By
Lemma~\ref{lem:tree-inclusion}, verifying the basic conditions (a), (b), and (c) suffices:
\begin{enumerate}
\item[(a)] $\mathrm{Skel}(\mathcal{G}_M) \subseteq \mathrm{Skel}(\mathcal{G}_t)$ holds by construction, since all edges in
$\mathrm{Skel}(\mathcal{G}_s) \setminus \mathrm{Skel}(\mathcal{G}_t)$ were removed.
\item[(b)] Let $u \to w \leftarrow v$ be a v-structure in $\mathcal{G}_M$. Both arrows are edges of $\mathcal D_s$, since $\mathcal D_M$ is a subgraph of $\mathcal D_s$. The vertices $u$ and $v$ are nonadjacent in $\mathcal D_s$, since $\mathrm{Skel}(\mathcal G_s)$ has no cycles. Hence $u \to w \leftarrow v$ is a v-structure in $\mathcal{G}_s$. Suppose for
contradiction that $u \to w \leftarrow v$ is not a v-structure in $\mathcal{G}_t$. Both legs lie in $\mathrm{Skel}(\mathcal G_M)\subseteq\mathrm{Skel}(\mathcal{G}_s)\cap\mathrm{Skel}(\mathcal{G}_t)$. Since $\mathcal{G}_t$ is a polytree, $\mathcal G_t$ has no shielding
edge $u - v$, and neither does $\mathcal G_s$. Hence $(u, w, v) \in \mathcal{S}(\mathcal{G}_s, \mathcal{G}_t)$. However, $H^\star$ hits
$M_{(u,w,v)}$ with an edge removal, and that edge removal acts on a leg of $(u,w,v)$. Hence $E_{\mathrm{conflict}}$
contains a leg of $(u,w,v)$, and that leg is absent from $\mathcal{G}_M$. The missing leg contradicts the presence of the v-structure
$u \to w \leftarrow v$ in $\mathcal{G}_M$. Thus every v-structure in $\mathcal{G}_M$ exists in $\mathcal{G}_t$.
\item[(c)] Let $u \to w \leftarrow v$ be a v-structure in $\mathcal{G}_t$ whose legs exist in $\mathcal{G}_M$. In $\mathcal D_M$, the legs carry the orientations of $\mathcal{D}_s$, and $u$ and $v$ are nonadjacent in $\mathcal D_s$. Hence the triple is a v-structure of $\mathcal G_M$ exactly when the triple is a v-structure of $\mathcal G_s$. Suppose $\mathcal{G}_M$ oriented this triple as a non-v-structure.
Then $\mathcal{G}_s$ also orients the triple as a non-v-structure. Since $\mathcal{G}_s$ has no cycle to shield $u$
and $v$, $(u, w, v)$ would belong to $\mathcal{S}(\mathcal{G}_s, \mathcal{G}_t)$. As in (b), one of its legs would then lie in $E_{\mathrm{conflict}}$ and would have been
deleted. Since both legs remain in $\mathcal{G}_M$, no conflict exists, and $u \to w \leftarrow v$ is a v-structure of $\mathcal{G}_M$.
\end{enumerate}
Thus $\mathcal{G}_M \preceq \mathcal{G}_t$ by Lemma~\ref{lem:tree-inclusion}.
\end{itemize}
 
The skeleton of $\mathcal{G}_M$ is
\[
\mathrm{Skel}(\mathcal{G}_M)=\bigl(\mathrm{Skel}(\mathcal{G}_s)\cap\mathrm{Skel}(\mathcal{G}_t)\bigr)\setminus E_{\mathrm{conflict}}.
\]
Since
$E_{\mathrm{conflict}}\subseteq\mathrm{Skel}(\mathcal{G}_s)\cap\mathrm{Skel}(\mathcal{G}_t)$ and $|E_{\mathrm{conflict}}|=\widetilde m_{\mathrm{op}}(\mathcal{M}_{\mathcal{S}})$,
\[
\mathrm{rank}(\mathcal{G}_M)=|\mathrm{Skel}(\mathcal{G}_s)\cap\mathrm{Skel}(\mathcal{G}_t)|
-\widetilde m_{\mathrm{op}}(\mathcal{M}_{\mathcal{S}}).
\]
The CPDAG poset is graded by rank, and $\mathcal{G}_M$ lies below both $\mathcal{G}_s$ and $\mathcal{G}_t$. Hence there is a
down-up path $P^\star$ that descends from $\mathcal{G}_s$ to $\mathcal{G}_M$ and then ascends from $\mathcal{G}_M$ to
$\mathcal{G}_t$. Each step of $P^\star$ changes the rank by one. Using $\mathrm{rank}(\mathcal{G})=|\mathrm{Skel}(\mathcal{G})|$, the length of $P^\star$ is
\[
\begin{aligned}
&\mathrm{rank}(\mathcal{G}_s)+\mathrm{rank}(\mathcal{G}_t)-2\,\mathrm{rank}(\mathcal{G}_M)\\
&\quad=|\mathrm{Skel}(\mathcal{G}_s)|+|\mathrm{Skel}(\mathcal{G}_t)|-2|\mathrm{Skel}(\mathcal{G}_s)\cap\mathrm{Skel}(\mathcal{G}_t)|+2\widetilde m_{\mathrm{op}}(\mathcal{M}_{\mathcal{S}})\\
&\quad=|\Delta|+2\widetilde m_{\mathrm{op}}(\mathcal{M}_{\mathcal{S}}).
\end{aligned}
\]
The last equality uses
$|\mathrm{Skel}(\mathcal{G}_s)|+|\mathrm{Skel}(\mathcal{G}_t)|-2|\mathrm{Skel}(\mathcal{G}_s)\cap\mathrm{Skel}(\mathcal{G}_t)|=|\Delta|$.
 
Every down-up path is a path, so
$d(\mathcal{G}_s,\mathcal{G}_t)\le d_{\downarrow\uparrow}(\mathcal{G}_s,\mathcal{G}_t)$. The path $P^\star$ is a down-up
path, so $d_{\downarrow\uparrow}(\mathcal{G}_s,\mathcal{G}_t)\le|\Delta|+2\widetilde m_{\mathrm{op}}(\mathcal{M}_{\mathcal{S}})$.
By Lemma~\ref{lem:forbidden-mismatch}, every member of $\mathcal{M}_{\mathcal{S}}$ is a mismatch for $(\mathcal{G}_s,\mathcal{G}_t)$. Theorem~\ref{thm:mismatch} with $\mathcal{M}=\mathcal{M}_{\mathcal{S}}$ therefore gives
$|\Delta|+2\widetilde m_{\mathrm{op}}(\mathcal{M}_{\mathcal{S}})\le d(\mathcal{G}_s,\mathcal{G}_t)$. Combining the three
inequalities gives
\[
|\Delta|+2\widetilde m_{\mathrm{op}}(\mathcal{M}_{\mathcal{S}})
\le d(\mathcal{G}_s,\mathcal{G}_t)
\le d_{\downarrow\uparrow}(\mathcal{G}_s,\mathcal{G}_t)
\le |\Delta|+2\widetilde m_{\mathrm{op}}(\mathcal{M}_{\mathcal{S}}).
\]
All three inequalities are therefore equalities:
\[
d(\mathcal{G}_s,\mathcal{G}_t)=d_{\downarrow\uparrow}(\mathcal{G}_s,\mathcal{G}_t)
=|\Delta|+2\widetilde m_{\mathrm{op}}(\mathcal{M}_{\mathcal{S}}).
\]
In particular, the down-up path $P^\star$ is a shortest path. The first two claims follow. Not every shortest path is necessarily a down-up path.

To show every graph on a down-up path from $\mathcal{G}_s$ to $\mathcal{G}_s$ is a polytree, let
\[
\mathcal{G}_s=\mathcal{G}_0\gtrdot\cdots\gtrdot\mathcal{G}_{i^*}\lessdot\cdots\lessdot\mathcal{G}_k=\mathcal{G}_t
\]
be any down-up path, for example $P^\star$. Each covering relation is an instance of $\preceq$, and $\preceq$ is transitive.
Hence $\mathcal{G}_i\preceq\mathcal{G}_s$ for $i\le i^*$, and $\mathcal{G}_i\preceq\mathcal{G}_t$ for $i\ge i^*$. By Fact~2 at the start of Appendix~\ref{app:distance-proofs}, $\mathcal{H}\preceq\mathcal{G}$ implies
$\mathrm{Skel}(\mathcal{H})\subseteq\mathrm{Skel}(\mathcal{G})$. Therefore
\[
\mathrm{Skel}(\mathcal{G}_i)\subseteq\mathrm{Skel}(\mathcal{G}_s)\ \text{ for } i\le i^*,
\qquad
\mathrm{Skel}(\mathcal{G}_i)\subseteq\mathrm{Skel}(\mathcal{G}_t)\ \text{ for } i\ge i^*.
\]
The skeletons of $\mathcal{G}_s$ and $\mathcal{G}_t$ have no cycles, and a subgraph of a graph without cycles has no
cycles. Hence every $\mathcal{G}_i$ is a polytree. In particular, the shortest down-up path $P^\star$ traverses through
polytrees.
\end{proof}

\section{Proofs for Section~\ref{sec:computation}}\label{app:computation-proofs}

\subsection{Consistency of mismatch heuristic}\label{app:consistency-proof}
 
For the rest of this subsection, fix the target $\mathcal{G}_t$, and let $\mathcal{M}(\mathcal{G})$ be the collection
of all mismatches for $(\mathcal{G},\mathcal{G}_t)$. Whether a step is a detour depends only on $\mathcal{G}_t$. A step on
the pair $e$ is a detour if the step moves $e$ away from its state in $\mathcal{G}_t$. A detour step on $e$ performs $\mathrm{ER}(e)$ if
$e\in\mathrm{Skel}(\mathcal{G}_t)$ and $\mathrm{EA}(e)$ otherwise.
 
\begin{lemma}\label{lem:consistency-steps}
Let $\mathcal{G}'$ be a neighbor of $\mathcal{G}$, let $e$ be the pair in which their skeletons differ, and let
\[
\Delta\widetilde m_{\mathrm{op}}=\widetilde m_{\mathrm{op}}(\mathcal{M}(\mathcal{G}))
-\widetilde m_{\mathrm{op}}(\mathcal{M}(\mathcal{G}')).
\]
Then
\begin{enumerate}
\item[(i)] if the step from $\mathcal{G}$ to $\mathcal{G}'$ adds $e\notin\mathrm{Skel}(\mathcal{G}_t)$, then
$\Delta\widetilde m_{\mathrm{op}}\le1$;
\item[(ii)] if the step deletes $e\in\mathrm{Skel}(\mathcal{G}_t)$, then $\Delta\widetilde m_{\mathrm{op}}\le1$;
\item[(iii)] if the step adds $e\in\mathrm{Skel}(\mathcal{G}_t)$, then $\Delta\widetilde m_{\mathrm{op}}\le0$;
\item[(iv)] if the step deletes $e\notin\mathrm{Skel}(\mathcal{G}_t)$, then $\Delta\widetilde m_{\mathrm{op}}\le0$.
\end{enumerate}
\end{lemma}
 
\begin{proof}
Let $P'$ be a path from $\mathcal{G}'$ to $\mathcal{G}_t$. The step from $\mathcal{G}$ to $\mathcal{G}'$ followed by
$P'$ is a path from $\mathcal{G}$ to $\mathcal{G}_t$. This path therefore executes an operation of every $M\in\mathcal{M}(\mathcal{G})$.
 
\emph{Cases (i) and (ii).} The first step of this path performs the detour operation $o$ on $e$, namely $\mathrm{EA}(e)$ in case (i) and
$\mathrm{ER}(e)$ in case (ii). If $M\in\mathcal{M}(\mathcal{G})$ does not contain $o$, then $P'$ itself executes an
operation of $M$. Since $P'$ was arbitrary, $M\in\mathcal{M}(\mathcal{G}')$. Adding $o$ to a minimum hitting set for
$\mathcal{M}(\mathcal{G}')$ therefore gives a hitting set for $\mathcal{M}(\mathcal{G})$. Hence $\widetilde
m_{\mathrm{op}}(\mathcal{M}(\mathcal{G}))\le1+\widetilde m_{\mathrm{op}}(\mathcal{M}(\mathcal{G}'))$ and
$\Delta\widetilde m_{\mathrm{op}}\le1$.
 
\emph{Cases (iii) and (iv).} The step moves $e$ to its state in $\mathcal{G}_t$, so the first step of this path is not a detour. Hence $P'$
itself executes an operation of every $M\in\mathcal{M}(\mathcal{G})$, and
$\mathcal{M}(\mathcal{G})\subseteq\mathcal{M}(\mathcal{G}')$. A hitting set for $\mathcal{M}(\mathcal{G}')$ also hits
every member of $\mathcal{M}(\mathcal{G})$. Hence $\widetilde m_{\mathrm{op}}(\mathcal{M}(\mathcal{G}))\le\widetilde
m_{\mathrm{op}}(\mathcal{M}(\mathcal{G}'))$ and $\Delta\widetilde m_{\mathrm{op}}\le0$.
\end{proof}
 
 \begin{proof}[Proof of Proposition~\ref{prop:consistency}]
\emph{All mismatches.}
The path of length zero executes no operation, so
$\mathcal M(\mathcal G_t)=\varnothing$.
Hence $h(\mathcal G_t)=0$.
For the distance search, $S_D(\mathcal G)=\mathcal N(\mathcal G)$.
We therefore prove
$h(\mathcal G)-h(\mathcal G')\le1$ for neighboring CPDAGs.

Let $\mathcal G'$ be a neighbor of $\mathcal G$, and let $e$ be
the pair in which their skeletons differ. Write
\[
\Delta\mathrm{sk}
=|\mathrm{Skel}(\mathcal G)\triangle\mathrm{Skel}(\mathcal G_t)|
-|\mathrm{Skel}(\mathcal G')\triangle\mathrm{Skel}(\mathcal G_t)|,
\]
and let $\Delta\widetilde m_{\mathrm{op}}$ be as in
Lemma~\ref{lem:consistency-steps}.
Only the status of $e$ changes, so $\Delta\mathrm{sk}\in\{-1,1\}$.

If $\Delta\mathrm{sk}=1$, the step adds a pair of
$\mathrm{Skel}(\mathcal G_t)$ or deletes a pair outside that skeleton.
Lemma~\ref{lem:consistency-steps}(iii)--(iv) gives
$\Delta\widetilde m_{\mathrm{op}}\le0$, so
\[
h(\mathcal G)-h(\mathcal G')
=\Delta\mathrm{sk}+2\Delta\widetilde m_{\mathrm{op}}
\le1.
\]
If $\Delta\mathrm{sk}=-1$, the step deletes a pair of
$\mathrm{Skel}(\mathcal G_t)$ or adds a pair outside that skeleton.
Lemma~\ref{lem:consistency-steps}(i)--(ii) gives
$\Delta\widetilde m_{\mathrm{op}}\le1$, so
\[
h(\mathcal G)-h(\mathcal G')
=\Delta\mathrm{sk}+2\Delta\widetilde m_{\mathrm{op}}
\le-1+2=1.
\]

\emph{Forbidden-triple mismatches.}
Write
\[
\mathcal M_{\mathcal S}(\mathcal G)
:=\{M_T:T\in\mathcal S(\mathcal G,\mathcal G_t)\}.
\]
We show that the four bounds of Lemma~\ref{lem:consistency-steps}
also hold with $\mathcal M_{\mathcal S}$ in place of $\mathcal M$.
Let $\mathcal G'$ be a neighbor of $\mathcal G$, with skeletons
differing on the pair $e$.
By Fact~1, there are DAG representatives $\mathcal D$ and
$\mathcal D'$ that differ only in the edge on $e$.
Fix $T=(u,w,v)\in\mathcal S(\mathcal G,\mathcal G_t)$.
\begin{itemize}
\item If $e$ is none of the three pairs of $T$, then
$\mathcal D$ and $\mathcal D'$ agree on those pairs.
The skeleton and v-structure status on $T$ are unchanged, so
$T\in\mathcal S(\mathcal G',\mathcal G_t)$.
\item If $e$ is a leg of $T$, then
$e\in\mathrm{Skel}(\mathcal G)\cap\mathrm{Skel}(\mathcal G_t)$.
The step therefore deletes $e$ and falls under
Lemma~\ref{lem:consistency-steps}(ii).
The detour operation $\mathrm{ER}(e)$ belongs to $M_T$.
\item If $e$ is the shield of $T$, then
$e\notin\mathrm{Skel}(\mathcal G)\cup\mathrm{Skel}(\mathcal G_t)$.
The step therefore adds $e$ and falls under
Lemma~\ref{lem:consistency-steps}(i).
The detour operation $\mathrm{EA}(e)$ belongs to $M_T$.
\end{itemize}
In cases (i)--(ii) of Lemma~\ref{lem:consistency-steps},
every member of $\mathcal M_{\mathcal S}(\mathcal G)$ either
contains the detour operation of the step or belongs to
$\mathcal M_{\mathcal S}(\mathcal G')$.
In cases (iii)--(iv),
$\mathcal M_{\mathcal S}(\mathcal G)
\subseteq\mathcal M_{\mathcal S}(\mathcal G')$.
The hitting-set arguments in the proof of the lemma therefore
give the same four bounds for $\mathcal M_{\mathcal S}$.

Since $\mathcal S(\mathcal G_t,\mathcal G_t)=\varnothing$,
the corresponding heuristic also satisfies $h(\mathcal G_t)=0$.
The preceding two-case calculation then gives
$h(\mathcal G)-h(\mathcal G')\le1$ for neighboring CPDAGs.
Thus the forbidden-triple bound of \citet{taeb2025distance}
is also consistent.

For either collection, summing the neighboring inequalities
along a shortest path from $\mathcal G$ to $\mathcal H$ gives
\[
h(\mathcal G)\le d(\mathcal G,\mathcal H)+h(\mathcal H),
\]
the consistency condition of \citet{hart1968formal}.
\end{proof}
 
\subsection{Certification of mismatches}\label{app:mismatch-certification}
We use the restriction $\mathcal{G}|_W$ and the distance $d_W$ of Section~\ref{sec:structured}.
 
\begin{lemma}[Restriction]\label{lem:restriction}
If $\mathcal{G}$ and $\mathcal{G}'$ are neighbors whose skeletons differ in the pair $e$, then $\mathcal{G}|_W$ and
$\mathcal{G}'|_W$ are neighbors when $e \subseteq W$ and are equal otherwise. Hence the operations on pairs inside $W$
along any path from $\mathcal{G}_s$ to $\mathcal{G}_t$ form a path from $\mathcal{G}_s|_W$ to $\mathcal{G}_t|_W$, and
there are at least $d_W(\mathcal{G}_s|_W, \mathcal{G}_t|_W)$ of them.
\end{lemma}
 
\begin{proof}[Proof of Lemma~\ref{lem:restriction}]
\emph{Well defined.} Let $\mathcal{D}$ and $\mathcal{D}'$ be DAGs in the same class. A triple inside $W$ is unshielded in the induced subgraph exactly when the triple is unshielded in the
DAG. Hence $\mathcal D[W]$ and $\mathcal D'[W]$ have the same skeleton and the
same v-structures, and $\mathcal{G}|_W$ is well defined.
 
\emph{Neighbors.} Let $\mathcal G$ and $\mathcal G'$ be neighbors, say $\mathcal{G} \lessdot \mathcal{G}'$. By Fact~1, there are DAGs $\mathcal{D}$ and
$\mathcal{D}' = \mathcal{D} + e$ in their classes. If $e \not\subseteq W$, then $\mathcal D'[W]=\mathcal D[W]$. Otherwise,
$\mathcal D'[W]=\mathcal D[W]+e$. By Fact~3, $\mathcal{G}|_W \preceq \mathcal{G}'|_W$, and the ranks of $\mathcal{G}|_W$ and $\mathcal{G}'|_W$
differ by one. The poset over $W$ is graded, so $\mathcal{G}|_W$ and $\mathcal{G}'|_W$ are neighbors.
 
\emph{Paths.} Along a path, consecutive restrictions
are therefore equal or neighbors. Removing repetitions gives a path from $\mathcal{G}_s|_W$ to $\mathcal{G}_t|_W$
whose steps are the operations on pairs inside $W$. By the definition of $d_W$, this path has at least $d_W(\mathcal{G}_s|_W, \mathcal{G}_t|_W)$ steps.
\end{proof}
 
\begin{proposition}[Local mismatches]\label{lem:local-mismatch}
Let $W\subseteq V$. If
\[
d_W(\mathcal{G}_s|_W,\mathcal{G}_t|_W)
>|\mathrm{Skel}(\mathcal{G}_s|_W)\triangle\mathrm{Skel}(\mathcal{G}_t|_W)|,
\]
then the set $M_W$ of detour operations on pairs inside $W$ is a mismatch for $(\mathcal{G}_s,\mathcal{G}_t)$.
\end{proposition}
 
\begin{proof}
By Lemma~\ref{lem:restriction}, the steps of a path from $\mathcal{G}_s$ to $\mathcal{G}_t$ that act on pairs inside $W$
form a path from $\mathcal{G}_s|_W$ to $\mathcal{G}_t|_W$, with at least $d_W(\mathcal{G}_s|_W,\mathcal{G}_t|_W)$ steps.
By assumption, this number exceeds the skeleton difference on $W$. By Equation~\eqref{eq:path-detour-operations} over $W$, a path with more steps than the skeleton difference on $W$ has
a detour step. That step moves a pair inside $W$ away from its state in $\mathcal{G}_t|_W$, which is its state in $\mathcal{G}_t$. The corresponding step of the original path therefore performs an operation of
$M_W$.
\end{proof}

Each pattern of Section~\ref{sec:distance-computation} gives a mismatch. A pattern occurs on a vertex set $W$ when the restrictions $\mathcal G_s|_W$ and $\mathcal G_t|_W$ are the two graphs of the pattern, up to a relabeling of the vertices. The pattern of
Figure~\ref{fig:mismatch-examples}(a) occurs on $W$ exactly when $W$ carries a forbidden triple, with either graph containing the v-structure. By Lemma~\ref{lem:forbidden-mismatch}, the set $M_1$ is then a mismatch. For the
four-vertex patterns, an enumeration of the 185 CPDAGs on four vertices gives distances 3, 4, 4 and 4 for the pairs in
Figures~\ref{fig:mismatch-examples}(b) and~\ref{fig:additional-mismatches}(a)--(c). The corresponding skeleton differences are 1, 2,
2 and 0. Each distance exceeds its skeleton difference, so Proposition~\ref{lem:local-mismatch} applies. Since $d_W$ is symmetric, Proposition~\ref{lem:local-mismatch} applies whichever of the two graphs plays the role
of $\mathcal{G}_t$. The resulting mismatch $M_W$ consists of the detour operations on pairs inside $W$, relative to $\mathcal{G}_t$. The sets $M_2,\dots,M_5$ of Section~\ref{sec:distance-computation} equal $M_W$ when $\mathcal{G}_t|_W$ is the lower graph of the figure. When $\mathcal{G}_t|_W$ is the upper graph, $\mathrm{ER}$ and $\mathrm{EA}$ are exchanged on the pairs in which the two skeletons differ.

\subsection{Disjoint stray edges}\label{app:disjoint-stray}
 
\begin{lemma}\label{lem:stray-dag}
Let $\mathcal{D}$ be a DAG containing a set of disjoint stray edges $\mathcal{E}_{\mathrm{stray}}$. Let $\mathcal{D}'$ be
the subgraph obtained by removing $\mathcal{E}_{\mathrm{stray}}$ from $\mathcal{D}$. Let $\mathrm{Op}$ be any valid
operation on an edge $h$ whose endpoints lie entirely within the conflict set $V_\Delta$. If $\mathrm{Op}$ is valid in
$\mathcal{D}$, then $\mathrm{Op}$ is also valid in $\mathcal{D}'$.
\end{lemma}
 
\begin{proof}
We examine the three types of operations.
 
\emph{Deletion.} Consider the deletion of an edge $h$. If $h \in E(\mathcal{D})$ and its
endpoints are in $V_\Delta$, then $h \notin \mathcal{E}_{\mathrm{stray}}$. Thus $h \in E(\mathcal{D}')$, and deleting an
edge from a DAG always yields a DAG.
 
\emph{Addition.} Validity here depends on acyclicity. If adding $h$ to $\mathcal{D}$ does not create a
cycle, then adding $h$ to a subgraph $\mathcal{D}' \subset \mathcal{D}$ cannot create a cycle.
 
\emph{Covered edge reversal.} Consider the reversal of a covered edge $h = a \to b$ in $\mathcal{D}$. The covering condition requires
\[
\mathrm{Pa}_{\mathcal{D}}(b) = \mathrm{Pa}_{\mathcal{D}}(a) \cup \{a\}.
\]
By the definition of disjoint stray edges, any
edge $e = \{u,v\} \in \mathcal{E}_{\mathrm{stray}}$ satisfies $\{u,v\} \cap V_\Delta = \emptyset$. Since
$a, b \in V_\Delta$, no stray edge is incident to $a$ or $b$. So removing $\mathcal{E}_{\mathrm{stray}}$ does not change
the parent sets of $a$ or $b$:
\[
\mathrm{Pa}_{\mathcal{D}'}(a) = \mathrm{Pa}_{\mathcal{D}}(a),
\qquad
\mathrm{Pa}_{\mathcal{D}'}(b) = \mathrm{Pa}_{\mathcal{D}}(b).
\]
Therefore, the covering condition holds in
$\mathcal{D}'$. The operation remains a covered edge reversal and is therefore valid in the subgraph.
\end{proof}
 
\begin{proof}[Proof of Proposition~\ref{prop:stray}]
We proceed by contradiction. Assume there exists a shortest path $P = (\mathcal{G}_0, \dots, \mathcal{G}_k)$ from
$\mathcal{G}_s = \mathcal{G}_0$ to $\mathcal{G}_t = \mathcal{G}_k$ in the poset $\mathcal{L}$ such that some intermediate graph contains a disjoint stray edge in
its skeleton.
 
By Lemma~\ref{lem:dag-cpdag}, there exists a valid sequence of DAGs $\mathcal{D}_0, \dots, \mathcal{D}_m$ representing
this path. Let $\mathcal{E}_{\mathrm{stray}}$ be the set of all disjoint stray edges appearing in any graph in this sequence. We
construct a new sequence $\widetilde{\mathcal{D}}_0, \dots, \widetilde{\mathcal{D}}_m$, where
\[
\widetilde{\mathcal{D}}_j = (V, E(\mathcal{D}_j) \setminus \mathcal{E}_{\mathrm{stray}}).
\]
By Lemma~\ref{lem:stray-dag}, all operations in the original sequence that acted on the conflict set $V_\Delta$ remain valid
in the new sequence. Operations that added or removed edges in $\mathcal{E}_{\mathrm{stray}}$ are omitted.
 
Since $\mathcal{E}_{\mathrm{stray}} \cap \mathrm{Skel}(\mathcal{G}_s) = \emptyset$ and
$\mathcal{E}_{\mathrm{stray}} \cap \mathrm{Skel}(\mathcal{G}_t) = \emptyset$, the new sequence still starts at a
representative of $\mathcal{G}_s$ and ends at a representative of $\mathcal{G}_t$. The original path $P$ must have
included rank-changing operations to add and subsequently delete every disjoint stray edge in $\mathcal{E}_{\mathrm{stray}}$,
since these edges are absent in both $\mathcal{G}_s$ and $\mathcal{G}_t$. In the pruned sequence $\widetilde{\mathcal{D}}$,
these operations are removed. Thus, the CPDAG path corresponding to $\widetilde{\mathcal{D}}$ is strictly shorter than
$P$, which contradicts the assumption that $P$ is a shortest path.
\end{proof}

\section{Additional experiments}
\label{sec:additional_experiments}

\subsection{Confidence regions for sparse graphs}\label{app:sparse_conf} 
\begin{figure}[h]
\centering
\includegraphics[width=0.85\textwidth,draft=false]{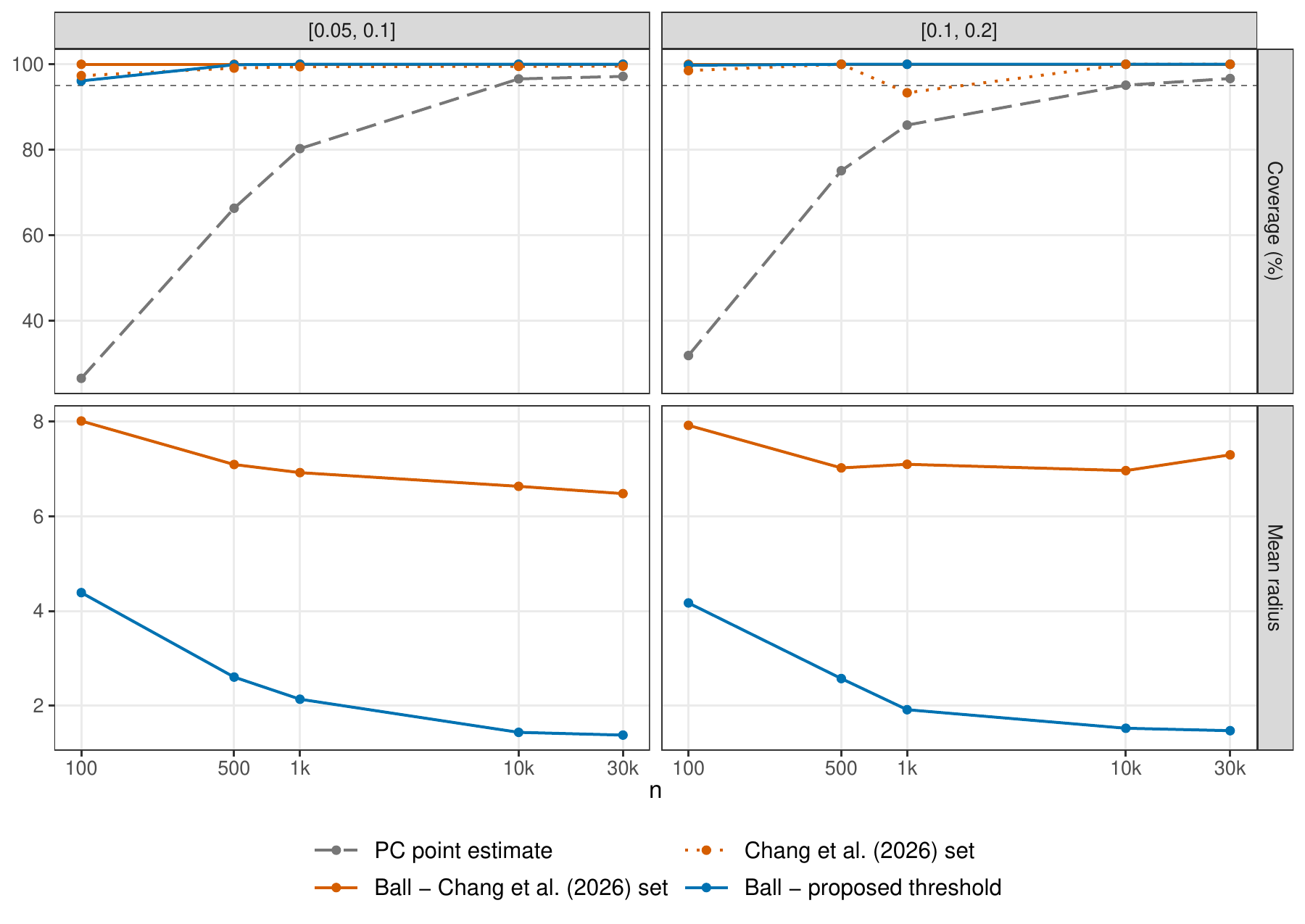}
\caption{Confidence ball results for the sparse $p=10$ regimes with minimum nonzero partial correlations in $[0.05,0.10]$ (left) and $[0.10,0.20]$ (right). Top: empirical coverage. Bottom: mean radius.}
\label{fig:confidence-ball-sparse-p10}
\end{figure}

\begin{figure}[h]
\centering
\includegraphics[width=0.85\textwidth,draft=false]{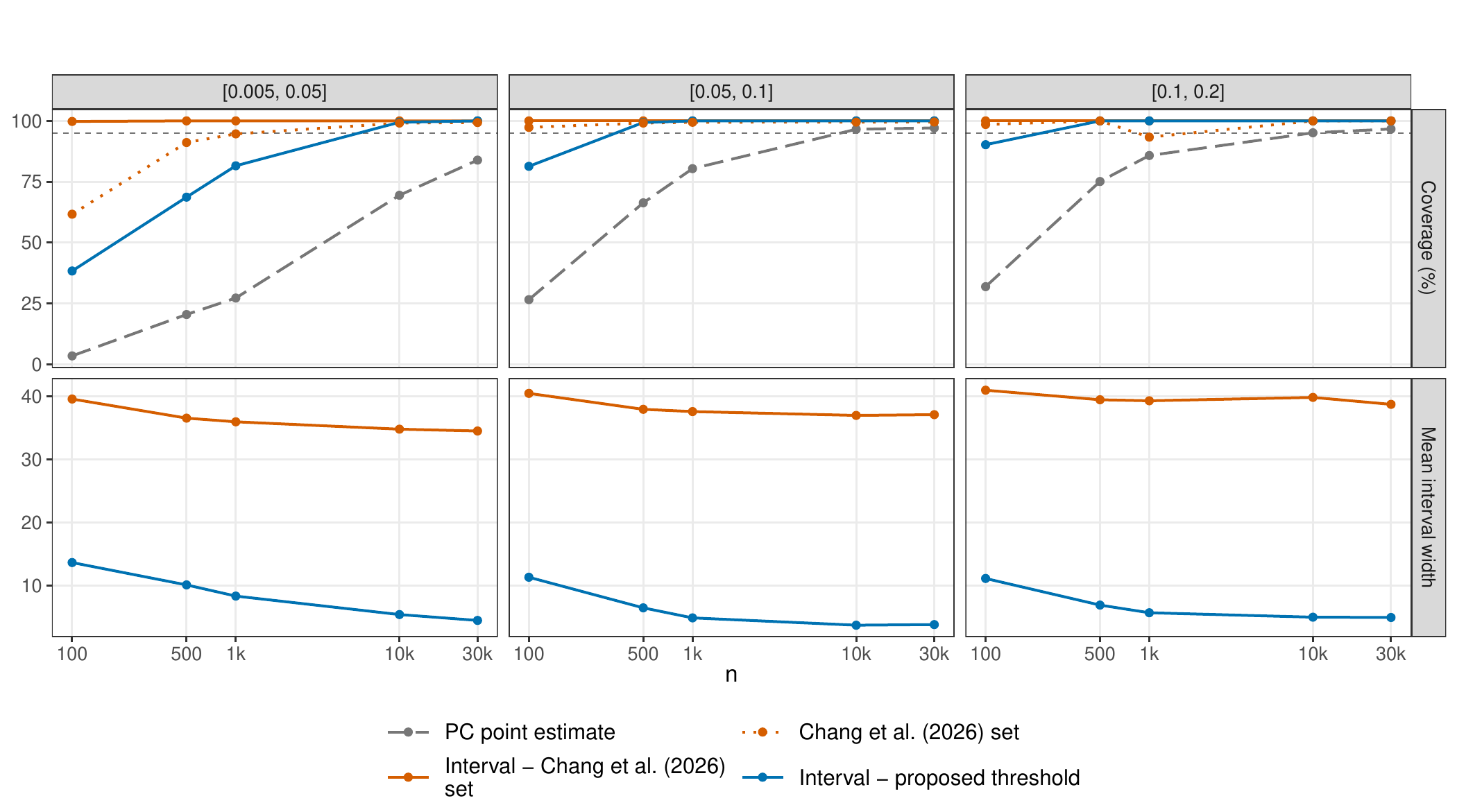}
\caption{Confidence interval results for the sparse $p=10$ regimes with minimum nonzero partial correlations in $[0.005,0.05]$ (left), $[0.05,0.10]$ (middle), and $[0.10,0.20]$ (right). Top: empirical coverage. Bottom: mean interval width. }
\label{fig:confidence-interval-sparse-p10}
\end{figure}

We sample 600 instances (20 population SEMs each with 30 independent datasets) for weak, moderate, and strong signaled sparse graphs (See Table~\ref{tab:confidence-regimes}). We construct confidence balls and intervals with our proposed threshold (See Section~\ref{sec:default}) and that of \cite{chang2026post} and report confidence ball mean radius and coverage in Figure~\ref{fig:confidence-ball-sparse-p10} and confidence interval mean width and coverage in Figure~\ref{fig:confidence-interval-sparse-p10} over these instances. Note that the confidence ball coverage and radius for the weak signal setting (minimum partial correlations in 0.005-0.05) is reported in Figure~\ref{fig:confidence-balls-sparse-p10-low-signal}. Table~\ref{tab:shared-edges-sparse} includes average common edge counts between the upper and lower graphs from the interval for each minimum partial correlation range over sample size.

\FloatBarrier

\FloatBarrier

\begin{table}[h]
\centering
\caption{Average number of edges shared by the lower and upper graphs in the sparse $p=10$ setting across sample sizes and partial-correlation ranges.}
\label{tab:shared-edges-sparse}
\begin{tabular}{lccccc}
\toprule
& $n=100$ & $n=500$ & $n=1000$ & $n=10000$ & $n=30000$ \\
\midrule
$[0.005,0.05]$ & 3.13 & 4.95 & 5.86 & 7.51 & 8.04 \\
$[0.05,0.1]$   & 3.07 & 5.09 & 6.07 & 6.83 & 6.86 \\
$[0.1,0.2]$    & 3.21 & 5.07 & 5.79 & 6.32 & 6.33 \\
\bottomrule
\end{tabular}
\end{table}

\FloatBarrier

\subsection{Additional experiments for model-oriented distance under structural proximity}\label{app:structural_prox} 
\begin{figure}[h]
\centering
\includegraphics[width=\textwidth,draft=false]{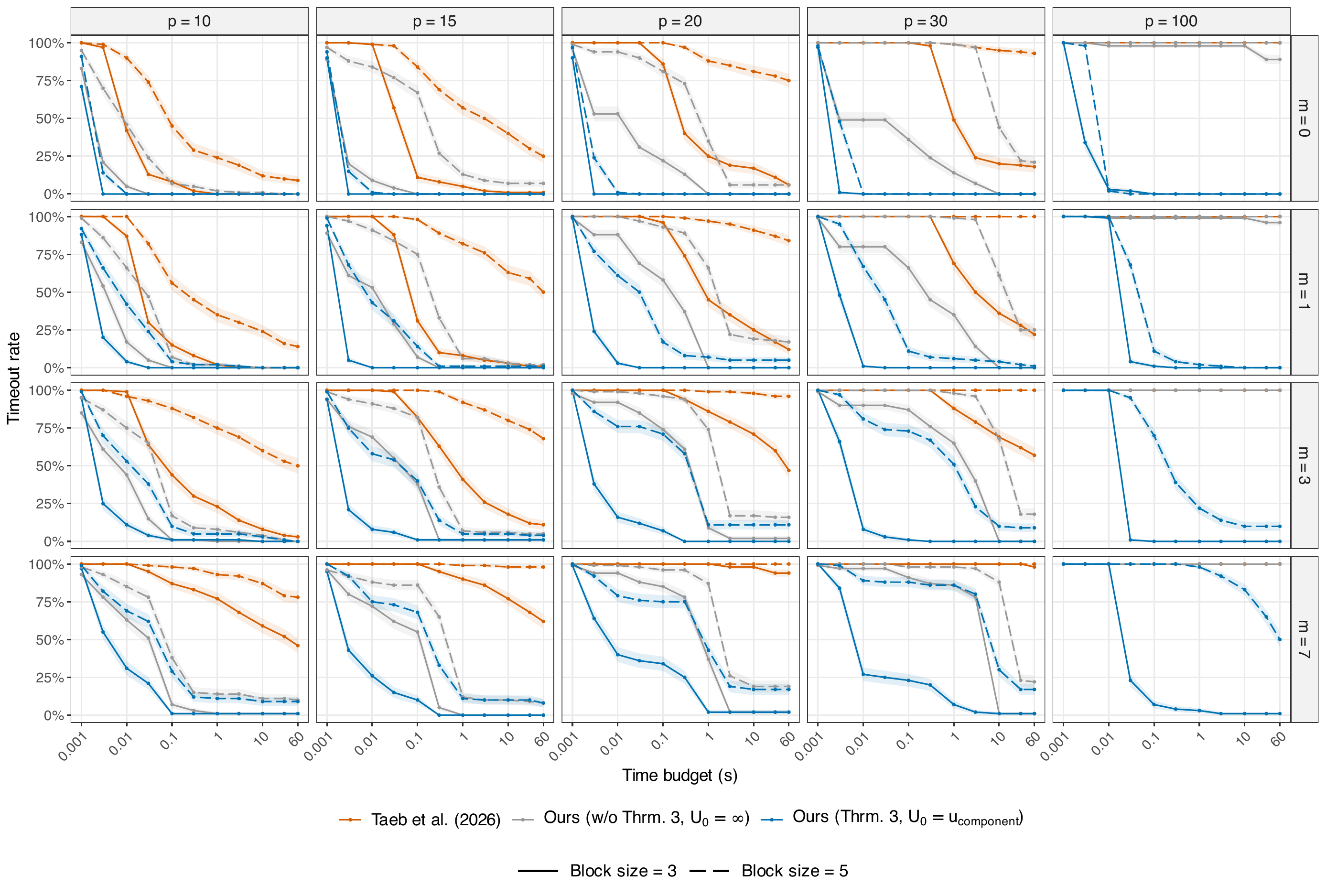}
\caption{Timeout rates for model-oriented distance searches between CPDAG pairs in the disconnected-component regime, with $q_{\mathrm{in}}=0.5$ and $m\in\{0,1,3,7\}$. Columns correspond to $p=10,15,20,30,$ and $100$; rows correspond to $m$. Solid and dashed curves use block sizes three and five, respectively. Each point is the proportion of the 100 pairs not solved within the indicated time budget; shaded regions show one standard deviation around that proportion. The curves compare the model-oriented-distance A$^\ast$ procedure of \citet{taeb2025distance} with our procedure without the disconnected-component upper bound and with our procedure using the disconnected-component upper bound.}
\label{fig:components-mpositive-timeout-qin0p5}
\end{figure}

\begin{figure}[h]
\centering
\includegraphics[width=\textwidth,draft=false]{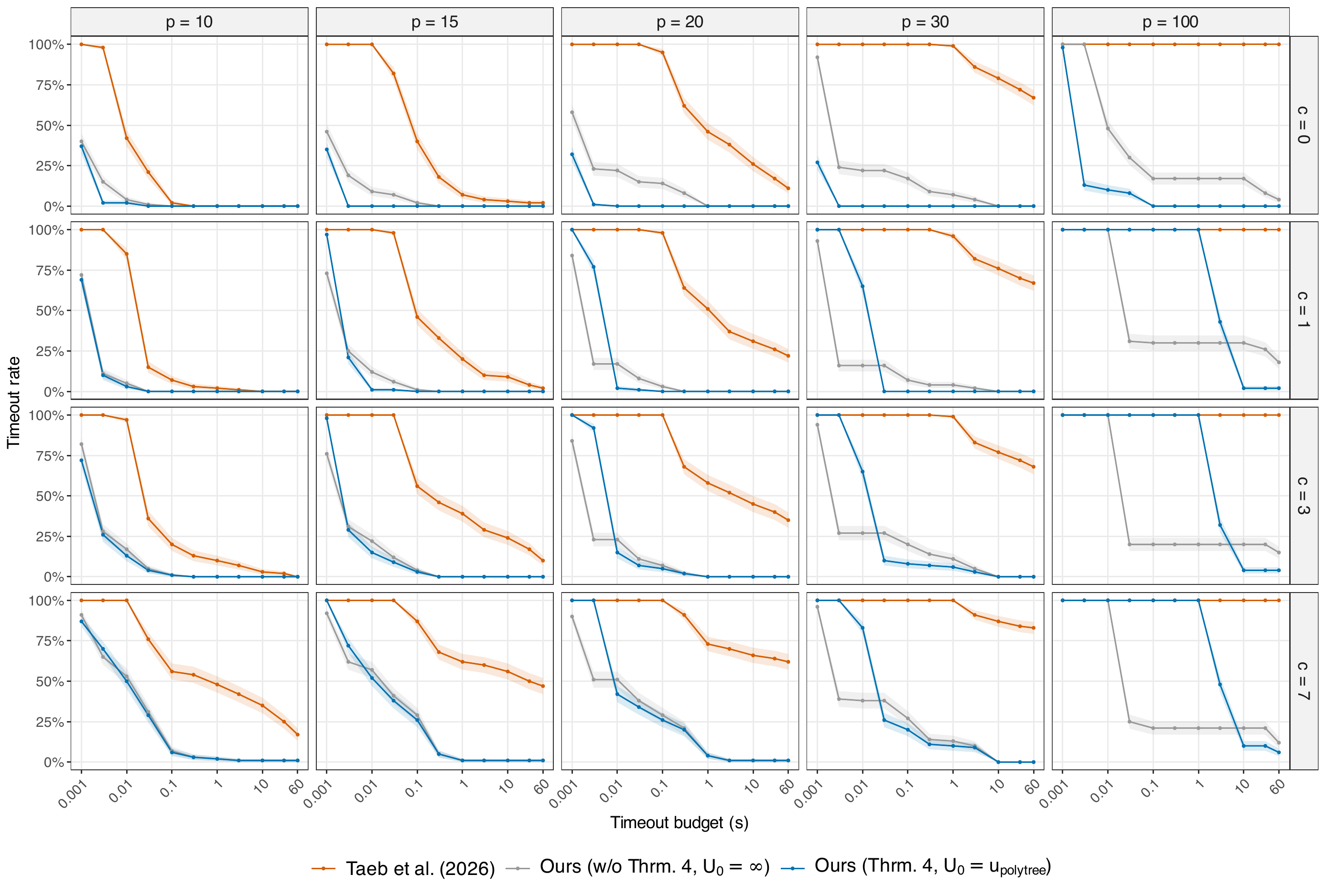}
\caption{Timeout rates for model-oriented distance searches between CPDAG pairs near the polytree class, with $q_{\mathrm{tree}}=0.5$ and $c\in\{0,1,3,7\}$. Columns correspond to $p=10,15,20,30,$ and $100$; rows correspond to $c$. Each point is the proportion of the 100 pairs not solved within the indicated time budget; shaded regions show one standard deviation around that proportion. The curves compare the model-oriented-distance A$^\ast$ procedure of \citet{taeb2025distance} with our procedure without the polytree upper bound and with our procedure using the polytree upper bound.}
\label{fig:polytree-cpositive-timeout-qtree0p5}
\end{figure}

\begin{figure}[h]
\centering
\includegraphics[width=\textwidth,draft=false]{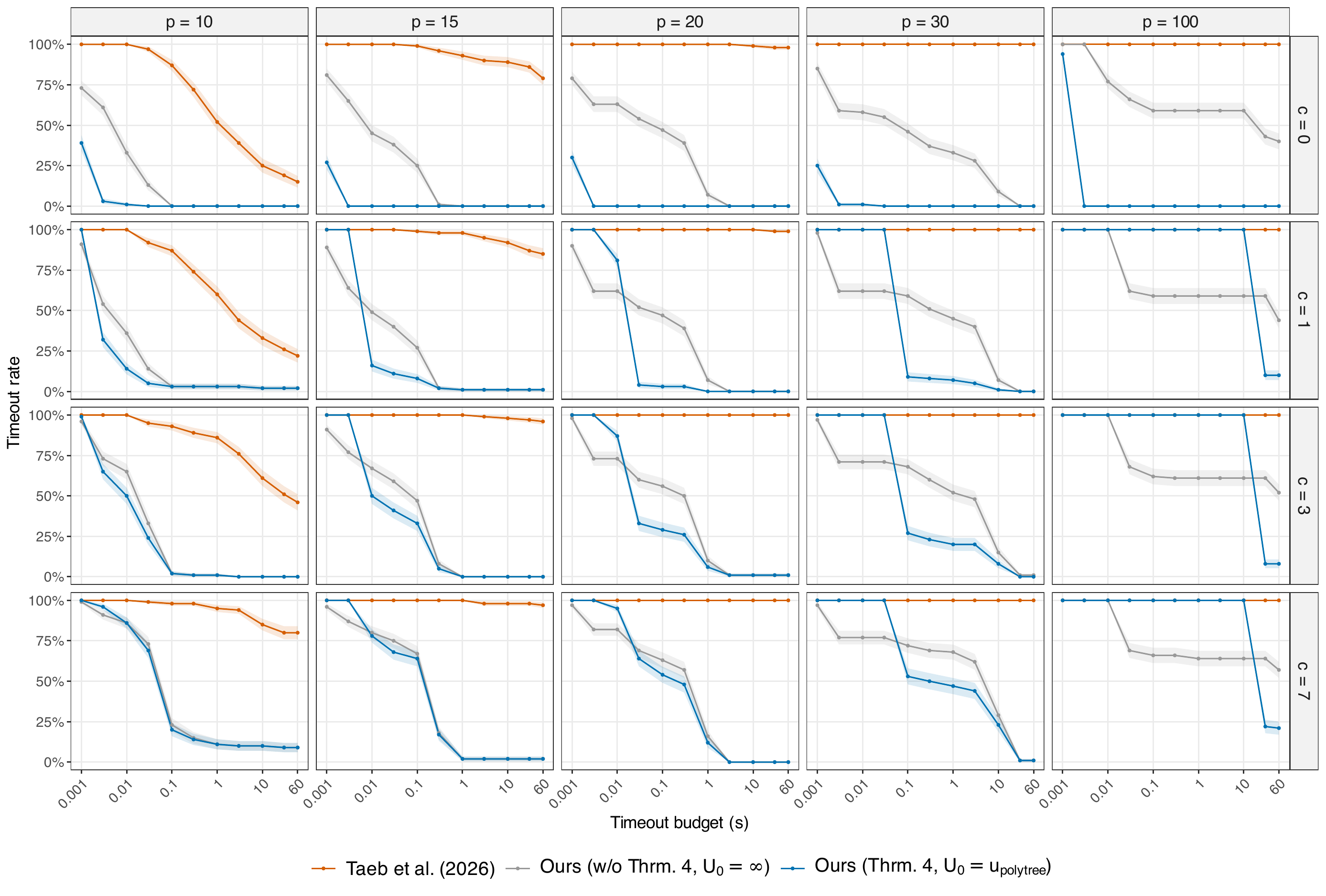}
\caption{Timeout rates for model-oriented distance searches between CPDAG pairs near the polytree class, with $q_{\mathrm{tree}}=1$ and $c\in\{0,1,3,7\}$. Columns correspond to $p=10,15,20,30,$ and $100$; rows correspond to $c$. Each point is the proportion of the 100 pairs not solved within the indicated time budget; shaded regions show one standard deviation around that proportion. The curves compare the model-oriented-distance A$^\ast$ procedure of \citet{taeb2025distance} with our procedure without the polytree upper bound and with our procedure using the polytree upper bound.}
\label{fig:polytree-cpositive-timeout-qtree1p0}
\end{figure}

We report timeout rates of our method for the model-oriented distance and that of \cite{taeb2025distance} under structural proximity in Section~\ref{sec:exp-components}. Here, we include additional timeout rates for the disconnected components regime for $q_{\text{in}}=0.5$ in Figure~\ref{fig:components-mpositive-timeout-qin0p5} and the polytree regime for $q_\text{tree}=0.5$ and $q_{\text{tree}}=1$ in Figures~\ref{fig:polytree-cpositive-timeout-qtree0p5} and ~\ref{fig:polytree-cpositive-timeout-qtree1p0}, respectively.

\FloatBarrier

\FloatBarrier

\FloatBarrier

\end{document}